\documentclass[12pt]{iopart}
\usepackage{iopams}

\usepackage{graphicx}
\usepackage{dcolumn}
\usepackage{bm}
\usepackage{hyperref}

\begin{document}

\title{Recent developments in Quantum Monte-Carlo simulations with applications for cold gases}
\author{Lode Pollet}
\address{Department of Physics, Arnold Sommerfeld Center for Theoretical Physics and Center for NanoScience, University of Munich, Theresienstrasse 37, 80333 Munich, Germany}

\eads{Lode.Pollet@lmu.de}

\date{\today}

\begin{abstract}
This is a review of recent developments in Monte Carlo methods in the field of ultra cold gases. For bosonic atoms in an optical lattice we discuss path integral Monte Carlo simulations with worm updates and show the excellent agreement with cold atom experiments. We also review recent progress in simulating bosonic systems with long-range interactions, disordered bosons, mixtures of bosons, and spinful bosonic systems.  For repulsive fermionic systems determinantal methods at half filling are sign free, but in general no sign-free method exists. We review the developments in diagrammatic Monte Carlo for the Fermi polaron problem and the Hubbard model, and show the connection with dynamical mean-field theory. We end the review with diffusion Monte Carlo for the Stoner problem in cold gases.
\end{abstract}


\tableofcontents

\section{Introduction}

Although the Schr{\"o}dinger equation describes the evolution of any quantum system, solving the many-body problem remains a daunting task. The growth of the Hilbert space is exponential with the number of particles. Exact diagonalization hence requires an exponential amount of computational resources, since an exponentially growing set of numbers needs to be stored in order to describe the state vector.

The premise of Monte Carlo methods is that, under very mild mathematical requirements, physical properties can be computed stochastically converging as $\sim 1 / \sqrt{N}$,
where $N$ is the number of independent samples, {\it irrespective of the dimension of the (Hilbert) space}. The task of a Monte Carlo developer is to find clever ways to generate independent samples for typically strongly interacting systems and arbitrary topologies. Quantum Monte Carlo simulations are classical Monte Carlo simulations of quantum systems that have been represented in a way amenable to Monte Carlo simulations. Such a quantum-to-classical mapping is necessary because the spectrum of a quantum system is a priori unknown. For example, using Feynman's path integral description for thermodynamic systems, a quantum particle is represented by a polymer or worldline, describing the propagation in imaginary time. We will see in Sec.~\ref{sec:PIMC_lattice} how this mapping of the $d$ dimensional quantum system to a classical system in $(d+1)$ dimensions is at the heart of {\it path integral Monte Carlo}, a method very successful for bosonic systems. In the best cases, an algorithm that scales linearly with the system size has been found; in other words, technical advances in computer hardware can directly be exploited to study larger systems and/or lower temperatures. We will see a couple of examples of this in Sec.~\ref{sec:PIMC_lattice} and Sec.~\ref{sec:PIMC_continuous}. However, the main drawback of stochastic methods is that they only work efficiently for a subclass of physical systems, namely those systems that are bosonic or can be mapped to bosonic systems. Generic fermionic systems cannot be sampled this way: the infamous sign problem occurs, in which a negative weight is associated with certain physical configurations. This does not prevent the application of the Monte Carlo method, but brings back the exponential scaling of the required resources with system size. That the sign problem is most likely unsolvable, has a deep physical meaning and follows from the proven NP-hardness of the problem~\cite{Troyer2005}. In such cases one resorts to approximations, or tries to find expansions (based on an analytical understanding of the problem) that converge sufficiently fast such that the sign problem remains manageable. We will also see a couple of  examples of this in Sec.~\ref{sec:diagrams} of this review. In the last section, we review the application of diffusion Monte Carlo to the Stoner problem in cold gases. 

Stochastic methods are used in every branch of research. Already in condensed matter physics alone the available literature is too vast and diverse for a single review. In order to have a better defined framework for this text, we restrict ourselves to those Monte Carlo algorithms and applications that have been used in the context of cold gases over the past 10 years. We will omit variational Monte Carlo and be very concise on diffusion Monte Carlo and auxiliary field Monte Carlo. This does not mean that these methods are unimportant or unsuccessful (the latter was recently able to identify a spin liquid in the Hubbard model on the honeycomb lattice for example~\cite{Meng2010}), but this decision is only determined by space constraints. Among the existing reviews on Monte Carlo methods we mention Ref.~\cite{Bajdic2009, Kolorenc2011} for variational and diffusion Monte Carlo methods for condensed matter physics combined with material descriptions, Ref.~\cite{Georges1992, Maier2005, Kotliar2006, Gull_review} for dynamical mean-field theory and impurity solvers, and Ref.~\cite{Assaad2001} for determinantal methods based on auxiliary field decompositions using discrete time. This review assumes that the reader is familiar with the basics of classical Monte Carlo simulations, as can be found in many excellent textbooks such as for example Refs.~\cite{LandauBinder2000, Krauth2006, JunLiu2001}. We are also focusing on three-dimensional (and occasionally two-dimensional) systems, omitting one-dimensional systems, for which the Density Matrix Renormalization Group (DMRG) is often the method of choice~\cite{Schollwoeck2011}. Ref.~\cite{Cazalilla2011} is a recent review covering the one-dimensional world of bosonic systems, including the relevant Monte Carlo methods.

The applications in this review are all situated in the field of ultra cold gases. These are dilute (a typical density is $\sim 10^{15}$ cm$^{-3}$), neutral atomic alkali-gases in a metastable state that has a lifetime of the order of a few seconds. Following the advances in laser trapping and cooling, bosonic atoms were cooled to quantum degeneracy in 1995. They were weakly interacting and well described by Bogoliubov's theory of the weakly interacting Bose gas. There are two ways to make the system strongly interacting: one is to tune a magnetic field close to a Feshbach resonance where the scattering length diverges; the other is to load them into an optical lattice where the physics is dominated by tunneling as the atoms become more localized when ramping up the lattice laser~\cite{Bloch2008}. Following the proposal of Jaksch {\it et al.}~\cite{Jaksch1998}, Greiner {\it et al.}~\cite{Greiner2002} demonstrated experimentally the superfluid to Mott insulator transition for ultracold bosonic atoms subject to an optical lattice. This demonstrated that condensed matter models could be implemented in cold gas experiments.  Compared to traditional condensed matter physics systems, ultracold gases are very clean, controllable, and have tunable system parameters~\cite{Bloch2008}.

The demonstration by Greiner {\it et al.}~\cite{Greiner2002}  has led to the paradigm of {\it quantum simulation}: an unsolvable model with competing interactions, believed to describe a real material and intractable by numerical means, can be implemented and realized in a cold gas experiment, and analyzed this way. By changing the interaction, dispersion, and density of the atoms the phase diagram of the model can be revealed. This allows to assess how well the model describes the material. An optical lattice system would thus be equivalent to an analog quantum computer, which is specifically tailored to one task (i.e., one particular model), but is more powerful than a classical computer in the sense that it can perform {\it quantum} operations (since typical scales in those systems are of the order of kHz, we do not claim that they are fast computers). Before the quantum optical lattice simulator can be trusted as a reliable device, it needs to be benchmarked and validated against known results. This is where exact numerical methods, and in practice these are quite often Monte Carlo simulations, come in: because of the cleanliness and control over optical lattice systems, their complete Hamiltonians can be simulated in cases where there is no sign problem, such as bosonic cold gases. Excellent agreement has been reached between experiments and simulations this way~\cite{Trotzky2010}, after taking a number of technical details into account. This has sparked renewed interest in developing numerically exact solutions to models in parameter regimes that were hitherto of lesser importance; for example, the three-dimensional Hubbard model for temperatures approaching the N{\'e}el temperature and interaction strengths up to $1.5$ times the bandwidth have now been fully and controllably mapped out~\cite{Fuchs2011}. For those, cluster dynamical mean-field theory simulations (so-called dynamical cluster approximation (DCA) simulations to be precise) were performed for cluster sizes of size $\sim 100$, enabling a reliable extrapolation in cluster size while still having a manageable sign problem. Now that optical lattice emulators have been validated and are gradually moving on to parameter regimes no longer tractable by traditional numerical means, will they be used as stand-alone machines?
In our view, the importance of numerical support for explanations of cold gas experiments will not diminish in the near future. Instead, we believe that computational physicists (for whom, ironically, {\it quantum simulation} is  historically understood as simulating quantum systems on classical computers) will develop a variety of new algorithms and convergence schemes tailored towards the problem of interest. The reward will be a direct test against those experiments, and the development of new algorithms that may be applicable in other fields.

It is my hope that newcomers to the field will find this text useful as a starting point in understanding the different expansion schemes, their similarities and differences, and learn how such schemes can be turned into powerful stochastic algorithms. Perhaps this text can lower the threshold to enter into this field. Codes, full algorithmic descriptions or expressions for detailed balance cannot be found in this review, but references are given. I also hope that this overview is useful for the researchers in the field of cold gases without numerical background, so that a single text can be used as an orientation platform that accurately describes the state of the art. I likewise hope that experts will  find this text useful for bringing together at first sight totally different methods, but which may ultimately lead to new ideas and crosstalk among aficionados of different subfields, and extend the double paradigm of quantum simulation to new models  and parameter regimes.

This paper is organized as follows. We start in Sec.~\ref{sec:PIMC_lattice} by reviewing path integral Monte Carlo methods for bosonic lattice systems. The key ideas behind the worm algorithm are explained. The physics of the Bose-Hubbard model is discussed in a nutshell, after which we compare the theoretically ideal case with the experimental realization of cold gases in a trap. We talk about time of flight interference images and single site in-situ resolution measurement tools. We proceed with disordered bosonic systems and the physics of the Bose glass phase. Polar molecules are next, which are systems with richer phase diagrams because of the long-range interaction between the molecules. Mixtures of bosonic atoms and bosonic atoms with an internal spin degree of freedom are also covered. In Sec.~\ref{sec:PIMC_continuous} we review path integral Monte Carlo simulations for bosonic systems in continuous space, covering the worm algorithm and the following applications: the weakly interacting Bose gas, disordered Bose gases and supersolids for atomic/molecular systems with long-range interactions. In Sec.~\ref{sec:diagrams} we leave the bosonic world for fermionic systems which in general do not have a sign-positive representation. We review determinantal methods which are often used when an additional symmetry allows for a positive expansion. In this context we discuss the critical temperature for the resonant Fermi gas at unitarity. Next we introduce diagrammatic Monte Carlo. Historically, this was first applied to the Fr{\"o}hlich Hamiltonian for electron-phonon interactions. This model is in fact sign positive, but the Fermi-polaron problem, which comes next, is not.  Also, the state of the art of diagrammatic Monte Carlo for the Hubbard model is reviewed. In the final parts of the review we show how dynamical mean-field theory (DMFT) and its cluster extensions have been used to map out the thermodynamics of the 3D Hubbard model for temperatures approaching the N{\'e}el temperature, and how DMFT can be combined with diagrammatic Monte Carlo, paving the way for future developments. Before concluding, we mention diffusion Monte Carlo and how it was used in understanding the Stoner transition (or absence thereof) for atoms on the repulsive branch of the Feshbach resonance. 


\section{Path Integral Monte Carlo: lattice models}
\label{sec:PIMC_lattice}

\subsection{Bose-Hubbard model}

Consider the Bose-Hubbard model~\cite{Fisher1989} describing scalar bosons on a lattice,
\begin{equation}
H = H_1 + H_0 = -t \sum_{\langle i,j \rangle} b_i^{\dagger}b_j + \frac{U}{2} \sum_i n_i(n_i-1) - \mu \sum_i n_i.
\label{eq:BoseHubbard}
\end{equation}
The operators $b_j$ satisfy the bosonic commutation relations, $[ b_i, b_j^{\dagger}] = \delta_{i,j}$ and zero for commutators with two creation or two annihilation operators. The
operator $n_i = b_i^{\dagger}b_i$ counts the number of bosons on site $i$.
The first term describes the hopping of bosons between neighbouring sites with tunneling amplitude $t$.
The second term describes the on-site repulsion with strength $U$, while the last term is proportional to the chemical potential $\mu$.
The latter two terms are diagonal in the Fock basis of occupation numbers, $\vert \{ n_i \} \rangle$, and are combined into $H_0$.
The kinetic term, which is a one-body operator, is not diagonal in this basis but will lead to transitions between Fock states with matrix elements $\langle \ldots n_i-1, n_j+1, \ldots \vert -tb_j^{\dagger}b_i \vert \ldots n_i, n_j \ldots \rangle = -t \sqrt{n_i(n_j+1)}$.
The lattice spacing is set to unity. Unless otherwise specified we have a cubic lattice in mind of linear size $L$ with periodic boundary conditions.

The Bose-Hubbard model has three phases, which can easily be identified in the limiting cases~\cite{Fisher1989, Sachdev99}:
First, in the limit of high temperature, the system is a normal liquid. At zero temperature and $t=0$, the system is a Mott insulator
with fixed, integer density, a gap, and zero compressibility. For finite hopping, stable Mott lobes around the $t=0$ insulators are found, which are surrounded by a gapless, compressible superfluid.
The superfluid phase also exists at finite temperature. The Bose-Hubbard model is the simplest model that describes a conductor-insulator transition for bosons. It can also describe the physics of the weakly interacting Bose gas.

\begin{figure}
\begin{center}
\includegraphics[width=0.8\columnwidth]{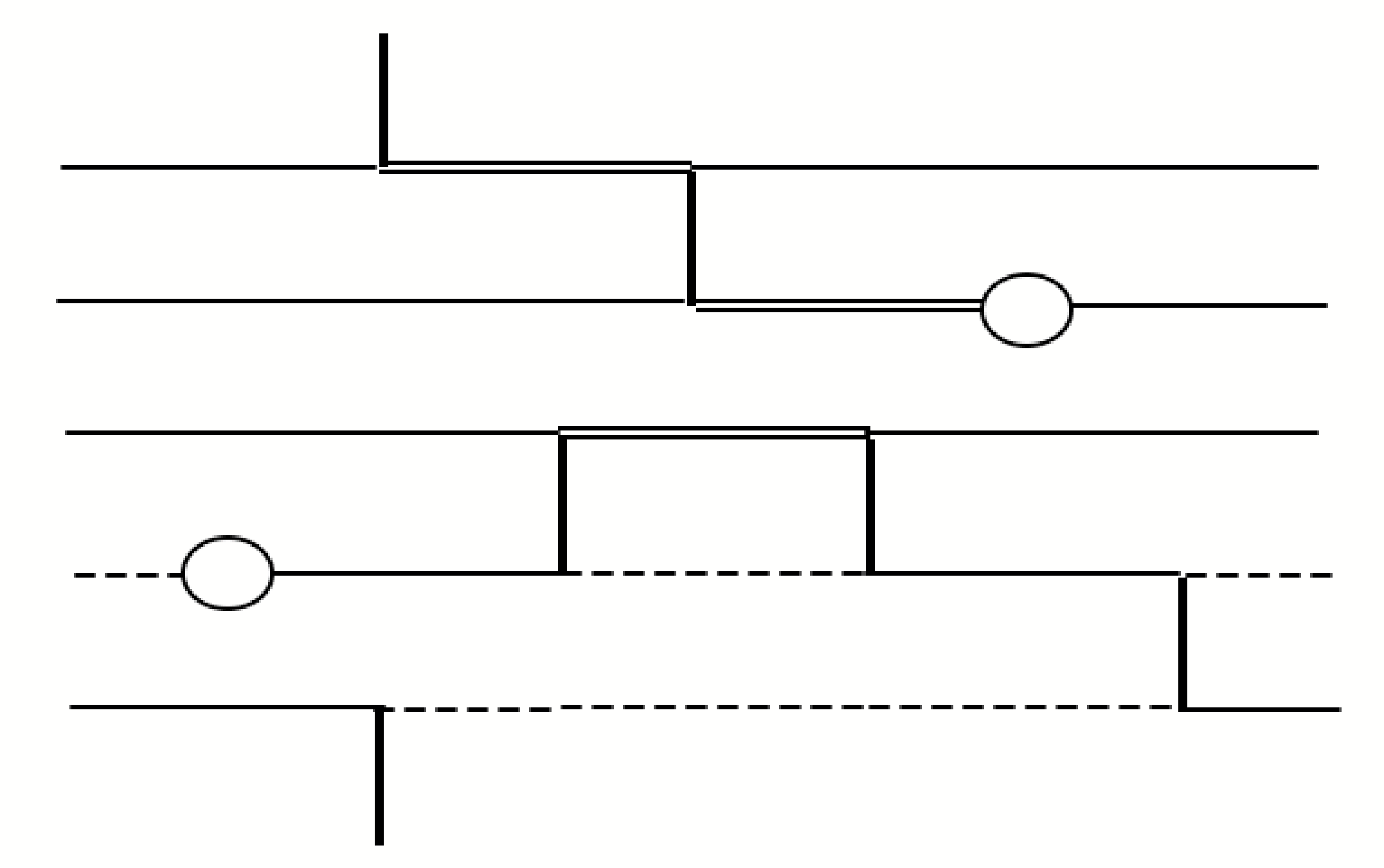}
\caption{Graphical representation of a typical configuration in the Green function sector. Imaginary time goes from left to right in the figure, there are five sites depicted. World lines are denoted by single lines (site is once occupied), double lines (site has occupancy two) or dashed lines (site is not occupied). Interactions (hopping of a particle) are denoted by vertical lines. The two circles mark a discontinuity in the world lines and correspond to the worm operators. One of them creates an extra particle, the other one annihilates it. Closed worldlines are formed when the worldlines annihilate each other. The figure is taken from Ref.~\cite{Pollet2007}. }
\label{fig:worldlines}
\end{center}
\end{figure}

\begin{figure}
\begin{center}
\includegraphics[width=0.8\columnwidth]{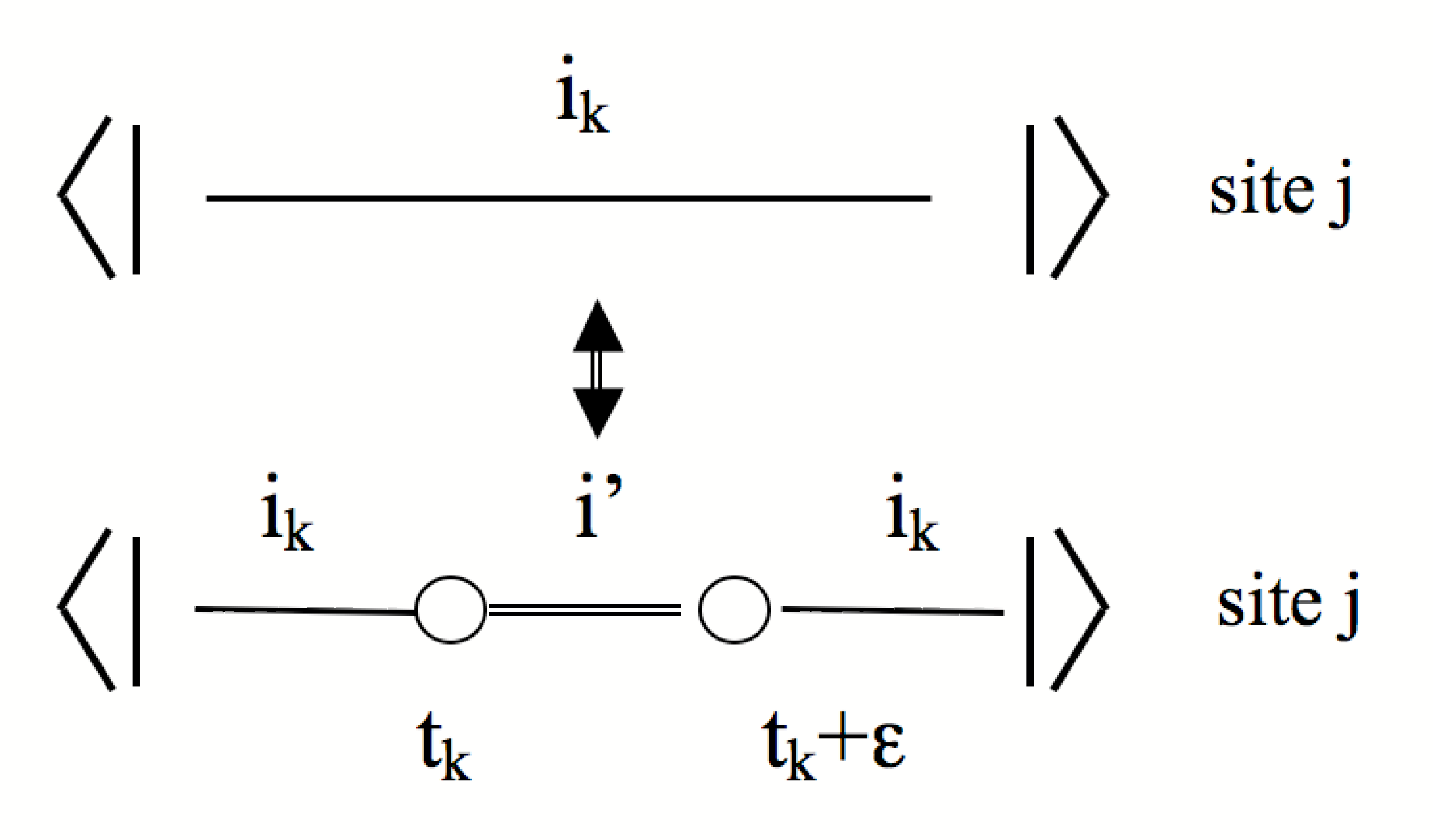}
\caption{Graphical representation of how a worm pair is inserted or removed, signaling a transition between the partition function sector and the Green function sector. For inserting the worm, an arbitrary site and arbitrary imaginary time are chosen. The occupation between the worm ends can in general be either  higher or lower than the occupation outside.  One of the worm ends remains stationary (called the worm tail), while the other one is mobile (called the worm head).  The figure is taken from Ref.~\cite{Pollet2007}.  }
\label{fig:LOWA_insert}
\end{center}
\end{figure}

\begin{figure}
\begin{center}
\includegraphics[width=0.8\columnwidth]{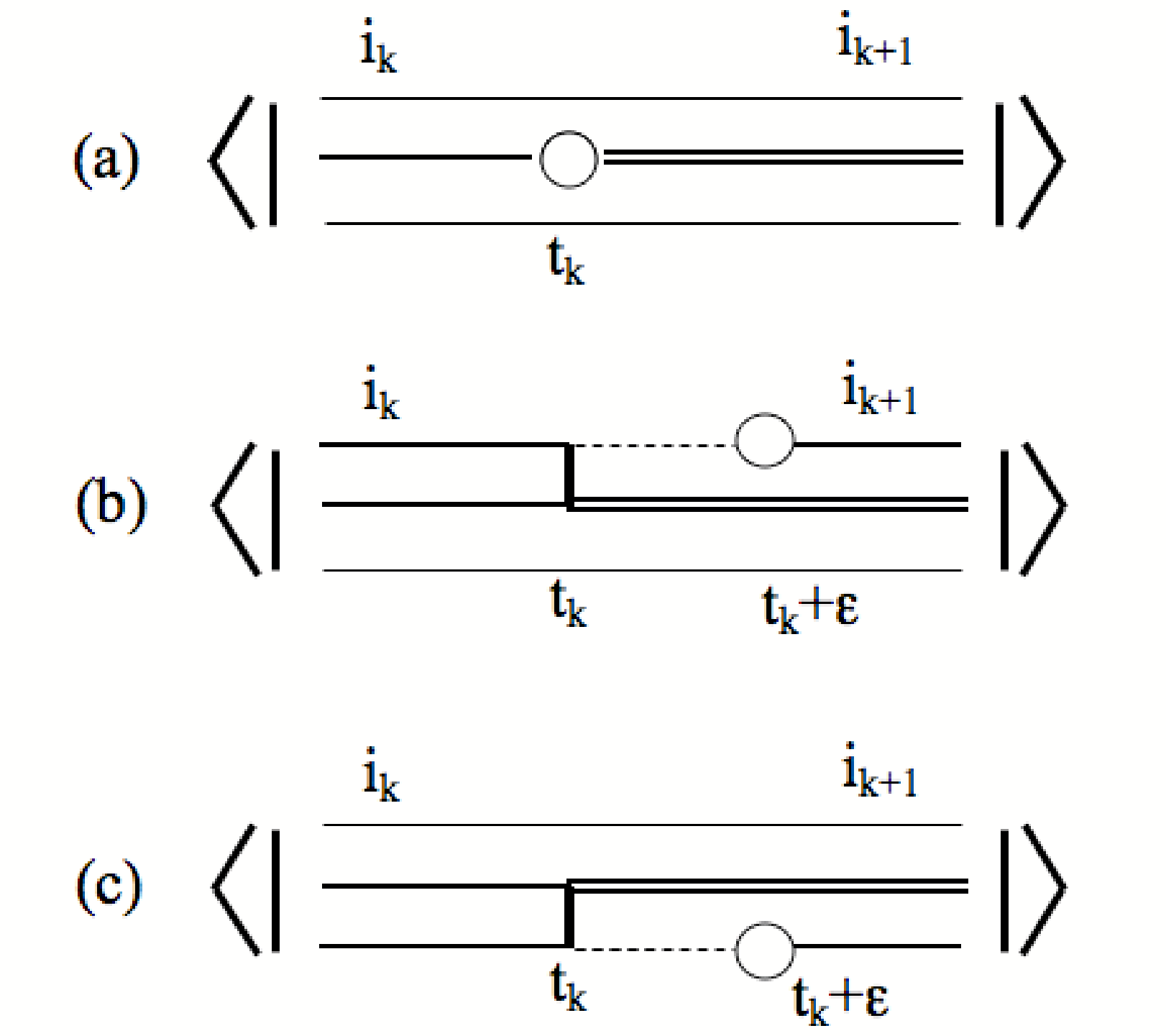}
\caption{Graphical representation of how a worm can insert/remove a kink ({\it i.e.,} a hopping of bosons), shown for a 1D lattice. When going from configuration (a) to configuration (b) or (c), we assume that the worm is moving to right (that is to greater imaginary times) and a kink is inserted to one of the neighbouring sites. Detailed balance also requires that we also stay sometimes in configuration (a). In algorithms with a fixed direction of propagation, the worm will reverse its propagation direction and start moving to lower imaginary times in such cases. For the reverse updates, when being in configuration (b) and moving to the left, the worm has the possibility of (a) removing the kink, (b) staying in the present configuration but changing its direction of propagation, (c) relinking the kink and changing the direction of propagation. All updates remain local : there are no changes to the configuration other than the ones shown over this small imaginary time interval and sites in the vicinity of the worm head.  The figure is taken from Ref.~\cite{Pollet2007}. }
\label{fig:LOWA_kink}
\end{center}
\end{figure}

\subsection{Continuous time expansion schemes and path integral Monte Carlo}

The starting point  is the following decomposition for the partition function,
\begin{equation}
Z = {\rm Tr} e^{-\beta H} = {\rm Tr} {\mathcal{T} }e^{-\beta H_0} \exp \left[ - \int_0^{\beta} d\tau H_1(\tau) \right],
\end{equation}
where $H_1(\tau) = e^{\tau H_0}H_1 e^{-\tau H_0}$. The trace is taken over all Fock basis states specified above (in which $H_0$ is diagonal). The exponential is expanded into a time-ordered product~\cite{Mahan, NegeleOrland, FetterWalecka},
\begin{equation}
Z =  {\rm Tr} {\mathcal{T} }e^{-\beta H_0} \left[ 1   - \int_0^{\beta} d\tau H_1(\tau) +  \int_0^{\beta} d\tau_1 \int_{\tau_1}^{\beta} d\tau_2 H_1(\tau_1) H_1(\tau_2) + \ldots   \right].
\label{eq:BH_expansion}
\end{equation}
This expansion can graphically be represented by worldlines (thanks to the $U(1)$ symmetry of the Bose-Hubbard model ), similar to as what is shown in Fig. ~\ref{fig:worldlines} except for the two circles (which are explained later). 
 The inverse temperature $\beta = 1/T$ is understood as an imaginary time, where the matrix elements of the operators $\exp [-\Delta \tau H_0]$ act as propagators
between the different states at $\tau_1, \tau_2, \ldots$, with $0 < \tau_1 < \tau_2 < \ldots < \tau_j < \ldots < \tau_n < \beta$. The perturbations $H_1$ change the states at those times.  A worldline then describes the trajectory of a particle propagating in imaginary time.

The expansion in path integral Monte Carlo (PIMC)  is always understood over a finite volume and finite imaginary time. There is no singularity caused by a phase transition, which can only be studied in a finite size scaling analysis. The expansion is written in terms of an entire function so that there are no non-physical divergencies in the PIMC formulation. 

The expectation value of an observable $A$ is given by  $\langle A \rangle = \frac{1}{Z} {\rm Tr} A e^{-\beta H}$. In PIMC, a statistical interpretation is given to Eq.~\ref{eq:BH_expansion} by introducing weights $w$ through $Z = {\rm Tr}_{\vert \{n_i \} \rangle} w({\vert \{ n_i \} \rangle})$. We now have to statistically generate all possible configurations according to their weights by generating all possible expansion orders and matrix elements, assign an (unnormalized) weight to each one of them, evaluate the observable $A$ in every configuration, and sum all these contributions. For instance, one can perform local updates by inserting pairs of hopping elements, in which a particle hops from a site to its neighbour, and back at a later time.

At high temperatures or deep in the Mott insulating phase, the kinetic energy is small compared to $U$ and/or $T$, and few perturbation orders are needed in Eq.~\ref{eq:BH_expansion}. Such is not the case when the contributions to the free energy coming from the hopping are large: There is no reason to expect that the local updates would be efficient. Even worse, they are not ergodic: The low energy states in a superfluid are given by states with a different winding number. These are paths in which a particle winds around the full length of at least one direction before closing on itself again. Configurations with different winding numbers are topologically distinct and cannot be transformed into each other by local updates alone. The winding number $W$ is directly related to the superfluid density through $\rho_s = \frac{\langle W^2 \rangle L^{2-d} } {d \beta}$~\cite{Pollock1987}, with $d$ the dimension of the system. Worldlines with non-zero winding number are caused by bosonic permutations.

\subsection{The worm algorithm}

The worm algorithm has completely solved those ergodicity problems~\cite{Prokofev1998_worm}. Instead of working with the partition function $Z$ alone one also works in the Green function sector $Z_G$,
\begin{equation}
Z_G = {\rm Tr} \mathcal{T} \{ b_i(\tau_0) b_j^{\dagger}(\tau) e^{-\beta H} \}.
\end{equation}
Graphically, the operators $b_i(\tau_0)$ $b_j^{\dagger}(\tau)$ are open ends delimiting a segment of a worldline (see Fig.~\ref{fig:worldlines}) and are called the worm head and the worm tail. The worms can be on any site and any time. A correct algorithm allows for the transition between the Green function sector and the partition function sector (where measurements of observables such as the energy and the superfluid stiffness are done), and allows to move the worms around in configuration space. This is shown in Fig.~\ref{fig:LOWA_insert}.  Worms also have the ability to insert and remove hopping elements (kinks), as shown in Fig.~\ref{fig:LOWA_kink}.
Since the worm operators correspond to open ends on a world line segment they have no problem in exploring configurations with different winding numbers. All updates are local in the Green function sector, which means that all acceptance factors can be made of order unity. In phases with off-diagonal long-range order (either true long-range order  such as seen in a Bose-Einstein condensate or quasi long-range order with correlation functions  decaying as a power-law) the worms will preferentially be far away from each other in configuration space, {\it i.e.}, we are efficient in describing the physics of those phases.

In the literature, different implementations of the above ideas can be found. The minimal requirements for the updates are however nothing more than the ones shown in the Figs.~\ref{fig:LOWA_insert} and ~\ref{fig:LOWA_kink}, combined with a move update which moves the position of the worm head forward or backward in imaginary time without changing the kinks. We will therefore only briefly mention the key ideas and refer to the relevant papers.  Prokof'ev, Svistunov and Tupitsyn were the first to introduce the worm algorithm~\cite{Prokofev1998_worm}. They formulated it for the Bose-Hubbard model in the grand-canonical ensemble in the path integral representation. Sandvik and Sylju{\aa}sen introduced worm operators for spin models in the stochastic series representation with directed loop updates~\cite{Sandvik1999, Syljuasen2003}. It is worth remarking that the loop algorithm was the first algorithm formulated in continuous imaginary time ({\it i.e., } free of Trotter discretization error) to overcome the critical slowing down near the second order phase transition~\cite{Evertz1993, Beard1996, Evertz2003}, even before the worm algorithm was invented. Pollet {\it et al.} tried to combine the two algorithms and formulated the worm algorithm in the path integral representation but with directed loops~\cite{Pollet2007}. Rombouts {\it et al.} formulated a worm algorithm in the canonical ensemble by letting the worm head and the worm tail propagate simultaneously in imaginary time, thereby creating and annihilating a particle on different sites simultaneously~\cite{Rombouts2006, VanHoucke2006}. These ideas were picked up by Rousseau {\it et al.} in formulating algorithms with multiple worms (called stochastic Green functions)~\cite{Rousseau2007, Rousseau2008}.  In ~\cite{Kawashima2004} a review of world-line Monte Carlo methods up to 2003 can be found with a discussion on the loop algorithm, directed loops algorithm and the worm algorithm. Classical worm algorithms have also been formulated~\cite{Prokofev2001, Prokofev2009, Alet2003}. These algorithms can, in particular, be used to efficiently simulate quantum criticality in cases when the universality class can be mapped onto a higher dimensional classical model (cf. Sec.~\ref{sec:WIBG_Tc}).

\subsection{The physics of the 3D Bose-Hubbard model in a nutshell}

\begin{figure}
\begin{center}
\includegraphics[width=0.8\columnwidth]{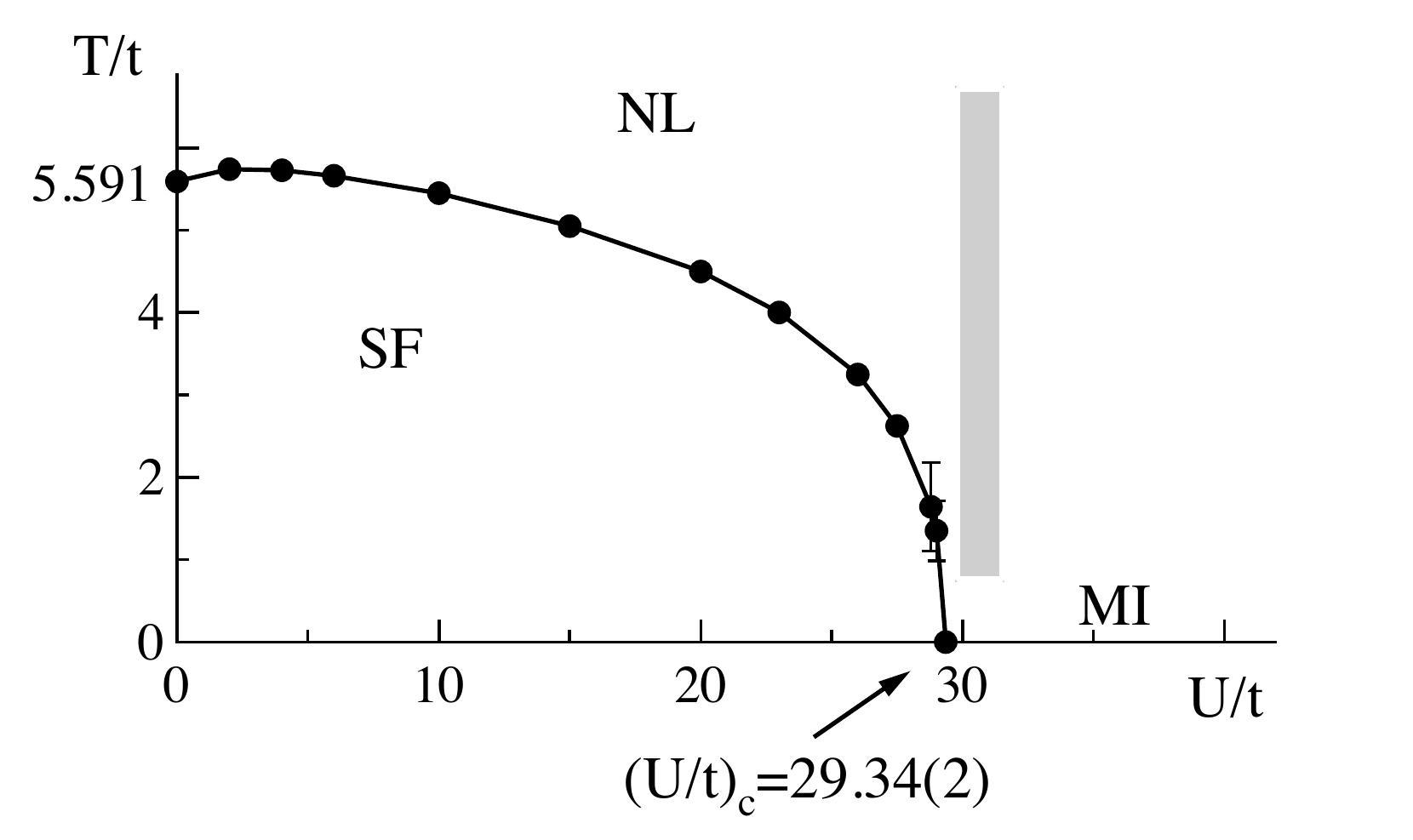}
\caption{Finite temperature phase diagram of the Bose-Hubbard model at unit filling on a simple cubic lattice. $T/t = 5.591$ is the critical temperature of a non-interacting Bose gas with the tight binding dispersion. Away from this point, the transition from the normal liquid (NL) to the superfluid phase (SF) belongs to the universality class of the  $d$-dimensional $XY$-model. Strictly speaking, the Mott insulator (MI) occurs only at zero temperature, but can loosely be defined to the right of the grey line at low enough temperature~\cite{Gerbier2007}. At commensurate densities, the transition from the SF to the MI belongs to the universality class of the $(d+1)$ dimensional $XY$-model. Reprinted figure with permission from Ref.~\cite{CapogrossoSansone07}. Copyright (2007) by the American Physical Society.
}
\label{fig:bose_hubbard}
\end{center}
\end{figure}

The phase diagram of the 3D Bose-Hubbard model, Eq.~\ref{eq:BoseHubbard}, is shown in Fig.~\ref{fig:bose_hubbard} at finite temperature and unit density. The transition from the normal liquid to the superfluid phase belongs to the 3D $XY$-model~\cite{Fisher1989}. At zero temperature, the superfluid phase undergoes a phase transition to the Mott insulating phase. This transition belongs to the $(d+1)$ dimensional XY model (and is hence of mean-field nature with logarithmic corrections): Near the tip of the Mott lobe, the particle and hole excitations have the same effective mass and become relativistic~\cite{Fisher1989}. Away from commensurability, the transition between the superfluid and the Mott insulator belongs to the Gaussian type universality class~\cite{WIBG, Fisher1989, Sachdev99}.

In the Mott insulator, the dispersion of the particle and hole excitations can be found by evaluating the Green function at zero momentum in the large imaginary time limit. 
Specifically, one finds, using the Lehmann representation, that $G({\mathbf p}, \tau) \to Z_{\pm} e^{{-\epsilon({\mathbf p})}_{\pm}\tau}, \tau \to \infty$, with $Z_{\pm}$ the quasiparticle weights and $\epsilon(\mathbf p)_{\pm}$ the energies of a single particle (+) and hole (-) excitation. 
This was used in Ref.~\cite{CapogrossoSansone07} to indeed see the emerging relativistic effects near the tip of the Mott lobe, but also to accurately determine the phase boundary of the Mott lobe. The semi-analytic predictions can also be used to study inhomogeneous models in the local density approximation~\cite{CapogrossoSansone07}.

The transition between the normal liquid and the superfluid can be studied by a finite size scaling analysis for the superfluid density~\cite{Fisher1989}. (note: Also the tip of the Mott lobe, where the critical exponents take mean-field values $\nu = 1/2$ and $z=1$,  is best found using a finite size scaling analysis of the superfluid density. No logarithmic corrections could be discerned in Ref.~\cite{CapogrossoSansone07}.)
Writing the distance to the critical point as $\theta = (T-T_c)/T_c$, the superfluid density $n_s$ obeys the scaling equation
\begin{equation}
n_s(\theta, L) = \xi^{-1} f_s(\xi/L) = L^{-1} g_s(\theta L^{1/\nu}),
\label{eq:FSS_gap}
\end{equation}
with $L$ the system size, $\xi$ the correlation length, and $f_s$ and $g_s$ universal scaling functions. The first equality follows from the theory of critical phenomena; in the second one we have used that the correlation length will be cut off by  the system size close enough to the transition point and that $\xi \sim \theta^{-\nu}$, with $\nu = 0.6717(1)$~\cite{Campostrini2006} for the universality class of the 3D XY model.
Hence, plotting $T/t$ as a function of $n_sL$ for different system sizes will display a crossing in a single point (to leading order), which is the critical temperature. Alternatively, data collapse can be obtained when plotting $n_sL$ as a function of $\theta L^{1/\nu}$.

\subsection{Bosons in a 3D optical lattice}

Although cold bosonic atoms subject to an optical lattice realize the Bose-Hubbard model, there are a few differences between experiments and the model discussed thus far. We list the most important ones:
\begin{itemize}
\item The confining potential to trap the neutral atoms is well described by a  parabolic potential. For a potential $V(r)$ with spherical symmetry, the local chemical potential is changed to $\mu - V(r)$, with $\mu$ the global chemical potential. Its influence on the phase diagram has been analyzed in great detail~\cite{Kashurnikov02, Batrouni02, Wessel04, Gygi06, Ma08, Mahmud11}. Because of the inhomogeneous chemical potential, spatial coexistence of superfluids and Mott insulating domains may occur. A sharp phase transition for a homogenous system is replaced by a gradual crossover in a trapped system. For shallow potentials, the local density approximation (LDA) is often used, where every site is treated as an independent system with its own chemical potential. It is a very good approximation when the correlation length in the system is small, i.e., away from phase transitions~\cite{Pollet10_crit} (see also Sec.~\ref{sec:singlesite}).
\item The fact that atoms are neutral particles makes detection much more difficult than in typical condensed matter systems. The most frequent detection tool is time-of-flight interference images, where the confining potential is switched off and the atoms expand ballistically, after which a laser shines on them and makes a picture on a camera. In the long-time limit the measured atomic interference pattern reflects the momentum distribution of the initial system. However, for typical expansion times of $15-20$ms, there remains a Fresnel diffraction term which is not negligible~\cite{Gerbier2008}, while interactions during the expansion can be neglected provided the atoms are well localized in the lattice and the density is low. This Fresnel diffraction term acts like a knife cutting off long-range correlations; typically correlations beyond 5-6 lattice spacings average out due to the fast oscillating phases. In particular, the $k=0$ condensate peak in an interference pattern is seriously affected (and cannot be used to infer the condensate fraction at all). The Fresnel term can fortunately be taken into account in Monte Carlo simulations~\cite{Gerbier2008}. For the weakly interacting Bose gas on the other hand {\it interactions} during the expansion cannot be neglected~\cite{Dalfovo1999}. 
  \item A major difference with condensed matter physics is that trapped cold gases are well described by an isolated system (recall for example the absence of a phonon bath). Despite the very low absolute temperatures, the involved scales are {\it not} orders of magnitude smaller than the bosonic condensation temperature (or the Fermi energy for fermionic systems). As a consequence, entropy is conserved but temperature changes when changing the system parameters in an ideal experiment, such as adiabatically ramping up the optical lattice~\cite{Schmidt2006, Ho2007, Pollet2008}. The entropy cannot directly be computed in quantum Monte Carlo simulations. Most often it is computed by integrating the specific heat over temperature, but also thermodynamic integration and quantum Wang-Landau sampling to obtain the full partition function have been tried~\cite{Pollet2008}.  For a uniform system the entropy is exponentially suppressed ($\sim e^{-U/T}$) in the Mott insulator, meaning that the Mott insulator cannot be reached under adiabatic changes~\cite{Ho2007} and that the system would heat into a normal phase. The normal phase, just as the Mott insulator, has no interference peaks in time-of-flight images. 
Fortunately, for a trapped system the entropy of the Mott lobes is not important since the bosons in the edges of the cloud are normal. The edges can hence serve as an entropy reservoir. In Ref.~\cite{Ho2007} the width of the superfluid rings between the Mott plateaus in the density wedding cake structure of the trapped Bose-Hubbard model was shown to be related to the total entropy in the system. It was argued that by ramping up the optical lattice, the confinement gets substantially tighter, the width of the superfluid regimes shrinks, and temperature may raise dramatically.  However, quantum Monte Carlo simulations showed that this scenario was too pessimistic and that Mott-like features (which require a temperature lower than the crossover scale $T/U < 0.2$~\cite{Gerbier2007})  can easily be reached for realistic parameter regimes~\cite{Pollet2008}.  Because the entropy is carried by the normal bosons and the largest volume fraction is found in the edges, one sees that for sufficiently deep lattices temperature scales with the on-site repulsion strength, $T \sim U$~\cite{Pollet2008, Gerbier2005, Ho2007}.  Note that temperature cannot be measured directly in the lattice system, although it can be accurately determined before the atoms are loaded in the lattice by time-of-flight images, provided the condensate fraction is not too high~\cite{Dalfovo1999}.
\item Since the system size is small and the number of measurement techniques limited, it is also difficult to determine 'transitions' accurately. It was argued in Refs.~\cite{Diener2007, Kato2008, Zhou09a} that an accurate determination of the critical temperature for the normal to superfluid temperature is complicated because of the existence of strong, though non-divergent, peaks already present in the normal phase. Together with the absence of a diverging length scale in the trapped system at the critical point, the precise determination of the critical point would be impossible. These arguments did not take the Fresnel diffraction term into account (which would rather enforce the argument). However, as we will discuss in detail below, Trotzky {\it et al.} could accurately determine the full finite temperature phase diagram at unit filling in the trap center by monitoring the visibility, the condensate peak and its width~\cite{Trotzky2010}, showing that the arguments of Refs.~\cite{Diener2007, Kato2008, Zhou09a} are presently not a limiting factor for a realistic system. 
\end{itemize}

\begin{figure}
\begin{center}
\includegraphics[width=0.8\columnwidth]{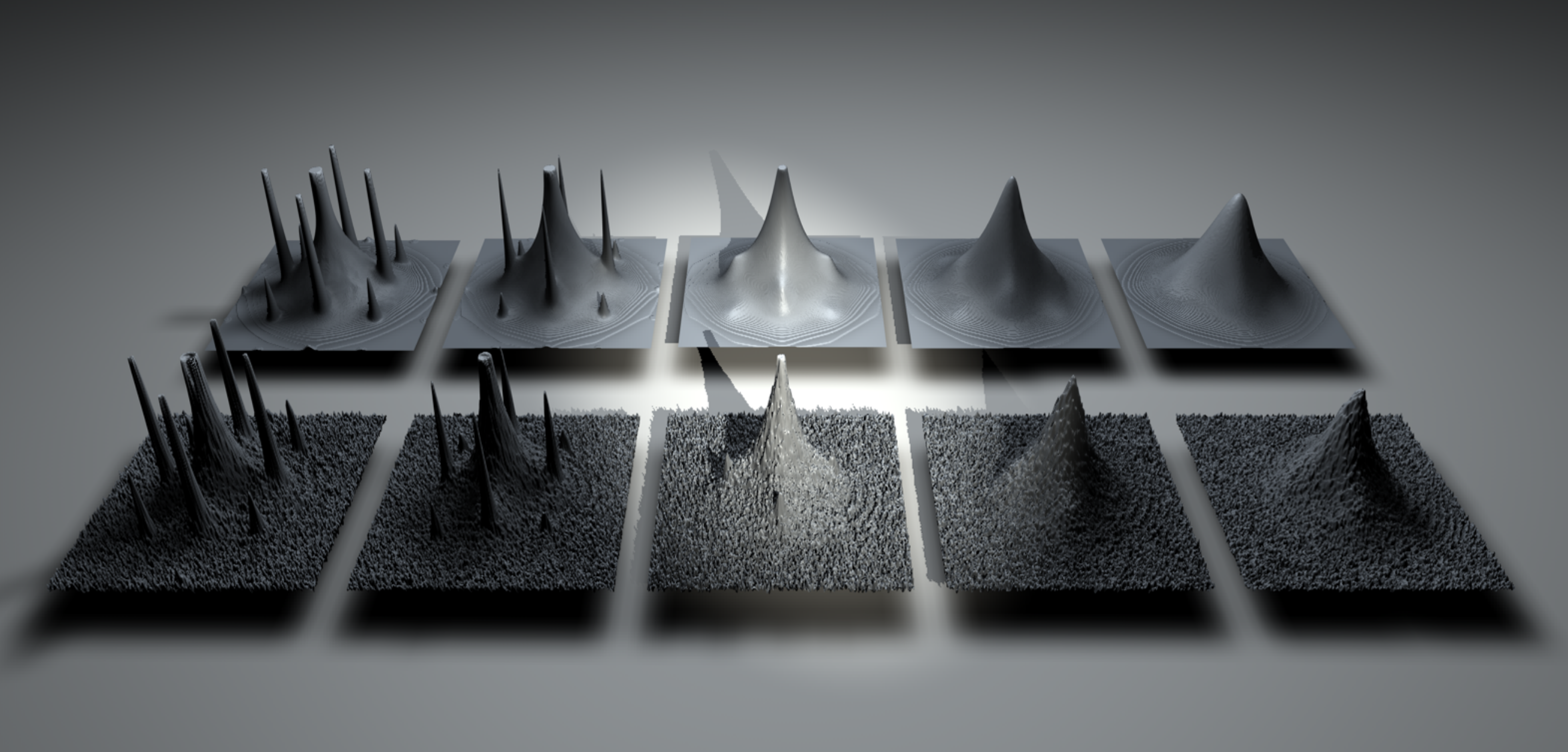}
\caption{ Comparison between the integrated column densities from time-of-flight images obtained experimentally (lower row) and by quantum Monte Carlo simulations (upper row) for fully realistic system parameters. Five different temperatures (these are input parameters in the Monte Carlo simulations) are shown, $T = 11.9$nK, $T=19.1$nK, $T=26.5$nK, $T=31.8$nK, and $T= 47.7$nK, from left to right. The lattice depth is 8 recoil energies ($U/t=8.11$) and there are about $N=2.8 \times 10^5$ particles in the system.   The figure is courtesy of Stefan Trotzky and results from the work reported in Ref.~\cite{Trotzky2010}. It was taken from the website of I. Bloch, \url{http://www.quantum-munich.de/media/finite-temperature-comparison-of-experiments-and-qmc-simulations/}. }
\label{fig:comparison_bosehubbard}
\end{center}
\end{figure}

In the introduction we mentioned that cold atom experiments can be seen as quantum simulators or quantum emulators, but before they can be trusted as such quantum analog computers, they should be benchmarked against known results. Trotzky {\it et al.} chose the superfluid-to-normal liquid transition of the Bose-Hubbard model at unit density in the middle of the trap~\cite{Trotzky2010} for this benchmarking. They mapped the transition out experimentally on the basis of time-of-flight images, and compared the results with quantum Monte Carlo simulations with  no free fitting parameter. The excellent agreement is shown in Fig.~\ref{fig:comparison_bosehubbard}. In this analysis, it was crucial to take the conservation of entropy and the finite duration of the time-of-flight images into account (the knotted peaks in Fig.~\ref{fig:comparison_bosehubbard} are a consequence of this). However, additional non-entropic heating was observable (and surprisingly large at the lowest temperatures), due to spontaneous emission, classical noise and lattice imperfections. This has stimulated studies from the quantum optics community to better estimate and find solutions for this problematic behaviour~\cite{Pichler2010, Gerbier2010}.

\subsection{Single-site resolution and addressability}
\label{sec:singlesite}

One of the most amazing advances in the last couple of years is the single site resolution and addressability of atoms in optical lattices~\cite{Gemelke09, Bakr09, Wuertz09, Bakr10, Sherson10, Weitenberg11a, Weitenberg11b, Endres11, Bakr11, Hung10, Hung11}. Individual atoms can now be observed and addressed, typically in a 2D setup. When the measurement in the quantum gas microscope~\cite{Bakr09} is performed, the atoms are quickly brought into a very deep lattice suppressing tunneling. The atoms are then illuminated with an optical molasses that serves to localize the atoms while fluorescence photons are collected by the high resolution optics. In addition, light assisted collisions immediately eject atoms in pairs from the lattice site, leaving behind an atom only if the initial occupation is odd. Remaining atoms scatter several thousand photons during the exposure time and can be detected with high fidelity. The density profiles were also shown to be in good agreement with Monte Carlo simulations~\cite{Bakr10, Endres11}, and hence provide further evidence that thermodynamics is a valid description for these systems. Also density-density correlation functions are now accessible experimentally, which for instance was demonstrated by measuring the parity operator for the 1D Mott insulator~\cite{Endres11}. We will now discuss two new opportunities for studying many-body physics which were made possible by the experimental advances in single-site resolution methods, namely the possibility to obtain the equation of state, and the study of critical phenomena.

\begin{itemize}
\item Single-site addressability could provide a means for obtaining the equation of state. In the edges the gas is in a normal phase, where measuring the density profile and comparing to high temperature series expansions~\cite{Zhou09a, Ho09, Ma10} gives access to all thermodynamic quantities. If this procedure cannot be followed, then temperature can still be determined by using the fluctuation-dissipation theorem~\cite{Zhou09b, Ma10} 
\begin{equation}
T \frac{\partial n( \bm{r})}{\partial \mu} = \int d \bm{r'} \langle n(\bm{r}) n(\bm{r'}) \rangle - \langle n(\bm{r}) \rangle \langle n(\bm{r'}) \rangle.
\end{equation}
which is valid in the local density approximation (LDA). The integral only needs to be evaluated over the density-density correlation length~\cite{Ma10}. 
We note that number fluctuations were previously suggested as an effective thermometer~\cite{Capo07, Gerbier06}.

\item It was also suggested that single-site detection tools can be useful for studying critical phenomena, which remained a controversial topic because of the parabolic confining potential~\cite{Zhou10, Fang11, Hazzard11, Guan10, Yin11, Batchelor10, Zhang11}. In Ref.~\cite{Pollet10_crit} the superfluid to normal transition was analyzed by finding regions where the LDA approximation breaks down. LDA implies quasi-homogeneity of the system when the change of thermodynamic properties of the system is negligible at the distance of the order of the correlation radius. LDA breaks hence down in the critical region. It was found numerically, and supported by analytical arguments, that LDA violations are tiny for the density profiles (because of the low value of the critical exponent $\alpha$) and by extension for the compressibility, which is the derivative of the density with respect to the chemical potential, in contrast to earlier findings~\cite{Zhou09a, Pollet10_comment, Zhou10_reply}. On the positive side, it was demonstrated that analyzing the transition using time-of-flight images can be done without fitting the time of flight images. One has to construct the amplitude of the critical signal as a function of temperature, $P_c(T) = n(k=0) - n(k_{\rm max})$ with $n(k)$ the interference patterns at momenta $k$ and $k_{\rm max}$ the momentum at which the absolute value of the first derivative $dn/dk$ has a maximum. By then plotting $Q(T) = P_c(T) k_{\rm max}^s(T)$ with some exponent $s > 2- \eta$ (with $\eta$ the critical exponent for the correlation function at the the critical point which is almost zero for the 3D XY transition), the critical temperature can be read out accurately where $Q(T)$ reaches a sharp minimum. The scheme exploits the fact that the momentum distribution behaves as $n(k) \sim \xi^{2 - \eta}$ for $k \ll \xi^{-1}$ and as $n(k) \sim k^{\eta - 2}$ for $\xi^{-1} \ll k \ll a^{-1}$ with $a$ the lattice spacing. In Monte Carlo simulations, this  scheme was shown to work even when taking the finite time-of-flight duration and other optical resolution limiting effects into account, but has not been tried experimentally yet.

\item Going into more detail about the issue of scaling theory, Campostrini and Vicari introduced the scaling exponent for the trap~\cite{Campostrini09, Campostrini10} which shows how critical phenomena for finite systems with a power-lap trap with potential $V(r) = v^p r^p = (r/l)^p$ can be analyzed when making the trap softer. $v$ and $p$ are positive constants, $l=1/v$ is the trap size. A harmonic potential corresponds to $p=2$. The authors focused mostly on the Mott insulator to superfluid transition driven by the chemical potential with dynamical exponent $z=2$ in 1d~\cite{Campostrini10, Campostrini2010b, Ceccarelli2012} and more recently also in 2d~\cite{Ceccarelli2012b}. Using hard-core bosons, there are two Mott insulators, namely for $\mu < -2t$ with $\langle n \rangle = 0$ and for $\mu > 2t$ with $\langle n \rangle = 1$ (in 1d), and a superfluid in between. The trap breaks particle-hole symmetry so that these two transitions are not necessarily  equal. Following scaling theory, the simplest trap-size scaling Ansatz one can make for the free energy at $T=0$ in $d$ dimensions is,
\begin{equation}
F(\mu, T, l, x) \to l^{-\theta(z+d)} \mathcal{F}( \bar{\mu}l^{\theta/\nu}, T l^{z\theta}, x l^{-\theta}).
\end{equation}
Here, $\bar{\mu} = \mu - \mu_c$ is the detuning, $\theta$  the trap scaling exponent, $x$ the distance to the trap center, and the function $\mathcal{F}$ a universal scaling function insensitive to the microscopic details.
Finite size effects (when using a finite box of length $L$) can be taken into account by adding an additional argument $L l^{-\theta}$. 
 A simple analysis leads to $\theta = p / (p+2)$~\cite{Campostrini09, Pollet10_crit}. Similar expressions can be written down for the density and density-density correlation function,
\begin{eqnarray}
\rho(x) = \langle n_x \rangle & \to & l^{-d\theta} \mathcal{D}(\bar{\mu} l^{\theta / \nu}, T l ^{z \theta}, x l^{-\theta}), \nonumber \\
G(x,y) = \langle n_x n_y \rangle - \langle n_x \rangle \langle n_y \rangle & \to & l^{-2 \theta d} \mathcal{G}( \bar{\mu} l^{\theta / \nu}, T l ^{z \theta}, x l^{-\theta}, y l ^{-\theta}),
\end{eqnarray}
and for the gap
\begin{equation}
\Delta = l^{-\theta z} \mathcal{K}(\bar{\mu} l^{\theta / \nu}).
\end{equation}
The above set of equations are often used in Monte Carlo simulations to study crtical phenomena: data collapse on the universal scaling functions can be oberved provided the data are rescaled with the correct RG exponents . Examples of this can be found in Eq.~\ref{eq:FSS_gap} as well as in Refs.~\cite{ Campostrini10, Campostrini2010b, Ceccarelli2012} for trap-size scaling.

The trap-size scaling theory for the low density limit $(\langle n \rangle = 0)$ in 1d for hard-core bosons (at $T=0$) can also be derived analytically from a mapping to quadratic spinless fermions~\cite{Campostrini2010b}. Peculiar behaviour such as discontinuities in the scaled particle density can be related to the quantum nature of the transition. When the filling of the corresponding homogeneous system is nonzero, an infinite number of level crossings occur when increasing the trap size~\cite{Campostrini2010b}. This leads to a modified trap-size scaling which is still controlled by the trap size exponent $\theta$ but shows modulations with a phase measuring the distance to the closest even level crossing. The asymptotic trap-size dependence of the superfluid phase, whose corresponding continuum theory is a conformal field theory with $z=1$, appears also to be modulated and is characterized by two length scales: one scaling as $\xi \sim l$ related to smooth modes, and one scaling as $\xi \sim l^{p/(p+1)}$ involving modes at the Fermi scale $k_F = \pi f$, with $f$ the filling of the homogeneous system~\cite{Campostrini2010b}.
At finite temperature, the periodic oscillations in the vicinity of the $ \langle n \rangle = 1$ transition found at $T=0$ vanished rapidly~\cite{Ceccarelli2012}.
Similar conclusions hold in 2d~\cite{Ceccarelli2012b}; the type of analysis discussed in this paragraph is expected to be applicable to other models as well.

The authors of Refs.~\cite{Campostrini2010b, Ceccarelli2012} also investigated the validity of the local density approximation. They find that the  LDA approximation of the particle density becomes exact in the large trap-size limit, and corrections are controlled by the trap exponent, but they  confirm that corrections to LDA become much larger
in the critical region.

\item The phase transition from vacuum to superfluid in 2D was studied experimentally in Ref.~\cite{Zhang11}. Previously, universal scaling behaviour was observed in interacting Bose gases in three~\cite{Donner07} and two dimensions ~\cite{Hung11b}, and in Rydberg gases~\cite{Loew09}. This transition is one of a dilute  (in the sense of~\cite{Sachdev99}, thus non-interacting in 2D and 3D and the Tonks-Girardeau limit in 1D) Bose gas, and is exactly at its upper critical dimension. One thus expects mean-field exponents $\nu = 1/2$ and $z=2$ with logarithmic corrections that could experimentally not be discerned however. The equation of state follows the scaling $\tilde{N} = F(\tilde{\mu})$ in which $F(x)$ is a universal function, and $\nu$ the inverse of the RG dimension of $\mu$. 
\begin{eqnarray}
\tilde{N} & = & \frac{N-N_r}{ (T/t)^{d/z + 1 - 1/(z\nu)} }  \nonumber \\    
\tilde{\mu} & = & \frac{ (\mu - \mu_0)/t}{ (T/t)^{1/(z\nu)}} 
\end{eqnarray}
are the scaled occupation number and scaled chemical potential. $N_r$ is the regular, non-universal part of the occupation number, which is zero for the vacuum to superfluid transition. Monte Carlo simulations for the superfluid to Mott insulator were found in excellent agreement with the experimental data~\cite{Fang11}, which allows to determine over which temperature range the zero temperature critical point can be felt. 
\end{itemize}

For completeness, we mention that the worm algorithm was also successfully applied to study the phase diagrams of the 2D Bose-Hubbard model~\cite{CapogrossoSansone08}. The reader can verify that the first path-integral Monte Carlo simulations without worm-type updates had large error bars on the determination of the critical point~\cite{Krauth1991b}. Even with current computers the algorithmic slowdown would prevent an accurate determination of properties when superfluid fluctuations are large (that is certainly for large system sizes) when not using the worm updates ).  In 1D, the phase diagram was accurately determined long ago by a density matrix renormalization group study~\cite{Kuehner98, Kuehner00}. State diagrams were calculated in the presence of a confining potential~\cite{Rigol09, Hen10, Mahmud11}. Excitation spectra across the superfluid to Mott insulator transition in 1D were studied by using maximum entropy~\cite{Jarrell96}, showing a gapped mode in the strongly interacting superfluid phase~\cite{Pippan09}.

\subsection{Disordered systems and the Bose glass phase}

The interplay between disorder and interactions is a long-standing problem in condensed-matter physics. Bosonic systems are especially difficult to handle since the limit of vanishing interactions is pathological: in the ground state, all bosons occupy the same lowest-energy localized state. In the presence of disorder, the concept of lakes is crucial: these are regions where the chemical potential is nearly homogeneous and mimic a uniform system. Such lakes can be arbitrarily large (so the quantization energy can be very low) but they are exponentially rare.
Nevertheless, they are crucial for such properties as the existence of a gap in the spectrum. In the presence of disorder, a new phase is possible at zero temperature~\cite{Giamarchi1987, Giamarchi1988}: the Bose glass phase (BG), which is gapless, compressible but insulating. It defies intuitive notions about conductivity based on Fermi liquid theory. Fisher {\it et al.}, building on the one-dimensional work by Giamarchi and Schulz~\cite{Giamarchi1987, Giamarchi1988}, argued the existence of the Bose glass phase in any dimension~\cite{Fisher1989}.

In the presence of a lattice, a commensurate system may be driven to a Mott insulator (MI). For a long time, there was controversy whether a direct transition between a superfluid (SF) and a Mott insulator (MI) in the presence of disorder was possible~\cite{Freericks1996,Scalettar1991,Krauth1991,Zhang1992,Singh1992,Makivic1993, Wallin1994,Pazmandi1995,Pai1996,Svistunov1996,Pazmandi1998,Kisker1997, Herbut1997,Sen2001, Prokofev2004, Wu_Phillips2008,Bissbort2009,Weichman2008_1,Weichman2008_2}. Fisher {\it et al.} argued that this was unlikely, though not fundamentally impossible~\cite{Fisher1989}. Curiously, a large number of direct numerical simulations and some approximate approaches observed the unlikely scenario. However, Fisher {\it et al} also showed that if the bound $\Delta$ on the disorder strength is larger than the half-width of the energy gap $E_{g/2}$ in the ideal Mott insulator, then the system is inevitably compressible and the transition is to the Bose glass phase. More recently, the {\it theorem of inclusions} proved that the Bose glass phase always has to intervene between the superfluid and the Mott insulator for generic disorder distributions~\cite{Pollet2009_disorder, Gurarie2009}. The proof proceeds in two steps by first showing that a direct transition between a gapless and a gapful phase is not possible, and second that the compressibility is finite at the superfluid to Bose glass transition (see Refs~\cite{Pollet2009_disorder, Gurarie2009} ). A direct consequence is that the transition of the Bose glass to the Mott insulator is of the Griffiths type: the vanishing of the gap at the critical point is due to a zero concentration of rare regions where extreme fluctuations of disorder mimic a regular gapless system, which was already conjectured by Fisher {\it et al.}~\cite{Fisher1989}. A Mott glass can occur for zero compressibility, {\it e.g.}, for disorder in the hopping of hard-core bosons at half filling, which has particle-hole symmetry~\cite{Pollet2009_disorder, Prokofev2004}.

\begin{figure}
\begin{center}
\includegraphics[width=0.8\columnwidth]{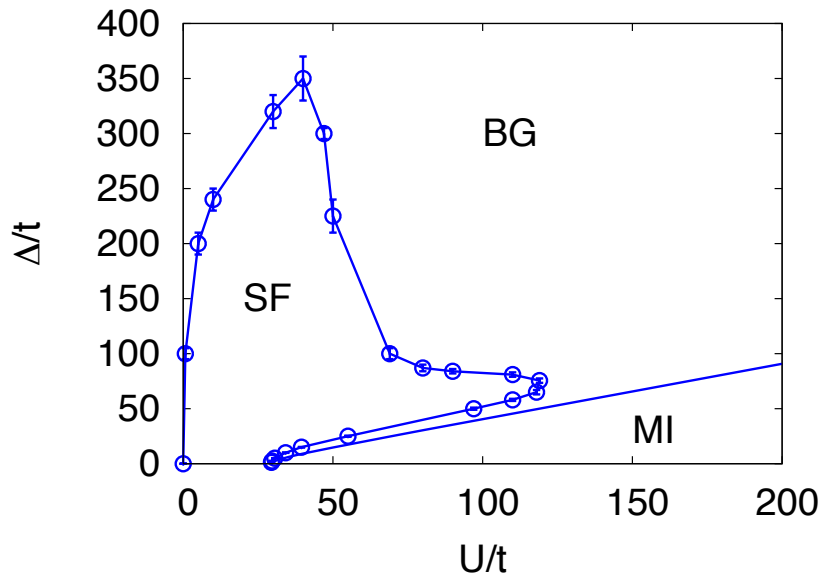}
\caption{ Phase diagram of the disordered 3D Bose-Hubbard model at unit filling, obtained by a finite-size analysis of winding numbers. In the absence of disorder, the system undergoes a quantum phase transition between SF and MI phases. The presence of disorder allows for a compressible, insulating BG phase, which always intervenes between the MI and SF phases because of the theorem of inclusions~\cite{Pollet2009_disorder, Gurarie2009}. The transition between MI and BG is of the Griffiths type, as an exception implied by the theorem of inclusions~\cite{Gurarie2009}. At $U/t \to 0$, the SF-BG transition line has an infinite slope~\cite{Falco2009}. 
Reprinted figure with permission from Ref.~\cite{Gurarie2009}. Copyright (2009) by the American Physical Society.
}
\label{fig:phasediagram_disorder}
\end{center}
\end{figure}

The phase diagram of the 3D Bose-Hubbard model with box disorder in the chemical potential and bound $\Delta$ is shown in Fig.~\ref{fig:phasediagram_disorder} for unit density. The BG phase always intervenes between the SF and MI, while the transition from the MI to the BG phase is of the Griffiths type. The transition line is then determined by measuring the gaps in the disorder-free MI.  At $U/t \to 0$, the SF-BG transition line has an infinite slope going like $\Delta \sim U^{1-d/4}$~\cite{Falco2009}, which is a consequence of Lifshitz-tails. Note that for low values of $U$ the superfluid region is extremely stable against disorder, and the transition to the BG phase reaches a maximum for $\Delta / t \approx 350$, a scale which was explained by percolation and local energy scales in Ref.~\cite{Gurarie2009}. The superfluid densities in the finger of the phase diagram were low, indicating low transition temperatures.

In two dimensions, the topology of the phase diagram is the same as in 3D. Here too, reentrant behaviour is seen and the superfluid phase can exist to very high disorder strengths~\cite{Soyler2011}. However, close to the superfluid to Mott-insulator transition for weak disorder, transition points for the SF-BG transitions could within error bars not be resolved from the transition point for a clean system~\cite{Soyler2011}.  In one dimension, Svistunov showed long ago that the BG phase always intervenes between the MI and SF~\cite{Svistunov1996}. The phase diagram was computed by Monte Carlo simulations~\cite{Prokofev1998_comment} and by the density matrix renormalization group~\cite{Rapsch1999}, with qualitative agreement.
Interestingly, controversy remains about the nature of the transition in one dimension. Giamarchi and Schulz performed a lowest order RG scaling and found that the transition between SF and BG occurs universally at $K = 3/2$  (instead of $K=2$ for the SF-MI transition in the clean system)~\cite{Giamarchi1988}.  In the strong disorder scenario put forward in Refs.~\cite{Altman2004, Altman2008, Altman2010} the transition for $\Delta/U \gg 1$ is believed to be non-universal with power law distributions of the Luttinger parameter. No simulations exist to date which distinguish between the two scenarios.

Experimentally, the Bose glass phase has never been detected unambiguously. The experiments of Helium-4 adsorbed on disordered substrates were not conclusive in establishing the existence of a Bose glass phase; the findings were better explained by a model which has a constant density of states for low and for high energies, with a gap in between~\cite{Crowell97}. In the cold atom experiments of Ref.~\cite{White09, Pasienski09} where optical speckles are used to generate the disorder (and in practice only a single disorder realization is used), no distinction between a Mott insulator and a Bose glass phase could be made; only insulating phases could be distinguished from superfluid phases on the basis of time-of-flight interference images and transport. They found insulating phases for disorder strengths several hundred times the tunneling amplitude, in agreement with the quantum Monte Carlo simulations. However, they did not find an insulator to superfluid transition (missing the 'finger' in Fig.~\ref{fig:phasediagram_disorder}). Although the disorder distributions are different in experiment and in simulations, the topology of the phase diagram should be the same. The discrepancy is attributed to the low transition temperature (or equivalently, the low superfluid density at zero temperature in the finger~\cite{Gurarie2009}) while the temperature in experiment is estimated to be well above it. 

Finally, we wish to mention that there have been a number of recent experiments on disordered quantum antiferromagnets~\cite{Yamada09, Hong09, Yu11}. They are, however, not without criticism either~\cite{Zheludev09, Wulff11}. 

\subsection{Polar molecules}

The influence of long-range interactions is currently also attracting a lot of interest in cold atomic and molecular gases: the first signatures of long-range interactions have been observed for magnetic interactions in $^{52}$Cr ~\cite{Griesmaier05_polar, Lahaye07_polar, Koch08_polar} with a dipole moment $D = 6\mu_B$  and $^{164}$Dy ~\cite{Lu11_polar}, which is the most magnetic atom with a dipole moment $d = 10\mu_{\rm B}$. Electric dipole and van der Waals interactions between Rydberg states give rise to intriguing collective phenomena~\cite{Heidemann07_polar}. In addition, there are huge experimental efforts towards the realization of quantum degenerate polar molecules~\cite{Sage05_polar, Ni07_polar, Deiglmayr08_polar, Aikawa10_polar}, where the permanent dipole moment of the molecules gives rise to a strong and highly tunable electric dipole-dipole interaction~\cite{Pupillo08_polar, Buechler07_polar, Menotti08_polar, Menotti08b_polar}. 

We will not review the physics of dense molecular samples here, for which excellent recent reviews are available~\cite{Carr09_polar, Menotti08_polar, Menotti08b_polar, Dulieu09_polar, Friedrich09_polar}, and in particular for lattice systems there is Ref.~\cite{Trefzger11_polar}. Rather, we will report on quantum Monte Carlo studies that explore the possibilities offered by current and future developments in this field. A central role so far was played by the possibility to observe a supersolid phase in these systems~\cite{Pollet10_polar, Capogrosso-Sansone10_polar}.

We consider bosonic polar molecules in a strong electric field along the z direction, which induces the dipole moment $d_z \le d$; here $d$ denotes the permanent dipole moment of the heteronuclear molecule. The dominant interaction between the polar molecules is then given by the dipole-dipole interaction $V(R) = \frac{d_z^2}{4 \pi \epsilon_0} \frac{R^2 - 3z^2}{R^5}$ where the strength of the dipole interaction can continuously be tuned by the strength of the electric field. In addition, the polar molecules are confined into the $xy$ plane by a strong transverse harmonic confinement as can easily be achieved by a strong one-dimensional standing laser along the $z$ direction. The combination of strong transverse trapping and dipole interactions creates a repulsive barrier, which prevents the collapse naturally present in bosonic dipolar gases. The effective long-range two-dimensional potential is then found by integrating over the $z$ direction, and reduces to $V^{2D}_{\rm eff} \sim D/r^3$. We refer to Refs.~\cite{Pupillo08_polar, Buechler07_polar} for a detailed discussion on how such a potential can be tailored.

\begin{figure}
\begin{center}
\includegraphics[width=0.8\columnwidth]{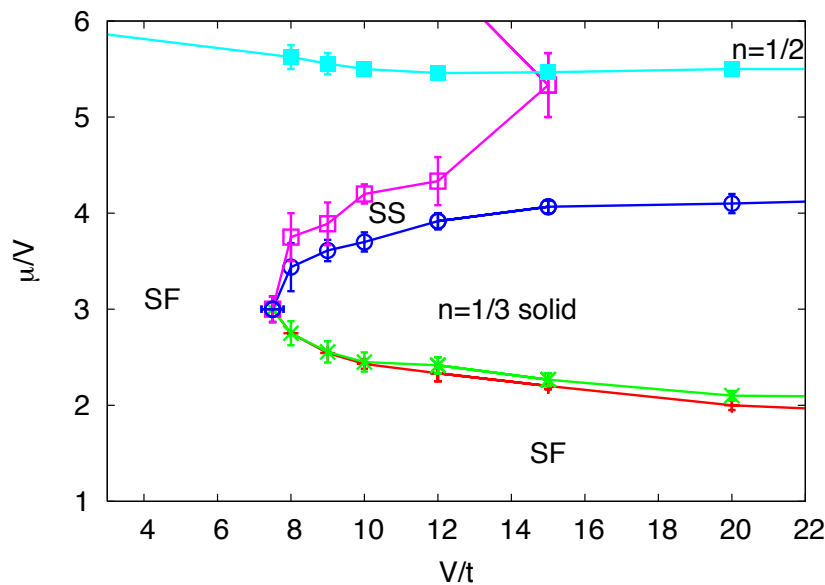}
\caption{Ground state phase diagram of the 2D Bose-Hubbard model with dipolar interactions with $t$ the tunneling amplitude and $V$ the interaction strength. The phases are a superfluid "SF", supersolid "SS", and a commensurate solid at filling factor $n=1/3$. With the double line we indicate a transition region of the Spivak-Kivelson bubble type (emulsions) gradually going over to a region of incommensurate, floating solids with increasing interaction strength. For large interaction strength, and starting around half filling, the supersolid phase is suppressed by emerging solid ordering (stripes at half filling and incommensurate, floating solids at other fillings). 
Reprinted figure with permission from Ref.~\cite{Pollet10_polar}. Copyright (2010) by the American Physical Society.
}
\label{fig:phasediagram_polar}
\end{center}
\end{figure}

The ground state phase diagram on the triangular lattice is shown in Fig.~\ref{fig:phasediagram_polar}.
For $V/t > 7.5$ there is an insulating commensurate solid at filling factor $n=1/3$.
For densities below $n < 1/3$ a superfluid phase is reached, similar to what is found for the short-range model with nearest-neighbour repulsion only~\cite{Boninsegni05_polar, Wessel05_polar, Heidarian05_polar, Melko05_polar, Melko06_polar} , but the transition here is different (it's first order in the short-range model because of domain-wall formation) and of the bubble type introduced by Spivak and Kivelson~\cite{Spivak04_polar}: over a finite but narrow range of chemical potentials, small crystallites form an emulsion of bubbles inside a liquid. Such emulsions are expected to be discernible only on astronomical scales; on finite lattices simulations reveal no difference from a first-order transition.
For $V \ge 30$ (not shown) we find the first evidence for additional plateaus at various fillings below $n=1/3$, which are not present in the short-range model. With increasing system size the number of plateaus grows and they are separated by small superfluid regions. It is expected that an incommensurate, floating solid is formed in the thermodynamic limit for strong interactions by analogy to the analysis of Ref.~\cite{Isakov07_polar}.  Note that in the classical limit of zero hopping the long-range model exhibits a devil's staircase (see Refs.~\cite{Bak82_polar, Burnell09_polar}  for 1D) of various solid phases.

Above the commensurate solid at $n=1/3$ we find a continuous second-order phase transition  to a supersolid phase belonging to the universality class of the 3D XY model, similar to what occurs in the short-range model. While near the tip the supersolid phase exists only over a narrow filling factor range, it quickly extends ($V/t=15$) all the way to half filling. For larger interactions ($V/t > 20$) and close to half filling, the structure factor and the superfluid density go down and supersolidity is lost for $V/t = 30$ at and near half filling. At finite temperature, the Kosterlitz-Thouless transition to the superfluid phase for weak interactions posed no problem, but the simulations became very difficult at larger interaction strengths. The optimal transition temperature $T_c/t = 0.53(8)$ for supersolidity was found for a density $n=0.4$ and an interaction strength $V/t = 12$, corresponding to a temperature $T=0.8(2)$nK for LiCs.

The model was also studied on the square lattice in Ref.~\cite{Capogrosso-Sansone10_polar}. They also found a superfluid, a solid, a supersolid and various incommensurate solid phases, but the supersolid transition temperature is generally lower than on the triangular lattice. A supersolid phase was found for both higher and lower densities than half filling, unlike short-range models with hard-core bosons~\cite{Boninsegni05_polar, Wessel05_polar, Heidarian05_polar, Melko05_polar, Melko06_polar}. For hard-core bosons on the square lattice with also next-nearest neighbour hopping~\cite{Chen08_polar, Dang08_polar}, a supersolid was found for densities $n < 1/4$ with star diagonal ordering,  and it was also found for densities $0.25 < n < 0.5$ between a star and stripe solid at half filling~\cite{Dang08_polar}.  For soft-core bosons on a square lattice a supersolid phase exists for $n> 0.5$ with nearest-neighbour interactions only.

Much more theoretical work has been done on polar molecules with mean-field and analytical methods. We list here a few examples.
Polar molecules in their ground state have been suggested to describe quantum magnetism~\cite{Gorshkov11}. The phase diagram of a  Bose-Einstein condensate  with dipolar interactions loaded into an optical lattice with a staggered flux was studied in Ref.~\cite{Thieleman11_polar}.  Apart from uniform superfluid, checkerboard supersolid, and striped supersolid phases, several supersolid phases with staggered vortices were identified, which can be seen as combinations of supersolid phases found in earlier work on dipolar BECs~\cite{Capogrosso-Sansone10_polar, Pollet10_polar} and a staggered-vortex phase found for bosons in optical lattices with staggered flux.

Finally, we note that fermionic dipolar systems have also received a lot of theoretical interest (in the continuum). They cannot be treated by path integral Monte Carlo methods, and a discussion is beyond the scope of this review. We list a few exotic proposals: detection of a  $p_x + ip_y$ fermionic superfluid~\cite{Cooper09_polar}, a  nematic non-Fermi liquid~\cite{Quantanilla09_polar, Fregoso09_polar, Carr10_polar}, and an unusual dimerized "pseudogap" state at intermediate temperature for a layered system of polarized fermionic molecules~\cite{Potter10}.

\subsection{Bose-Bose and Bose-Fermi mixtures}

The Mott phase of a single species of atoms suppresses any low-energy transport. If the Mott phase consists of at least two species, then the net number-of-atoms transport is still suppressed. However,  counterflow, in which the currents of the two species are equal in absolute values but in opposite directions, can survive and become superfluid under certain conditions~\cite{Kuklov03_twocomponent}. In case of a 2-component Bose-Bose mixture, the super-counter-flow (SCF) corresponds to an easy-plane ferromagnet in the strong coupling limit (or an easy-plane anti-ferromagnet for Fermi-Fermi mixtures). Such a phase has a resulting $U(1)$ symmetry, compared to the $U(1) \times U(1)$ symmetry of the full Hamiltonian. The symmetry breaking and phase transitions at double commensurate filling for various inter and intra-species interactions have been discussed in Ref.~\cite{Kuklov04_twocomponent} for the $J-$current model, which is the $(d+1)$ dimensional classical analog of the $d$ dimensional quantum system. The possible phases are (1) $SF_A$ + $SF_B$  (or 2SF) , which are 2 miscible superfluids with non-zero order parameters $\langle \psi_A \rangle \neq 0$ and $\langle \psi_B \rangle \neq 0$, (2) $SF_A$ + $MI_B$ with  $\langle \psi_A \rangle \neq 0$ and $\langle \psi_B \rangle = 0$ and vice versa, (3) $MI_A$ + $MI_B$ with  $\langle \psi_A \rangle = 0$ and $\langle \psi_B \rangle = 0$ , and (4) SCF with $\langle \psi_A \rangle = 0$ and $\langle \psi_B \rangle = 0$ but $\langle \psi_A \psi_B^{\dagger} \rangle \neq 0$. For attractive intra-species interactions, it is possible to form a pair-superfluid phase (PSF) with order parameter  $\langle \psi_A \psi_B \rangle \neq 0$ for equal densities (directly translating in a SCF phase for repulsive interactions at commensurate fillings). It was shown that the universal properties of the $2SF-PSF$ and $2SF-SCF$ quantum phase transitions can be mapped onto each other and that their universality class is identical to the one of th$(d+1)$ dimensional normal-superfluid transition in a single-component liquid~\cite{Kuklov04_twocomponent}. Off-diagonal long-range order for the PSF and SCF phases is seen in the two-body density matrix. It is hence crucial to allow for updates in the Monte Carlo sampling such that the paths can wind together around the system volume, which can be accomplished by having worm operators for each species simultaneously in configuration space hereby sampling the two-body Green function. The worm operators $Q_{2b}$ in the two-body channel take the form (the worm operators in the one-body channel for each component separately can be kept, but do not lead to an ergodic algorithm on their own)
\begin{equation}
Q_{2b} = \sum_{i,j} \int_0^{\beta} d\tau \int_0^{\beta} d\tau' (a_i^{\dagger}(\tau) b_j(\tau') + a_i^{\dagger}(\tau) b_j^{\dagger}(\tau') + a_i(\tau)b_j^{\dagger}(\tau') + a_i(\tau) b_j(\tau')
\end{equation}
Note that this will result in an algorithm that is a factor $\beta L^d$ slower than the original worm algorithm for the single species case. However, off-diagonal long-range order in the two-body channel matters most when the imaginary times of the worm operators are close to each other. Hence, when inserting the worm-pair of the second species, one should insert this pair closely to either the worm head or tail of the other species. One can further restrict the sampling of the two-body Green function (and reducing the algorithmic slowdown) by considering a reduced worm pair~\cite{Ohgoe11_twocomponent} $Q = \sum_i \int_0^{\beta} d\tau ( a_i^{\dagger}(\tau) b_i(\tau) + a_i(\tau) b_i^{\dagger}(\tau))$ (for SCF and analogously for PSF) and adding an appropriate set of one-body worm operations of the same type ($a$ or $b$).

\begin{figure}
\begin{center}
\includegraphics[width=0.8\columnwidth]{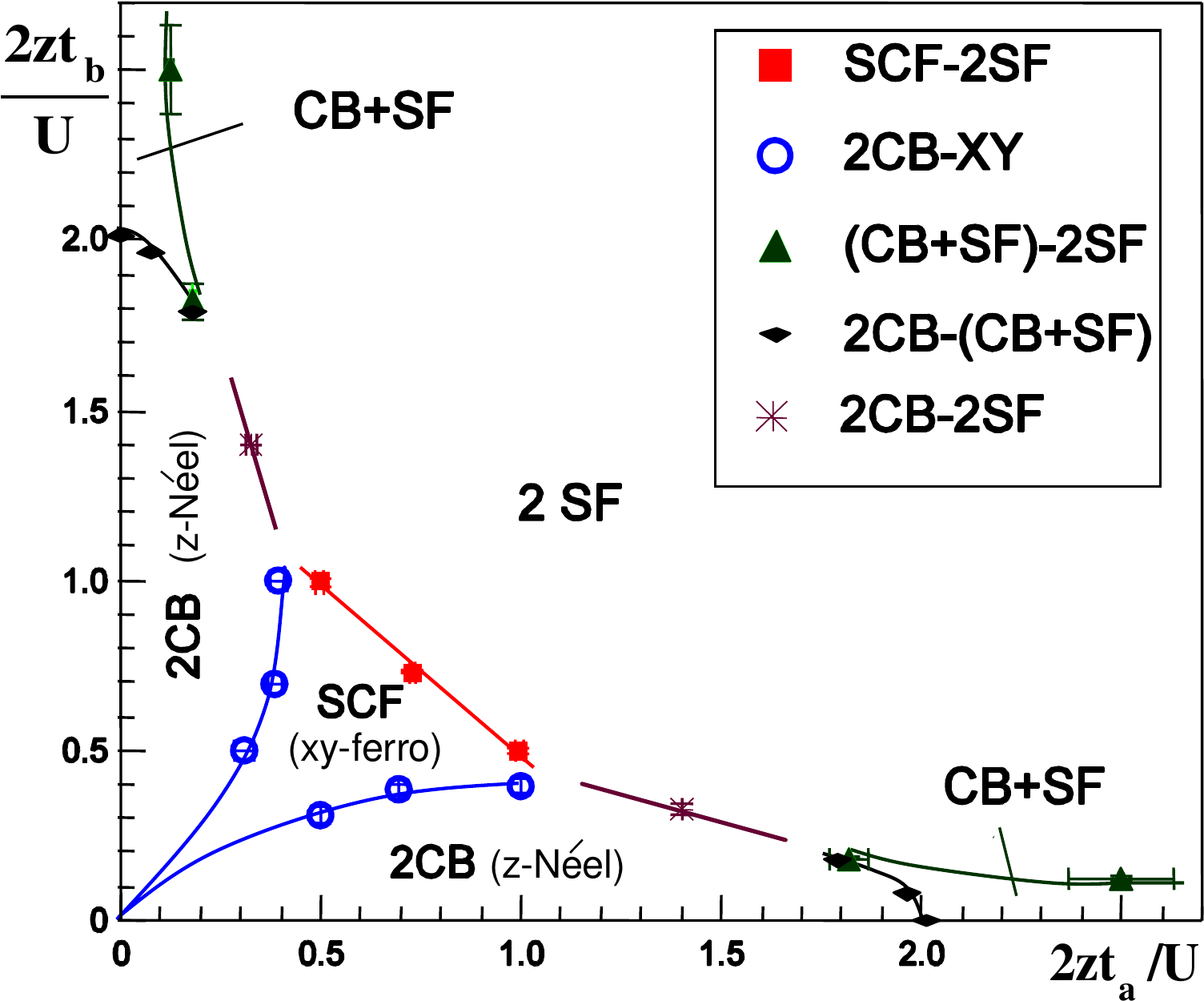}
\caption{Ground state phase diagram of the 2D two-component Bose-Hubbard model at double half-integer filling. The intra-species on-site interactions are infinitely strong (hard-core limit), while the inter-species interaction, $U$, is finite. The respective hopping amplitudes are $t_a$ and $t_b$, and $z = 4$ is the coordination number. The revealed phases are (i) a checkerboard solid in both components or z-N{\'e}el phase (2CB), (ii) a checkerboard solid in one component and superfluid in the other  (CB+SF), (iii) a superfluid in both components (2SF), (iv) a super-counter-fluid or XY-ferromagnet (SCF). The observed transition lines are: 2CB-SCF (first-order), SCF-2SF (second-order), 2CB-2SF (first-order), 2CB- CB+SF (second-order), and CB+SF-2SF (first-order).  
Reprinted figure with permission from Ref.~\cite{Soyler09_twocomponent}. Copyright (2009) by the American Physical Society.
}
\label{fig:twocomponent_phasediagram}
\end{center}
\end{figure}

The ground state phase diagram of the quantum two-component ('a' and 'b') Bose-Hubbard model for hard-core bosons with nearest-neighbour hopping and on-site intra-species interaction in the Hamiltonian
\begin{equation}
H  =  -t_a \sum_{\langle i,j \rangle} a_i^{\dagger} a_j - t_b \sum_{\langle i,j \rangle} b_i^{\dagger} b_j + U \sum_i n^a_i n^b_i
\end{equation}
at double half filling has been analyzed in Refs.~\cite{Soyler09_twocomponent, Capogrosso-Sansone10_twocomponent} and is shown in Fig.~\ref{fig:twocomponent_phasediagram}. It displays all the insulating and superfluid phases anticipated in the previous paragraph. The transitions at finite temperature have also been analyzed in 2D and 3D~\cite{Capogrosso-Sansone10_twocomponent}. For the Ising case in 3D (2CB), $T_c$ was found to be $T_c / t_b = 0.175(15)$ for $U/t_b = 11, t_a / t_b = 0.1$, which is where $T_c$ is expected to be the largest. This corresponds to an entropy per particle of $S/N = 0.5(1)$.
For the melting of the $xy-$ferromagnet (SCF) in 3D, $T_c/t_b = 0.208(7)$ for $U/t_b = 21, t_a / t_b = 1$ was reported, with a critical entropy per particle of $S/N = 0.35(4)$~\cite{Capogrosso-Sansone10_twocomponent}. In 2D, the critical entropy for the Ising transition is about a factor of 2 lower than in 3D, for the $xy-$ferromagnetic transition it is an order of magnitude lower because of the Kosterlitz-Thouless nature of the transition. The entropies in 3D are within reach of current experimental setups; however the time scales needed for establishing magnetic ordering are large compared to the lifetime and remain challenging to achieve~\cite{Capogrosso-Sansone10_twocomponent}. The phase diagram of the two-component Bose-Hubbard model was also obtained in the  dynamical mean-field theory (see Sec.~\ref{sec:bdmft}) approximation~\cite{Hubener09_twocomponent}. The change of the shape of the Mott lobe upon increasing the concentration of the second Mott lobe was studied in Ref.~\cite{Guglielmino10_twocomponent}, and the realization of such phases in a system with a parabolic trap was discussed in Ref.~\cite{Guglielmino11_twocomponent}.

Mixtures including conversion terms have been studied on a 2D lattice in Ref.~\cite{Forges11_twocomponent}.  When interspecies interactions are smaller than the intraspecies ones, the system is unpolarized, whereas in the opposite case the system is unpolarized in even Mott insulator lobes and polarized in odd Mott lobes and also in the superfluid phase. In the latter case, the transition between the Mott insulator of total density $2$ and the superfluid can be of either second or first order depending on the relative values of the interactions, whereas the transitions are continuous in all other cases.

The one-dimensional phase diagram of a two-component bosonic system where additionally pairs of type A can locally be annihilated and simultaneously pairs of type B be created was studied in Ref.~\cite{Forges10_twocomponent}.  The phase diagram of the ferromagnetic case (positive pair hopping) was found in close agreement to the phase diagram found with spinor bosons (see Sec.~\ref{sec:spinor}). In this case, the population is always balanced.  In the antiferromagnetic case however, the superfluid phase is always polarized. The second Mott lobe has population balance, but a transition inside the first Mott lobe from a balanced population for strong interactions toward a polarized Mott insulator for weak interactions was found.

One-dimensional Bose-Fermi mixtures (with spin polarized fermions) are, after a Jordan-Wigner transformation on the fermions, also amenable for path integral Monte Carlo studies. Most of the reported works were done in the canonical ensemble using the worm algorithm variants of Ref.~\cite{Pollet06_twocomponent}. The possible phases translate directly to their higher dimensional analogs  mentioned above~\cite{Pollet06_twocomponent, Pollet08_twocomponent, Hebert08_twocomponent, Hebert07_twocomponent, Zujev08_twocomponent, Varney08_twocomponent}. In particular at double half filling with equal hopping amplitudes, phases consisting of two miscible Luttinger liquids (corresponding to 2SF), a spin density wave (corresponding to SCF), and a ferromagnet (phase separation) were found~\cite{Pollet06_twocomponent}. Here too, in case the boson-boson and the fermion-boson repulsion is very large, a mapping to a spin model can be done. The {\it XXZ} Hamiltonian $H^{\rm XXZ} = \sum_i J (\sigma^x_i \sigma^x_{i+1} + \sigma^y_i \sigma^y_{i+1}) + J_z \sigma_i^z + \sigma_{i+1}^z$ with $J = - \frac{t_B t_F}{U_BF}$ and $J_z = \frac{t_B^2 + t_F^2}{2U_{\rm BF}} - \frac{t_b^2}{U_{BB}}$ has a first order transition from the gapless SDW phase toward the ferromagnet for $J=J^z$ or $U_{\rm BF} / U_{\rm BB} = 2$ for $t_b = t_F = 1$.  When the hopping amplitudes are unequal, a transition between the gapless SDW phase and an Ising Neel phase is also possible~\cite{Pollet06_twocomponent}.

Cold atom experiments with a $^{87}$Rb - $^{41}$K Bose-Bose mixture~\cite{Catani08_twocomponent} or a $^{87}$Rb - $^{40}$K Bose-Fermi mixture~\cite{Guenter06_twocomponent, Ospelkaus06_twocomponent} focused on the loss of coherence of the $^{87}$Rb bosonic atoms when adding the second species, notwithstanding their different statistics and interspecies interactions (even the sign!). By using a Feshbach resonance the interaction strength between the $^{87}$Rb bosons and the  $^{40}$K  fermions could be tuned~\cite{Best09_twocomponent}, where a strong asymmetry between the repulsive and the attractive side was found. For a bosonic mixture of the $\vert F=1, m_F = -1 \rangle$ and $ \vert F=2, m_F = -2 \rangle$ hyperfine levels of $^{87}$Rb atoms, it was found that the presence of a second component can reduce the apparent superfluid coherence, most significantly when the second component either experiences a strongly localizing lattice potential or none at all~\cite{Gadway10_twocomponent}. Mixtures with different bosonic and fermionic isotopes of Yb were also recently studied experimentally in an optical lattice~\cite{Sugawa11_twocomponent}. It was found that an interspecies interaction between bosons and fermions leads to drastic changes, causing effects that include melting, generation of composite particles, an anti-correlated phase and complete phase separation. A number of theoretical explanations which include heating in order to conserve entropy~\cite{Guenter06_twocomponent, Pollet06_twocomponent, Cramer08_twocomponent, Cramer11_twocomponent} , the multiband model~\cite{Tewari09_twocomponent},  the self-trapping effect~\cite{Luehmann08_twocomponent, Best09_twocomponent}, changes in the chemical potential due to the presence of external harmonic confinement~\cite{Buonsante08_twocomponent}, and polaronic effects~\cite{Guglielmino10_twocomponent, Gadway10_twocomponent} have been put forward to explain the experimental observations. In general, it is fair to state that for multi-component systems a similar quantitative understanding as for the single species Bose-Hubbard model is still lacking.

In a series of experiments at MIT~\cite{Weld09_twocomponent, Weld10_twocomponent, Medley11_twocomponent} a magnetic field gradient was used to initially separate the two components spatially. The region of overlap forms a domain wall, and its width can be used as a thermometer. It can be shown that the average magnetization $ \langle s \rangle = {\rm tanh}( - \beta \Delta {\mathbf \mu B}(x)/2)$, with $\Delta {\mathbf {\mu}}$ the difference in magnetic moment between the two components, and ${\mathbf B}(x)$ the position-dependent magnetic field. The temperature is hence directly measurable, and temperatures of the order of a few tens of pK have been reported. The latter is obtained by spin gradient demagnetization: the magnetic field gradient is slowly reduced, so that the two components can mix over a wider region where more states become available hereby reducing the temperature~\cite{Medley11_twocomponent}. 

A different route to quantum magnetism was followed in Ref.~\cite{Simon11_twocomponent}, following a proposal by S. Sachdev {\it et al.}~\cite{Sachdev02_twocomponent} in the context of the first superfluid-to-Mott insulator transition experiments~\cite{Greiner2002} with single species. If a tilt is applied to the lattice, then atoms prepared in a Mott insulator, can only tunnel if the linear tilt $E$ equals the potential energy. Hence, if $E = U$, an atom can tunnel to a neighbouring site, provided the atom there has not tunneled yet. This creates an effective spin-spin interaction within the resonant subspace. The quantum phase  transition from a paramagnetic to an antiferromagnet was observed in Ref.~\cite{Simon11_twocomponent} in 1D. The main advantage of this scheme is that the relevant time scale is set by the hopping $t$ and not by the superexchange scale $\sim t^2/U$ and that the initial entropies in the Mott insulator can be very low. More sophisticated lattice geometries will produce frustrated systems with novel quantum liquid and dimer covered ground states~\cite{Pielawa11_twocomponent}.

\subsection{Spinor bosons}
\label{sec:spinor}

\begin{figure}
\begin{center}
\includegraphics[width=0.8\columnwidth]{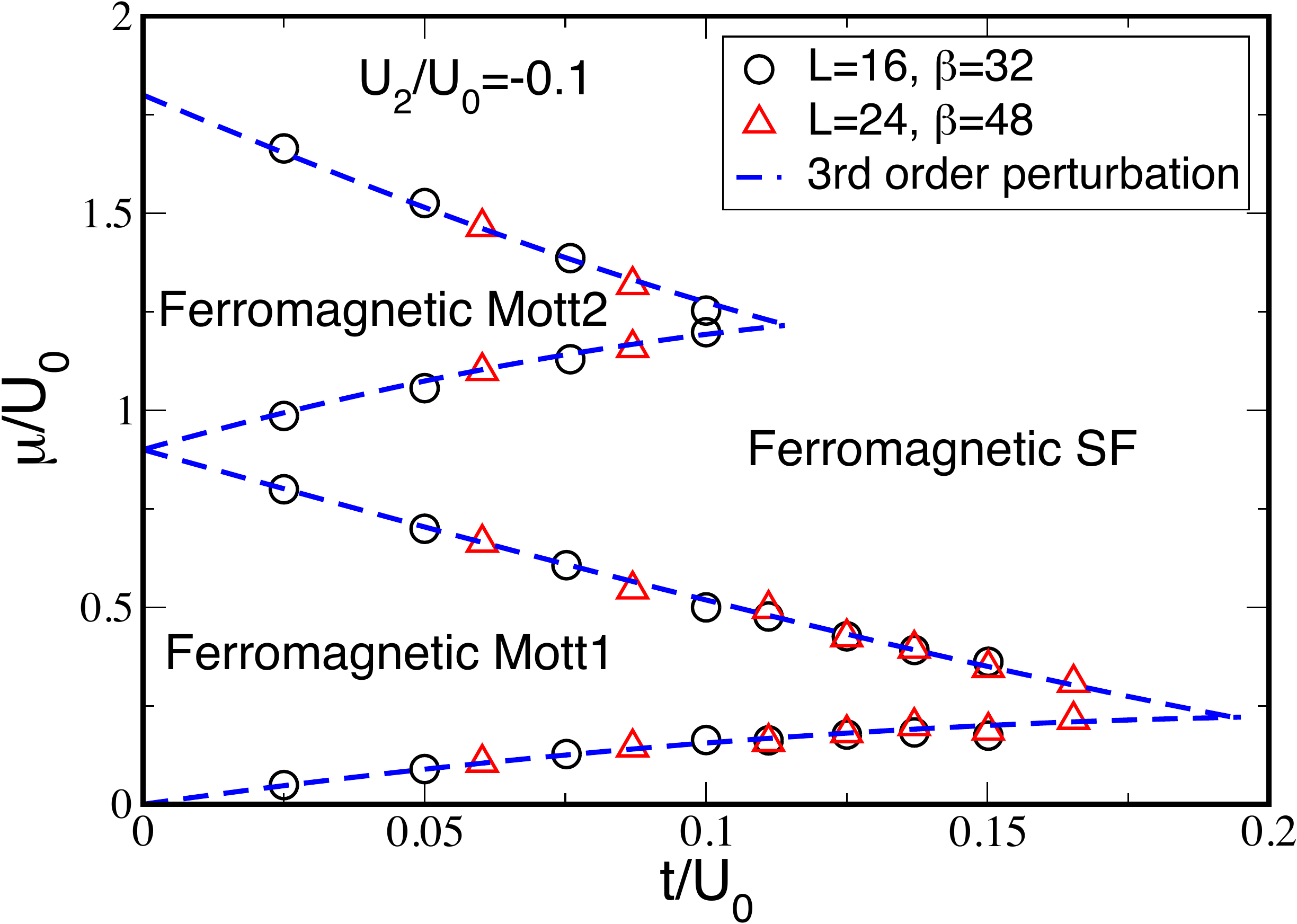}
\caption{Phase diagram of the spin-1 Bose-Hubbard model in 1D for $U_2 / U_0 = -0.1$. 
Reprinted figure with permission from Ref.~\cite{Batrouni09}. Copyright (2009) by the American Physical Society.
}
\label{fig:spinor_phasediagram}
\end{center}
\end{figure}

Bosons with an internal spin degree of freedom on a lattice are described by the following spin Bose-Hubbard model,
\begin{equation}
H = -t \sum_{\langle i,j \rangle, \sigma} b_{i, \sigma}^{\dagger}b_{j,\sigma} + \frac{U_0}{2} \sum_i n_i(n_i-1) + \frac{U_2}{2} \sum_i ( F_i^2 - 2 n_i),
\label{eq:BoseHubbard_spin}
\end{equation}
in standard lattice notation.
The spin operator $F_i = \sum_{\sigma}^{\dagger} F_{\sigma, \sigma'} a_{\sigma'}$ with $F_{\sigma, \sigma'} $ the standard spin-F matrices contain contact interactions as well as interconversion terms between the spins. This model has been studied for $F=1$ in 1D in the ground state by using QMC~\cite{Apaja04, Batrouni09} and DMRG~\cite{Rizzi05, Bergkvist06}.
When $U_2 < 0$ ferromagnetism is favoured. The resulting phase diagram is shown in Fig.~\ref{fig:spinor_phasediagram}. Compared to the phase diagram of scalar bosons, the main difference is the shrinking of the base of the Mott lobes on the $y$-axis. The transition between the ferromagnetic Mott lobes and the ferromagnetic superfluid is continuous. When $U_2 > 0$ ({\it e.g.,} $^{23}$Na), low total spin states are favored. In the absence of hopping, the base of the even Mott lobes grow at the expense of the odd ones, which disappear entirely for $U_0 = 2U_2$. The transition between even(odd) lobes is first(second) order. Mean-field theory~\cite{Demler02, Snoek04, Imambekov03, Imambekov04, Krutitsky04, Kimura05, Pai08} predicts for $2dU_2/U_0 < 0.1$ in $d=2,3$ that, when $t/U_0^{c1} \sim \sqrt{U_2/4dU_0}$, the Mott lobes of even order are comprised of two phases: (a) a singlet phase for $t/U_0 \le t/U_0^{c1}$ and (b) a nematic phase for $t/U_0^{c1} \le t/U_0 \le t/U_0^{c}$ where $t/U_0^{c}$ is the tip of the Mott lobe. The nematic phase breaks spin rotational symmetry but preserves time reversal symmetry and has gapless spin-wave excitations. The spin-singlet phase does not break spin symmetry and has a gap to all excitations. Inside the Mott lobe, mean-field theory predicts that the nematic-to-singlet transition is first order. Mean-field theory also predicts for $2dU_2/U_0 > 0.1$ in $d=2,3$ that even Mott lobes are entirely in the singlet phase and odd Mott lobes entirely nematic. The superfluid polar phase has broken spin rotational symmetry without breaking time reversal symmetry.

The 1D QMC simulations of~\cite{Batrouni09} however find only signs of a crossover between the nematic and singlet phase inside the even Mott lobes by investigating the behaviour of $\langle F^2 \rangle \to 0$. The first Mott lobe was entirely dimerized, as was also found in an earlier study~\cite{Apaja04} and in mean-field theory~\cite{Imambekov03}. 

In experiments, the dipolar ferromagnetic interactions of $^{87}$Rb have given rise to spin textures after a rapid quench across the ferromagnetic phase transition~\cite{Sadler06, Vengalattore08}, when simultaneous magnetic ordering and atomic superfluidity was observed. So far, no experiments have been performed on spinor bosons in an optical lattice.

It would be very interesting to extend the present QMC studies to higher dimensions and check in particular if there is a true nematic-to-singlet transition inside the even Mott lobes for $U_2 > 0$. Also the nature of the transitions between the phases remain poorly understood. Algorithmic advances are needed to study bigger system sizes than the $\sim 30$ sites studied so far. It would also be interesting to compute with QMC the entropy at the phase transitions (so far only mean-field results exist~\cite{Mahmud10}), both with and without a confining potential, to see under what conditions an experimental realization of $F=1$ and possibly higher spin $F$ systems is feasible.


\section{Path Integral Monte Carlo : continuous space models}
\label{sec:PIMC_continuous}

After having described path integral Monte Carlo simulations on a lattice with continuous imaginary time, we move on to path integral Monte Carlo methods in continuous space, which can only be formulated with discrete imaginary time. The applications will be similar to the ones discussed for lattice systems: scalar bosons and the superfluid to normal transition, disordered scalar bosons and supersolids for bosons with long-range interactions.

\subsection{Methods}

\begin{figure}
\begin{center}
\includegraphics[width=0.8\columnwidth]{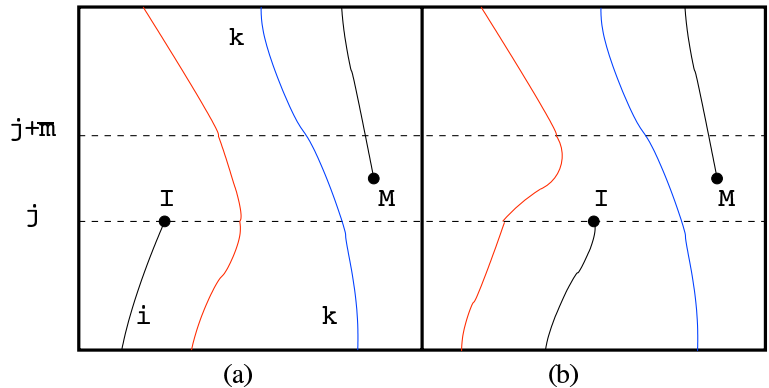}
\caption{Illustration of the "swap" or "reconnect" update in the worm algorithm for continuous-space models. The x-axis denotes space, the y-axis imaginary time. In (a) there are two full paths, and two segments which terminate on the worm head (I) and the worm tail (M). The worm head is at time slice $j$. A slice $j + \bar{m}$ is chosen. After the update (b), the worm head has jumped to a neighbouring path, and a new segment is generated from time slice $j$ to time slice $j + \bar{m}$, connecting the old position of the worm head with the previously existing path. The previously existing path between $j$ and $j + \bar{M}$ is erased. 
Reprinted figure with permission from Ref.~\cite{Boninsegni05}. Copyright (2006) by the American Physical Society.
}
\label{fig:PIMC_swap}
\end{center}
\end{figure}

We consider a many-body system with Hamiltonian
\begin{equation}
H = H_1 + H_2 = -\frac{\hbar^2}{2m} \sum_i \nabla^2_i + \sum_{i < j} V( \vert {\mathbf r}_i - \mathbf{r}_j \vert ),
\end{equation}
where $V$ is a pairwise interaction depending only on the relative distance between the particles. Note that the algorithm can straightforwardly be formulated for more general interactions, but we restrict ourselves to the standard case of a central potential for simplicity. $H_1$ is the kinetic energy term; $H_2$ the potential energy term.
The notation ${\mathbf R} \equiv ({\mathbf r}_1, \mathbf{r}_2, \ldots \mathbf{r}_N)$ will be used to denote the positions of all particles in the system.

The partition function is the trace over the density operator, $Z = {\rm Tr} e^{-\beta H}$. The position-space density matrix is $\rho( {\mathbf R}, {\mathbf R'}, \beta) = \langle {\mathbf R} \vert e^{-\beta H} \vert {\mathbf R'} \rangle$. Note that the product of two density matrices is again a density matrix since $e^{(-\beta_1 + \beta_2)H} =e^{-\beta_1H} e^{-\beta_2H}$. When repeating this $M$ times, and defining the time step $\tau = \beta / M$, one has a discrete path. In general, the kinetic energy operator and the potential energy operator do not commute, but for $M$ large enough, the primitive approximation
\begin{equation}
e^{-\tau (H_1 + H_2)} \approx e^{-\tau H_1} e^{-\tau H_2}
\end{equation}
can be used. It has an error of order $\tau^2$ which can be neglected. According to the Trotter formula, $e^{-\beta (H_1 +H_2)} = \lim_{M \to \infty} [ e^{-\tau H_1 }e^{- \tau H_2}]^M$, valid for self-adjoint operators $H_1$, $H_2$ and $H_1 + H_2$ with a spectrum bounded below, it becomes exact for $M \to \infty$ ~\cite{Trotter59}. In practice one uses a higher-order scheme such as the Chin formula~\cite{Chin02}, but we continue with the primitive approximation in order not to overload the notation.

For the evaluation of the position-space density matrix in the primitive approximation,
\begin{equation}
\rho( {\mathbf R}_0, {\mathbf R}_2, \tau) \approx \int d {\mathbf R}_1 \langle {\mathbf R}_0 \vert e^{-\tau H_1} \vert {\mathbf R}_1 \rangle \langle {\mathbf R}_1 \vert e^{-\tau H_2} \vert {\mathbf R}_2 \rangle, 
\end{equation}
we need to know the kinetic and the potential energy density matrices. Since the potential energy is diagonal in position space, its matrix element is trivial, 
$\langle {\mathbf R}_1 \vert e^{-\tau H_2} \vert {\mathbf R}_2 \rangle = e^{-\tau H_2 ( {\mathbf R}_1) } \delta({\mathbf R}_1 - {\mathbf R}_2)$.  The kinetic energy term is a one-body operator and can be diagonalized by a Fourier transform. Replacing the finite sum over the momenta of a finite box by a continuous integral (see also Ref.~\cite{Ceperley95}),
\begin{equation}
\langle {\mathbf R}_0 \vert e^{-\tau H_1} \vert {\mathbf R}_1 \rangle = (4 \pi \hbar^2 \tau / (2m) )^{-dN/2} e^{-\frac{2m({\mathbf R}_0 - {\mathbf R}_1)^2}{4 \hbar^2 \tau} }.
\label{eq:PIMC_gauss}
\end{equation}
It is the appearance of $\tau$ in the denominator of the exponential that prevents a formulation of the algorithm without Trotter error.
We now arrive at a discrete path integral expression for the density matrix in the primitive approximation,
\begin{eqnarray}
\rho( {\mathbf R}_0, {\mathbf R}_M, \beta) & = & \int d  {\mathbf R}_1 \ldots d {\mathbf R}_{M-1} (4 \pi \hbar^2 \tau / (2m))^{-dNM/2} \nonumber \\
{} & {} & \exp \left( -\sum_{m=1}^{M} \left[ \frac{ 2m({\mathbf R}_{m-1} - {\mathbf R}_m)^2 }{4 \hbar^2 \tau} + \tau V({\mathbf R}_m) \right] \right).
\end{eqnarray}
Because of the indistinguishability of bosons, the density matrix is always understood as the sum over all permutations,
\begin{equation}
\rho( {\mathbf R}_0, {\mathbf R}_M, \beta) = \frac{1}{N!} \rho( {\mathbf R}_0, {\mathcal P}  {\mathbf R}_M, \beta) 
\end{equation}

Evaluation of physical observables requires the sampling over all possible paths: all possible positions of the particles and all possible exchanges~\cite{Ceperley95}. Superfluid properties for large systems can only be simulated efficiently using the same 'worm'-idea we have seen in Sec.~\ref{sec:PIMC_lattice}: one works with open segments of paths (or worldlines) instead of closed worldlines, a procedure which directly samples the Green function. Off-diagonal long-range order can then be sampled efficiently since the one-body density matrix (that is the equal-time Green function) has a large weight there. For the updates we refer the reader to the full discussion in Refs.~\cite{Boninsegni05, Boninsegni06}. The crucial update is the "swap" or "reconnect" update, in which the worm head jumps to a neighbouring worldline, hereby performing a bosonic exchange. This is illustrated in Fig.~\ref{fig:PIMC_swap}. The update can be done with acceptance factors of order unity~\cite{Boninsegni05, Boninsegni06}. An additional advantage of the worm algorithm is that it works directly in the grand-canonical ensemble.

When the worm algorithm for continuous space was introduced in Refs.~\cite{Boninsegni05, Boninsegni06} it was illustrated for the U(1) transition between normal and superfluid $^4$He in 2D (which is a Kosterlitz-Thouless transition) and 3D (which is the so-called {\it lambda} transition) with a few thousand particles, which is roughly a factor 100 higher than what was previously achievable. The transition temperature $T_c$ agreed better than $0.5\%$ with experiment. The deviation is explained by the neglect of three-body (and higher)  interactions in the Aziz potential~\cite{Aziz79}. Access to larger particle numbers was crucial in the study of solid $^4$He and the existence of a possible supersolid phase. Path integral Monte Carlo simulations have shown that the ideal hcp solid is an insulator~\cite{Boninsegni2006_helium_1, Clark2006} and that vacancies phase separate~\cite{Boninsegni2006_helium_2}, but that defects such as grain boundaries~\cite{Pollet2007_helium} and dislocations~\cite{Boninsegni2007_helium} may be superfluid depending on the elastic properties~\cite{Pollet2008_helium}. The controversy on the issue of supersolid defects in Helium still continues but this discussion is beyond the scope of this review. For a recent review on the properties of (super)solid Helium, see Ref.~\cite{Prokofev2007}, and for a trend article see~\cite{Kuklov2011_helium}.

\subsection{Weakly Interacting Bose gas}
\label{sec:WIBG_Tc}

\begin{figure}
\begin{center}
\includegraphics[width=0.6\columnwidth]{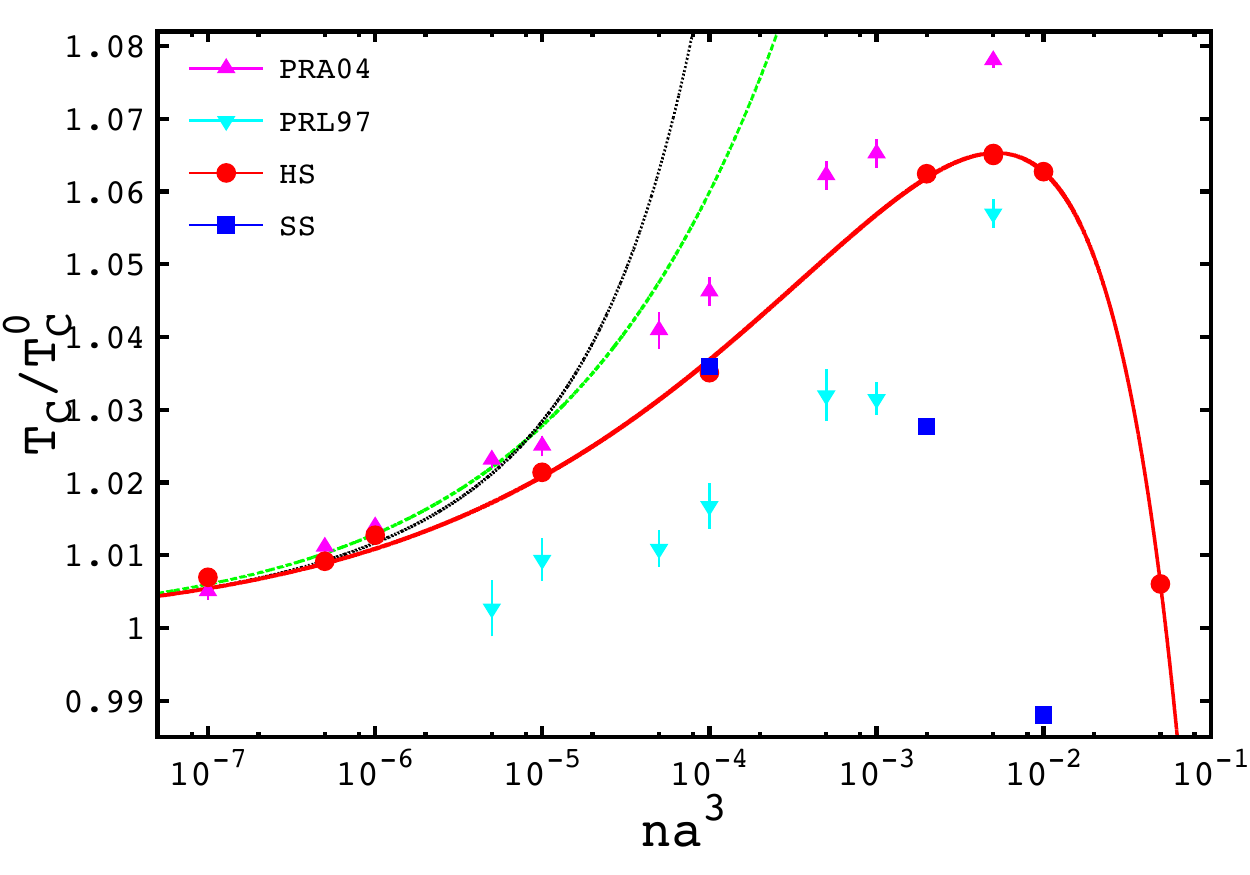}
\caption{ Critical temperature of the 3D dilute bose gas as function of the gas parameter $na^3$. The symbols labeled by PRA04 correspond to the results of Ref.~\cite{Nho04}, the ones labeled by PRL97 correspond to Ref.~\cite{Grueter97}. The dashed line (green) is the expansion (1) of Ref.~\cite{Kashurnikov01} and the dotted line (black) is the expansion of Ref.~\cite{Arnold01a} including logarithmic corrections. The solid line (red) is a guide to the eye. 
Reprinted figure with permission from Ref.~\cite{Pilati08}. Copyright (2008) by the American Physical Society.}
\label{fig:PIMC_dilutegas}
\end{center}
\end{figure}

For a system such as superfluid $^4$He or solid $^4$He, the pairwise potential is typically the Aziz-potential~\cite{Aziz79}, which looks very similar to a Lennard-Jones potential.
For cold atoms however, the true interatomic potential is often poorly known. Because of the diluteness of the system and the low momenta and energies involved in typical collisions, only the $s-$wave scattering length matters, provided the effective range is small enough.
Different model potentials can then be used such that the low-energy scattering properties are reproduced. Typical choices include a hard-sphere potential ($V^{\rm HS}(r) = \infty, r < a$ and zero otherwise), a soft-sphere potential  ($V^{\rm SS}(r) = V_0, r < R_0$ and $V_0 > 0$,  and the potential is zero otherwise), or a negative power potential ($V^{\rm NP}(r) = \alpha/r^p$ with $\alpha > 0$ and $p > 3$). In all cases the scattering length $a$ can be computed analytically; in particular for hard-sphere potentials the cutoff is the same as the scattering length while for soft-sphere potentials it is always larger~\cite{LandauLifshitz}.

Instead of the primitive approximation (or a higher order scheme) it is often advantageous to use the pair-product approximation,
\begin{equation}
\rho( {\mathbf R}, {\mathbf R}', \tau) =  \prod_{i=1}^{N} \rho_1({\mathbf r}_i, {\mathbf r}'_i, \tau) \prod_{i<j} \frac{\rho_{\rm rel}({\mathbf r}_{ij}, {\mathbf r}'_{ij}, \tau )} {\rho^0_{\rm rel} ({\mathbf r}_{ij}, {\mathbf r}'_{ij}, \tau ) },
\end{equation}
where $\rho_1$ is the single-particle density-matrix (Eq.~\ref{eq:PIMC_gauss}) and $\rho_{\rm rel}$ is the two-body density matrix of the interacting system which depends only on the relative coordinates ${\mathbf r}_{\rm ij} = {\mathbf r}_i - {\mathbf r}_j$. The latter is divided by the corresponding ideal-gas term, which is given by Eq.~\ref{eq:PIMC_gauss} with the replacement $m \to m/2$. The computation of the two-body density matrix requires only solving a (radial) Schr{\"o}dinger equation~\cite{Pollock84_PPA, Pollock87_PPA, Ceperley95, Pilati06}, but in the special case of hard-sphere potential the analytic approximation introduced by Cao and Berne~\cite{CaoBerne} is very accurate and therefore used in practice,
\begin{equation}
\frac{\rho^{\rm CB}_{\rm rel}({\mathbf r}_{ij}, {\mathbf r}'_{ij}, \tau } {\rho^0_{\rm rel} ({\mathbf r}_{ij}, {\mathbf r}'_{ij}, \tau )} = 1 - \frac{a(r+r') - a^2}{rr'}  e^{- [rr' + a^2 -a(r+r') ] (1 + \cos \theta) m / (2 \hbar^2 \tau)},
\end{equation}
with $a$ the cut-off (equivalent to the scattering length) of the hard-sphere potential and $\theta$ the angle between ${\mathbf r}$ and ${\mathbf r}$'. 
For a hard-sphere and soft-sphere potential there is no choice but to use the pair-product approximation since the primitive approximation is not valid when the potential has discontinuities. 

This method has been used to compute the equation of state of the Bose gas both in the normal and the superfluid phase~\cite{Pilati06}. It was also used in Ref.~\cite{WIBG} to compare the properties of the dilute gas in 1D, 2D and 3D against an improved Beliaev diagrammatic technique.
The worm algorithm also allowed for a more precise determination of the critical temperature of the 3D dilute Bose gas~\cite{Pilati08}. Without interactions, the universality class of Bose-Einstein condensation belongs to the Gaussian complex-field universality class, but with interactions this changes to an XY model. Thus, the critical temperature $T_c$ with interactions cannot perturbatively be obtained from $T_c^0 = (2\pi\hbar^2/mk_B)[n/\zeta(3/2)]^{2/3}$ with $\zeta(3/2) = 2.612)$, which is the transition temperature for the non-interacting model~\cite{Baym99}. The deviation from $T_c^0$ is parametrized as
\begin{equation}
T_c = T_c^0(1 + c(an^{1/3})).
\label{eq:bosegas_tc}
\end{equation}
The linear change in the scattering length was predicted by Lee and Yang in 1958~\cite{LeeYang}, but no information on the numerical coefficient $c$ was provided, not even its sign. Rigorous upper bounds on $T_c$ can be proved~\cite{Seiringer08}, but they are much weaker (going as the root of $na^3$). Ref.~\cite{Seiringer08} also provides an overview of different predictions for $T_c$. 
The numerical coefficient $c$  was calculated in Refs.~\cite{Kashurnikov01, Arnold01b} by solving the effective 3D classical $\vert \psi \vert^4$ model using lattice Monte Carlo simulations, clearly establishing the linear behaviour in Eq.~\ref{eq:bosegas_tc}.
The reported (universal) value is $c=1.29(5)$.	The	same classical model was used in Ref.~\cite{Arnold01a} to calculate higher order logarithmic corrections to Eq.~\ref{eq:bosegas_tc}. In Refs.~\cite{Grueter97, Nho04} conventional PIMC was used, but simulations suffered from small particle numbers and an inefficient calculation of superfluid properties, and even large discrepancies at high densities between the two simulations were found. Conventional path integral Monte-Carlo was also applied to the trapped system with in two dimensions~\cite{Krauth1996}. The worm algorithm simulations of Ref.~\cite{Pilati08} addressed considerably larger particle numbers (up to $10^6$ for $na^3 = 5 \times 10^{-3}$) than Refs.~\cite{Grueter97, Nho04} and number of time slices. For low values of the gas parameter (in the universal regime), the results agree with the classical field calculations. For higher densities $T_c$ first increases, reaches a maximum, and then decreases below $T_c^0$. Simulations with a hard-sphere and a soft-sphere potential gave the same answer for $na^3 \le 10^{-4}$, which is higher than the estimate $na^3 \le 10^{-6}$ for the validity of Eq.~\ref{eq:bosegas_tc}. This is illustrated in Fig.~\ref{fig:PIMC_dilutegas}. Also the Kosterlitz-Thouless transition of the dilute Bose gas in 2D was analyzed for large system sizes in Ref.~\cite{Pilati08}, with good agreement with classical $\vert \psi \vert^4$ theories up to quite large densities. The weakly interacting 2D Bose gas in a harmonic trap was investigated in Ref.~\cite{Holzmann2008, Holzmann10} with conventional path-integral Monte Carlo simulations, addressing the Kosterlitz-Thouless transition and the fluctuation regime and universal properties, respectively.

\subsection{Disordered systems}

Disordered bosons in continuous space were studied in Refs.~\cite{Pilati_disorder1, Pilati_disorder2}. Without a lattice there are no commensurability effects and hence the interplay between a Bose glass phase and a Mott insulator is absent. In Refs.~\cite{Pilati_disorder1, Pilati_disorder2} the suppression of $T_c$ caused by the disorder was addressed. The disorder was modeled by an isotropic 3D speckle potential. Unlike the disorder-free system $T_c$ changed considerably between $na^3 = 10^{-4}$ and $na^3 = 10^{-6}$ (An older study found no substantial drop in $T_c$~\cite{Gordillo00}). Agreement with a  perturbative approach for $\delta$-correlated disorder could not unambiguously be established since the precision in the weak disorder limit was not high enough, and where deviations from $T_c^0$ become appreciable, the perturbative approach is no longer valid~\cite{Lopatin02}. In the regime of weak interactions and strong disorder, the superfluid transition turns out to be well characterized by the existence of a mobility edge, separating localized from extended states, which is largely independent of temperature and interaction strength. In the regime of strong disorder, strong interactions and low temperatures, a phase where the gas is both normal and highly degenerate (with an energy scaling as $\sim T^2$) was identified, which should be related to the Bose glass phase predicted at $T = 0$.

\subsection{Supersolids and long-range interactions}

\begin{figure}
\begin{center}
\includegraphics[width=0.6\columnwidth]{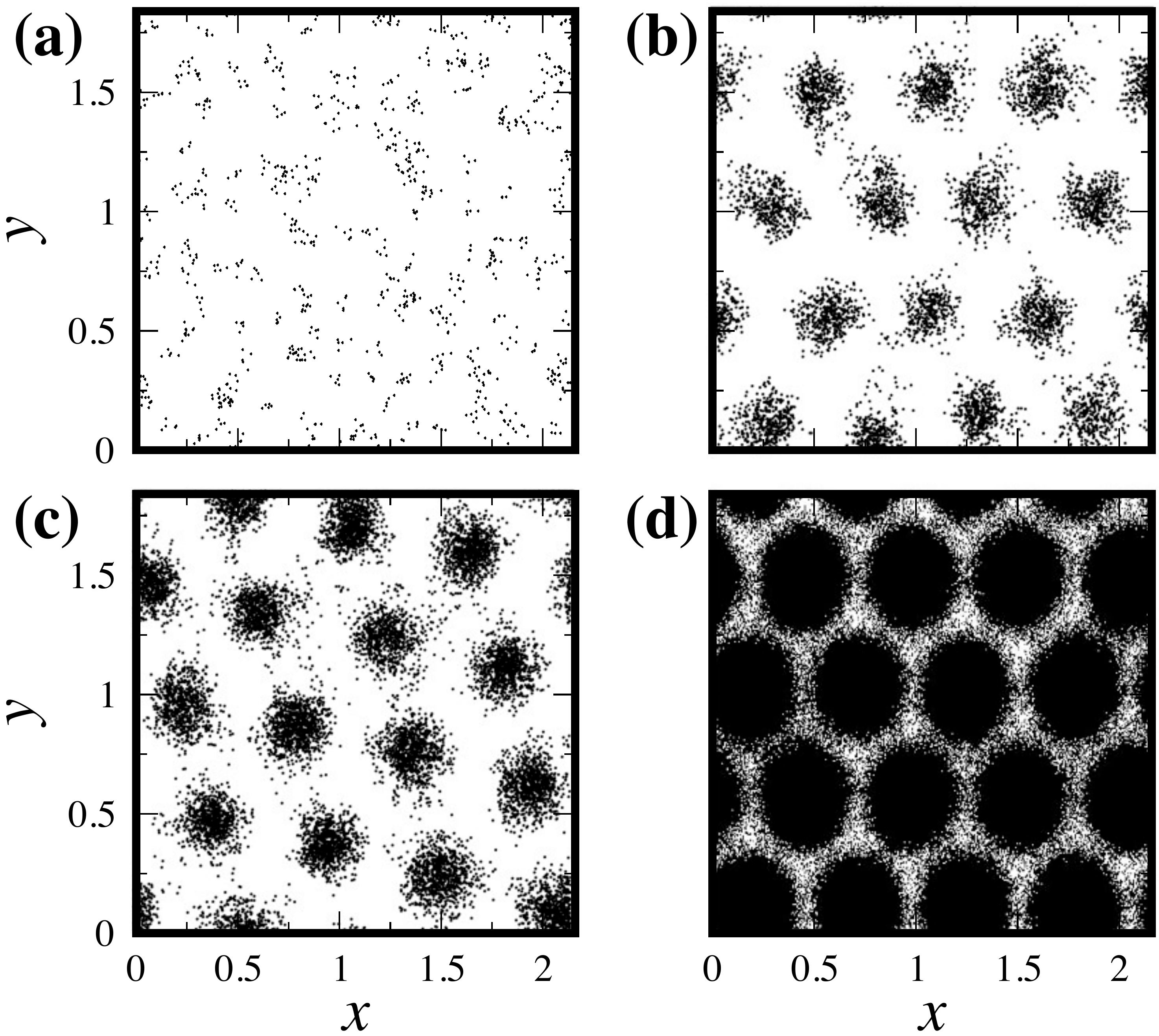}
\caption{ Snapshots of a system of bosons interacting via the potential $V(r) = D/a^3, r \le a$ and $V(r) = D/r^3, r > a$. 
The dimensionless interparticle distance is $r_s = 1/\sqrt{nr_0^2} = 0.14$ with $n$ the density and $r_0$ the characteristic length given by $r_0 = mD/\hbar^2$. 
The cutoff of the potential is set by $a/r_0 = 0.3$. Snapshots for four temperatures are shown, (a) $T = 200  D/r_0^3$, (b) $T = 20 D/r_0^3$, (c) $T = 1.0 D/r_0^3$, and (d) $T = 0.1 D/r_0^3$.  
At high temperature a classical gas is observed. When lowering the temperature, droplets form that become phase coherent at the lowest temperature. 
Reprinted figure with permission from Ref.~\cite{Cinti10_SS}. Copyright (2010) by the American Physical Society.}
\label{fig:droplet}
\end{center}
\end{figure}

Since the experiments by Kim and Chan on solid $^4$He~\cite{KimChan04a, KimChan04b}, interest in the supersolid phase has never faded. It is now generally accepted that the ground state of $^4$He is an insulating, commensurate solid. The question can hence be asked whether there exist supersolids for a single species in continuous, uniform space interacting via a pair potential. Going back to the work of Gross~\cite{Gross57, Gross58}, it is expected on the basis of a Gross-Pitaevskii picture~\cite{Josserand07_SS, Henkel10_SS} that a model with interactions $V(r) = V_0, r \le a$, and zero otherwise, can be supersolid for certain densities (the interparticle distance has certainly to be smaller than the cutoff $a$ and of the order of $a/2$). This is indeed the case. Several potentials with a smoother behaviour were also tried~\cite{Cinti10_SS} such as $V(r) = 1/ (a+r)^3$ and $V(r) = V_0, r \le a$ and $V(r) = V_0/ (r/a)^{\alpha}, r > a$ with $\alpha = 3$ or $6$ all lead to supersolid phases. The picture is one of droplets that become phase coherent as one lowers the temperature, as can be seen in Fig.~\ref{fig:droplet}. All of these potentials can be tailored using polar molecules or Rydberg atoms~\cite{Henkel10_SS, Pupillo10_SS}.

A pure $\sim 1/r^3$ interaction in 2D does not have a supersolid phase~\cite{Buechler07_polar, Astrakharchik07_SS, Mora07_SS}. At weak interactions, a normal fluid is found at high temperature which undergoes a Kosterlitz-Thouless transition to a superfluid~\cite{Filinov10_SS}. At stronger interactions in the ground state there is a commensurate triangular solid~\cite{Buechler07_polar} which is reached from the superfluid phase by a Spivak-Kivelson type transition~\cite{Spivak04_polar}, but which has never directly been seen in simulations because of the large system sizes that are required. Note that emulsions of solid immersed in the liquid phase {\it can} be interpreted as a supersolid phase~\cite{Kuklov2011_helium}.

Dipolar systems have also been studied inside a parabolic potential~\cite{Lozovik04_SS, Pupillo10_SS, Jain11_SS}. It was also shown that a mixture of equal-mass dipolar isotopes, in such a configuration, de-mixes at finite temperature due to quantum statistical effects~\cite{Jain11b_SS}.


\section{Diagrammatic methods}
\label{sec:diagrams}

A diagrammatic expansion for some relevant quantity $Q$ ({\it e.g.}, a Green function) is a series of integrals with an ever increasing number of integration variables,
\begin{equation}
Q(y) = \sum_{m=0}^{\infty} \sum_{\xi_m} \int \mathcal{D} (\xi_m, y, x_1, \ldots, x_m) dx_1 \ldots dx_m.
\label{eq:diagram_expansion}
\end{equation}
Here, $y$ is a set of parameters on which the quantity $Q$ can depend, $\xi_m$ are indices for different terms of the same expansion order $m$, and the $x$ are integration variables~\cite{Prokofev98_froehlich, VanHoucke08}. 
In diagrammatic Monte Carlo, these integrals are not performed explicitly (unlike analytical methods) but configurations are generated for specific values of $m, \xi_m$ and $x_1 \ldots x_m$. The summation over these variables is done by Monte Carlo sampling. Almost all methods considered in this paper can be written in the form of Eq.~\ref{eq:diagram_expansion}; they only differ in the choice of the expansion parameter, the representation, and the presence or absence of a positive-definite expansion (Frobenius theorem). 

Path integral methods are based on a strong coupling expansion, which is an expansion in the kinetic energy. For bosonic systems this led to a sign-free expansion, but for fermionic or frustrated magnetic systems such an expansion is usually not sign positive. The sign problem is usually so bad (because of the scaling with the system volume) that such algorithms make no sense. This chapter is dedicated to methods based on an expansion in the two-body potential energy term $H_1$ of the Hamiltonian, or in $U$ for short, which offers new opportunities: (i) it is a standard perturbative many-body expansion which can be combined with textbook many-body techniques, and also formulated directly for real time and frequencies~\cite{Mahan, NegeleOrland, FetterWalecka}, (ii) the one-body problem $H_0$ is quadratic, so we can use Wick's theorem, (iii) the problem can be formulated for a finite lattice or immediately in the thermodynamic limit. On the negative side, (i) the series convergence is not guaranteed and this is problem and parameter dependent ({\it e.g.}, for bosons in the thermodynamic limit, a naive expansion in $U$ diverges because attractive bosons lead to a collapse), (ii) the sign problem, which is often needed for establishing series convergence, makes the evaluation of higher order diagrams problematic, (iii) technical aspects are more difficult than in the case of path integral Monte Carlo (e.g., storage of a four-point vertex function can be a real issue). 


Let's consider the statistical operator expressed in the real space - imaginary time representation,
\begin{equation}
\exp(- \beta H) = \exp(-\beta H_0) {\mathcal T} \exp \left( -\int_0^{\beta} d\tau H_1(\tau) \right),
\label{eq:diagram_stat_op}
\end{equation}
with $\beta$ the inverse temperature, the Heisenberg operator $H_1(\tau) = \exp(\tau H_0) H_1 \exp(- \tau H_0)$, and $\mathcal{T}$ the  time ordering operator.
Expanding Eq.~\ref{eq:diagram_stat_op} in powers of $H_1$, the partition function takes the form (for the Hubbard model)
\begin{eqnarray}
Z & = & \sum_{n=0}^{\infty} (-U)^n \sum_{x_1 \ldots x_n} \int_{0<\tau_1 < \tau_2 < \ldots < \beta} \prod_{j=1}^{n} d\tau_j {\rm Tr} \left[ \ldots \right] \nonumber \\
\left[ \ldots \right] & = & \left[ \e^{-\beta H_0} c_{\uparrow}^{\dagger} (x_j \tau_j) c_{\uparrow} (x_j \tau_j) c_{\downarrow}^{\dagger}  (x_j \tau_j) c_{\downarrow}(x_j \tau_j) \right].
\label{eq:diagram_series}
\end{eqnarray}
This expansion generates the diagrams that consist of the four-point vertices $U$ connected by single-particle propagators  $G_{\sigma}^{(0)} (x_i - x_j, \tau_i - \tau_j) = - {\rm Tr} [  \mathcal{T} e^{-\beta H_0} c^{\dagger}_{\sigma} (x_i \tau_i) c_{\sigma} (x_j \tau_j) ]$. The $p$-th order diagram is graphically given by a set of $(p!)^2$ possible connections of vertices by propagators shown in the top row of Fig.~\ref{fig:determinant}. Historically, this expansion was only used in connection with determinantal methods where the sign problem was either absent (due to an additional particle-hole symmetry) or manageable. This is discussed in Sec.~\ref{sec:determinant} and illustrated for the resonant Fermi gas in Sec.~\ref{sec:resonant_fermions}. In Sec.~\ref{sec:diagmc} we will see how, in cases the sign problem is too bad, a sampling over all diagrams can be tried, with applications for the Fr{\"o}hlich polaron, the Fermi polaron and the Hubbard model.
Also dynamical mean-field theory (DMFT) can be understood in this way (Sec.~\ref{sec:dmft}) and combined with diagrammatic Monte Carlo algorithms.

\subsection{Determinantal methods}
\label{sec:determinant}

\begin{figure}
\begin{center}
\includegraphics[angle=90, width=0.8\columnwidth]{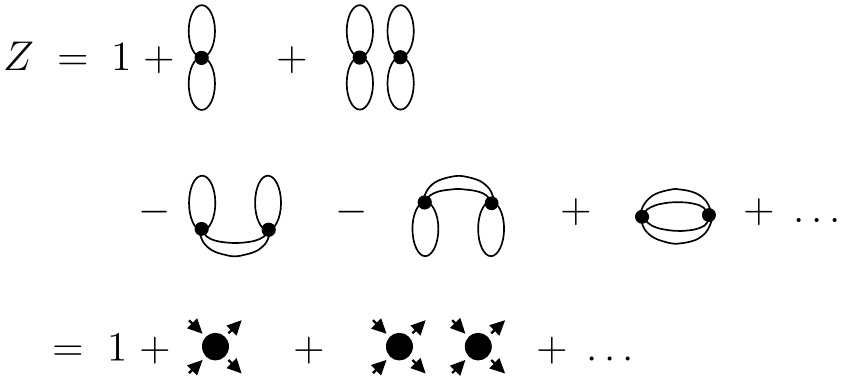}
\caption{Diagrammatic series for the partition function. The upper line is the graphical representation of the series Eq.~\ref{eq:diagram_series}, lower line depicts Eq.~\ref{eq:diagram_determinant}. The diagram signs are shown explicitly. The figure is taken from Ref.~\cite{Burovski2006b}. }
\label{fig:determinant}
\end{center}
\end{figure}

Determinantal methods are used to alleviate the sign problem of Eq.~\ref{eq:diagram_series}. In case of attractive fermions with equal spin population, or for repulsive fermions at half filling on a bipartite lattice (which has an additional particle-hole symmetry), the sign problem is absent. These methods are formulated on a finite lattice, so just like in path integral Monte Carlo there is no real symmetry breaking. Also in DMFT (see Sec.~\ref{sec:dmft}) the use of determinantal methods is widespread to solve the impurity problem.
Let's consider all diagrams of a given order $p$ with a fixed vertex configuration~\cite{Burovski2006, Burovski2006b}, 
\begin{equation}
\mathcal{S}_p = \left\{ (x_j, \tau_j), j = 1, \ldots p \right\},
\end{equation}
and sum over all $(p!)^2$ ways of connecting vertices with propagators. Then Eq.~\ref{eq:diagram_series} takes the form
\begin{equation}
Z = \sum_{p=0}^{\infty} (-U)^p  \sum_{x_1 \ldots x_n} \int_{0<\tau_1 < \tau_2 < \ldots < \beta} \prod_{j=1}^{n} d\tau_j \det A^{\uparrow}(\mathcal{S}_p) \det A^{\downarrow} (\mathcal{S}_p),
\label{eq:diagram_determinant}
\end{equation}
with $A^{\sigma}(\mathcal{S}_p)$ the $p \times p$ matrices whose elements are the single-particle propagators,  $A_{ij}^{\sigma} (\mathcal{S}_p) = G_{\sigma}^{(0)}(x_i - x_j, \tau_i - \tau_j), i,j = 1, \ldots p $.
For an equal number of spin-up and spin-down particles, $\det A^{\uparrow} \det A^{\downarrow} = \vert \det A \vert^2$ is positive. The Feynman diagrams are represented by a collection of vertices shown in the bottom row in Fig.~\ref{fig:determinant}. 

The simplest Monte Carlo scheme consists just of inserting and removing vertices and reevaluating the determinant (using so-called fast updates)~\cite{Rubtsov03, Rubtsov04, Rubtsov05, Burovski2004}, but also worm-type updates have been devised for the dilute gas regime in order to simultaneously measure the correlation function~\cite{Burovski2006b}. (note: Worm-type updates are here not related to winding numbers as in path integral Monte Carlo, but the term is used in the weaker sense of going to an extended configuration space where physical constraints are broken. By going back to the physical configuration space, all physical constraints are restored, and a configuration that is strongly decorrelated from the previous one is reached.) We refer to the cited papers for a detailed description of the algorithm. 

Since the single-particle propagators depend on the Manhattan distance between lattice sites, a finite lattice is used in these methods. A continuous-space version can also be formulated~\cite{Burovski2008}. Also note that such methods are useless for bosons because the evaluation of a permanent (instead of the determinant for fermions)  is a $\#P$-problem (note that this is circumvented by {\it sampling} over all bosonic permutations in path integral Monte Carlo).

\subsection{Application: The balanced spin$-1/2$ Fermi gas at unitarity}
\label{sec:resonant_fermions}

\begin{figure}
\begin{center}
\includegraphics[width=0.8\columnwidth]{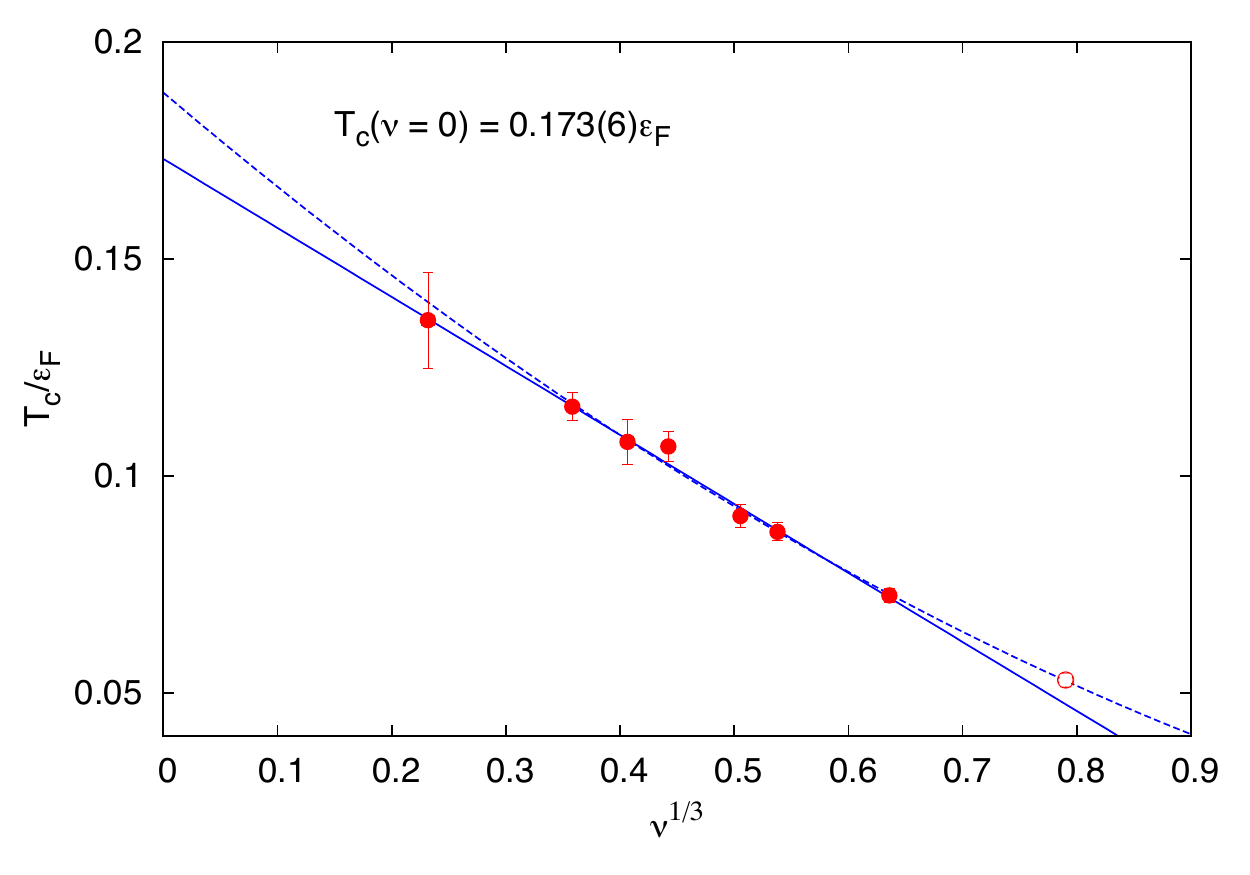}
\caption{The critical temperature $T_c$ in units of the Fermi energy $\epsilon_F$  versus filling factor $\nu$. The continuum limit corresponds to $\nu \to 0$. The linear extrapolation (solid line) of the seven data points at lowest filling factors (filled circles) yield a critical temperature $T_c/ \epsilon_F = 0.173(6)$. The dashed line corresponds to a quadratic fit through all data points. 
Reprinted figure with permission from Ref.~\cite{Goulko2010}. Copyright (2010) by the American Physical Society.}
\label{fig:detqmc_tc}
\end{center}
\end{figure}

For Fermi gases in the vicinity of a Feshbach resonance, the scattering length diverges.  Universal results are found in the zero-range resonant limit: when the effective range $r_0$ of the interaction goes to zero, $r_0 \to 0$, the $s$-wave scattering length remains finite, {\it ie,} $k_Fa$ remains fixed for $k_F r_0 \to 0$. The nature of the interaction potential is then irrelevant and the only remaining energy scale is the Fermi energy. The chemical potential is related to the Fermi energy by a universal number, $\mu = \xi \epsilon_F$, where $\xi$ is known as the Bertsch parameter. Up to this parameter, the  equation of state is the same as that of an ideal gas. A value smaller than one indicates attractive interactions. The unitary regime was studied in great detail in BCS-BEC crossover studies (for a review, see Ref.~\cite{Giorgini2008, Varenna}). We are focusing on the determination of the critical temperature at unitarity and the equation of state.

Using determinantal Monte Carlo simulations, Goulko and Wingate found the critical temperature at $T_c / \epsilon_F = 0.173(6)$~\cite{Goulko2010} using a linear extrapolation shown in Fig.~\ref{fig:detqmc_tc}. They put their final value at $T_c / \epsilon_F = 0.171(5)$, a  value a bit higher than the previous result from~\cite{Burovski2006, Burovski2006b}, where $T_c/ \epsilon_F = 0.152(7)$ was found with an identical method, illustrating the difficulties with the extrapolations.  However, the  continuous space version of the method~\cite{Burovski2008} found the same result as in~\cite{Burovski2006}. Bulgac {\it et al.} used an auxiliary field Monte Carlo approach (see Sec.~\ref{sec:auxiliary}) which extracted the critical temperature from the finite-size scaling of the condensate fraction, using the same procedure as in~\cite{Burovski2006, Burovski2006b}, found an upper bound $T_c / \epsilon_F \le 0.15(1)$ ~\cite{Bulgac2008}.  Previously,  the same group~\cite{Bulgac2006, Bulgac2007} claimed $T_c / \epsilon_F = 0.23(2)$. There are also results obtained with the restricted path integral Monte Carlo method~\cite{Akinenni2007}, $T_c / \epsilon_F \approx 0.245$. Results for the imbalanced case (which has a sign problem) are also available~\cite{Goulko2010}. We thus see that different Monte Carlo methods agree with each other, but others produced results for $T_c$ that vary over more than $70\%$, well outside error bars. This shows that interpretation of data, extrapolations, and providing correct error bars remain a hard task, for which the utmost care is needed.

Using diagrammatic Monte Carlo (see Sec.~\ref{sec:diagmc}), Van Houcke {\it et al.}~\cite{VanHoucke2011} determined the equation of state of the balanced Fermi gas at unitarity for temperatures up to 5 times $T_c$. The results were in excellent agreement with the MIT experiments~\cite{VanHoucke2011, Ku2011}  and in fair agreement with the ENS experiments~\cite{Nascimbene2010}. With this method, the difficult double extrapolation in system size and density of the lattice determinant methods~\cite{Burovski2006, Goulko2010} can be avoided, but an extrapolation in expansion order is needed (and these simulations are not sign-free). The MIT experiments put $T_c/T_F = 0.167(13) $~\cite{Ku2011}. 
Discrepancies of a similar magnitude existed for the chemical potential at unitarity. The MIT experiments find $\xi = 0.376(5)$~\cite{Ku2011} (in disagreement with the ENS experiments which found $\xi = 0.415(10)$).  The MIT value is consistent with the  upper bound $\xi < 0.383$~\cite{Forbes2011} and is close to $\xi = 0.36(1)$ from a selfconsistent T-matrix calculation~\cite{Haussmann2007} (cf. the good agreement between the selfconsistent T-matrix calculation and the exact result for the Fermi polaron problem at unitarity discussed in Sec.~\ref{sec:fermi_polaron}). It lies below the earlier estimates $\xi = 0.44(2)$~\cite{Carlson2003} and $\xi = 0.42(1)$ ~\cite{Astrakharchik2004} found in fixed-node quantum Monte-Carlo simulations which provide upper bounds.

\subsection{Sampling of  all Feynman diagrams}
\label{sec:diagmc}

Prokof'ev and Svistunov have introduced, in a number of different contexts, a diagrammatic Monte Carlo scheme with a scalar representation in which all diagrams are sampled instead of a determinant evaluated. The idea is straightforward though audacious: an algorithm is devised such that all Feynman diagrams (involving all topologies, all expansion orders and all allowed momenta and frequencies) are sampled in a Monte Carlo scheme. An advantage of Feynman diagrams over strong coupling expansions is the absence of symmetry factors and lattice embedding coefficients. Other  advantages of the method include the flexibility of the scheme (it is in principle applicable to any Hamiltonian), the connection with analytical tools, the possibility to directly formulate the method in the thermodynamic limit, and the possibility to work directly with real time or real frequencies, sidestepping the ill-conditioned analytical continuation problem inherent to methods formulated in imaginary time. The drawbacks are the lack of a convergence guarantee, the sign problem (meaning that only low expansion orders are accessible), and the fact that broken phases require a separate treatment. Prokof'ev and Svistunov argue that these drawbacks are however acceptable: The method does not try to alleviate the sign problem; in fact, the sign 'blessing' is often crucial for the series convergence. When the system is deep inside a well defined phase a few diagram expansion orders suffice to accurately describe the physics, and it is expected that the sign problem is still tolerable for these orders. Note that the sign problem does not scale with the system volume for these methods.

An often employed strategy is to reduce the space of the diagrams. While the method of Sec.~\ref{sec:determinant} sums up all possible diagrams for the full Green function, including disconnected ones, the present method focuses typically on the selfenergy $\Sigma(k,\omega)$ in combination with the Dyson equation,
\begin{equation}
G( k, \omega)^{-1} = G_0(k, \omega)^{-1} - \Sigma(k,\omega),
\end{equation}
In particular, disconnected diagrams should not be generated for the selfenergy. 

Instead of considering an expansion with bare propagators $G_0$, one can also consider a skeleton expansion in which the bare propagators are replaced by fully dressed propagators $G$, which should then be determined selfconsistently~\cite{Prokofev07_bold}.
This further reduces the space of diagrams which should be sampled (but does not prevent the exponential growth of the number of diagrams with expansion order) : if a diagram is one-particle reducible, {\it i.e.}, it contains a  subdiagram obtained by cutting two propagator lines and is disconnected otherwise from the rest of the diagram, then this diagram has to be discarded in a skeleton expansion. This scheme is known as bold diagrammatic Monte Carlo~\cite{Prokofev07_bold}, and requires a selfconsistency loop since the unknown propagator $G$ has to be determined selfconsistently. In practice, the method is started with bare propagators after which statistics for the selfenergy are collected. Then, a Dyson equation is performed (involving (fast) Fourier transforms if the coordinate and the (imaginary) time representation are used) and a new propagator is obtained. This procedure is  iterated until selfconsistency is reached. For the final run with a fixed propagator $G$ the usual Markov chain and Monte Carlo convergence properties hold. For the selfconsistency loop however there is always the possibility that a metastable solution is found (e.g., in the vicinity of a first order transition). This is inherent to any selfconsistency problem and not a property of a diagrammatic Monte Carlo process.

As previously mentioned, there is no mathematical guarantee that the series expansion in Eq.~\ref{eq:diagram_expansion} is convergent. The series may well be asymptotic or even divergent. Weak divergences can be overcome by applying resummation techniques. One constructs the partial sums up to the maximum expansion order $N_*$,
\begin{equation}
\Sigma(N_*) = \sum_{n=1}^{N_*} D_N F_N^{(N_*)},
\end{equation}
where the requirement on the factors $F_N^{(N_*)}$ is such that they approach unity for $N \ll N_*$ for large $N_*$ and suppress higher order contributions rendering the series $\sum_{N=1}^{\infty} D_N F_N^{(N_*)}$ convergent. The crossover region from where the function is close to unity to where it is approximately zero also has to increase with $N_*$. There are infinitely many way of satisfying these conditions and we list only a few examples: Ces\'aro-Riesz, Borel and Lindel{\"o}f resummation. Depending on the nature of the divergence, not all may work, but final results have to be independent of the choice of $F$ provided $F$ is strong enough to compensate the divergence~\cite{VanHoucke08, Prokofev07_bold}.

From the generality of the above discussion, the reader will appreciate the potential of these methods but also understand that they are still under development and that their true potential is only gradually being discovered. Ther are still many unexplored ideas which may turn this method into a versatile tool and which could lead to interesting research.
We now give a few examples of where these ideas have already been implemented successfully. For historical reasons, we start with Fr{\"o}hlich polarons, proceed with the Fermi-polaron problem, and finally show results for the Hubbard model. 

\subsubsection{Fr{\"o}hlich polarons} 

\begin{figure}
\begin{center}
\includegraphics[width=0.8\columnwidth]{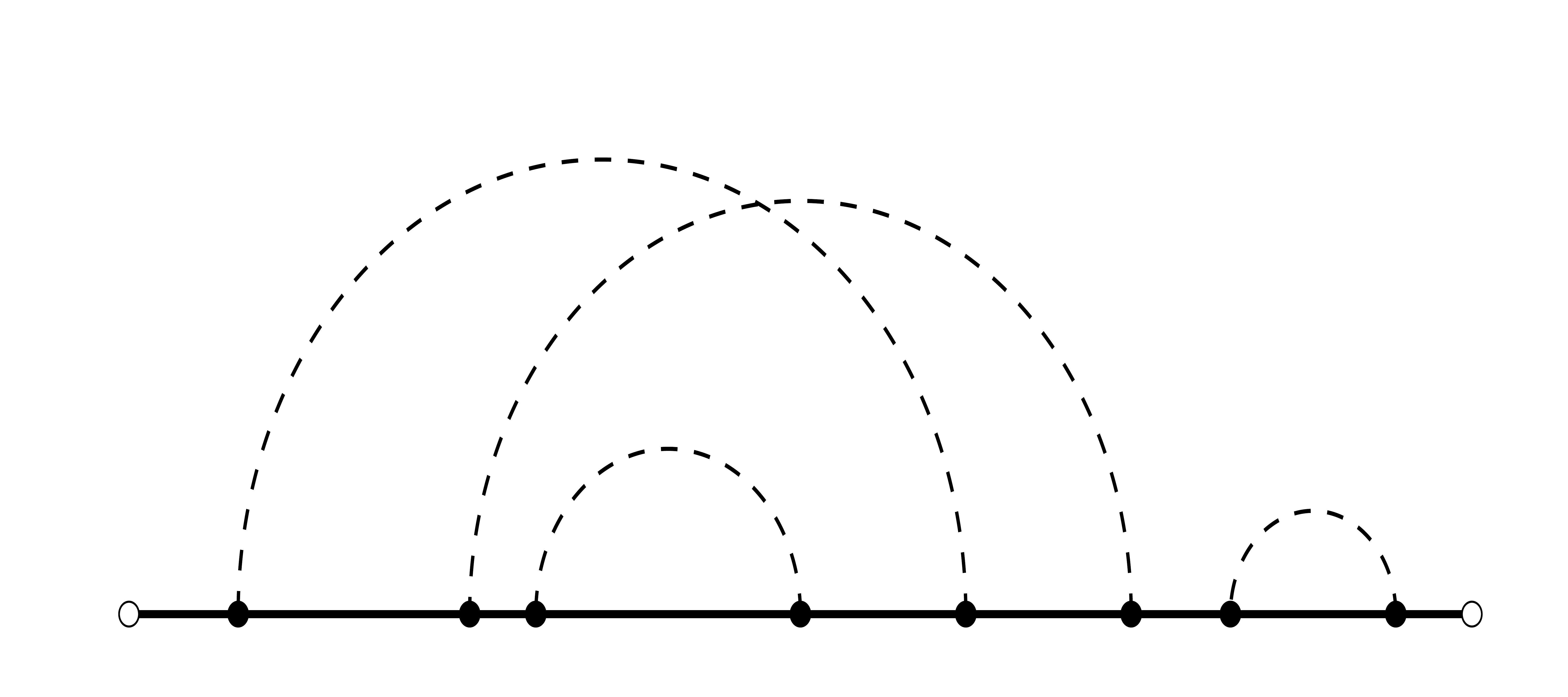}
\caption{A typical diagram contributing to the polaron Green function of the Fr{\"o}hlich Hamiltonian. Dashed lines denote phonon propagators, full backbone lines bare electron propagators. The length of the diagram is the total imaginary time of the Green function. In this linear representation, the first and last electron Green function are necessarily uncovered by phonon propagators. 
Reprinted figure with permission from Ref.~\cite{Prokofev98_froehlich}. Copyright (1998) by the American Physical Society.}
\label{fig:froehlich_diagram}
\end{center}
\end{figure}

The Fr{\"o}hlich Hamiltion describes optical phonons coupled to electrons via
\begin{equation}
H_{\rm e-ph} = \sum_{k,{\bm q}} V(\bm{q}) (b_{\bm q}^{\dagger} - b_{-{\bm q}}) a_{\bm {k-q}}^{\dagger} a_{\bm k},
\end{equation}
with $a_{\bm k}$ and $b_{\bm q}$ the electron and phonon annihilation operators with momenta $\bm{k}$ and $\bm{q}$, $V(\bm{q}) = i(2 \sqrt{2} \alpha \pi)^{1/2} / q$ the coupling strength and $\alpha$ a dimensionless coupling constant. The phonon propagator is independent of momentum, $D(\bm{q}, \tau) = \exp( - \omega_p \tau)$, with $\omega_p$ the frequency of the optical phonon. Electron propagators are given by $G_0(\bm{p}, \tau) = e^{- (p^2 / (2m) - \mu) \tau}$, with $\mu$ the chemical potential (here a tuning parameter). 

The expansion for the polaron Green function in terms of bare electron Green functions and phonon propagators turns out to be positive definite. A typical diagram is shown in Fig.~\ref{fig:froehlich_diagram}. An ergodic set of updates consists of just inserting and removing phonon propagators at arbitrary points in imaginary time, but other updates such as shifting the position in imaginary time of a vertex or changing the topology of the diagram by reconnecting the phonon propagators with different vertices can improve the sampling efficiency~\cite{Prokofev98_froehlich, Mishchenko00}. As always, one can think of many good ways of performing the sampling.

On a present-day laptop one can sample expansion orders up to $\sim 100$ in a couple of minutes with an accuracy of $\sim 1$\%. In Ref.~\cite{Mishchenko00} a cyclical representation was used, which allows for improved estimators.

The polaron Green function is the central quantity in this problem. It follows from the Lehmann representation that, if
\begin{equation}
G(\bm{k}, \tau \gg \omega_p^{-1}) \to Z_k \exp [ - (E(\bm{k}) - \mu) \tau ],
\label{eq:polaron_green}
\end{equation}
then for $k=0$ the energy $E_0$ is the ground state energy of the system and the factor $Z$ shows the fraction of the bare-electron state in the true eigenstate of the polaron. It also follows that one must choose $\mu < E_0$, but the closer it is tuned to $E_0$ the more accurate the exponential decay of the Green function can be resolved. 

Since the physics of electron-phonon interactions is beyond the scope of this review, we refer to Refs.~\cite{Prokofev98_froehlich, Mishchenko00} for a general physics discussion and quantitative results for the binding energy, effective mass, structure of the polaronic cloud, and the spectral analysis (after analytic continuation) for any coupling strength (both small and large polarons).  For completeness we mention that polarons in the Su-Schrieffer-Heeger model have also been studied in detail~\cite{Marchand10}.

\subsubsection{The Fermi-polaron problem in three dimensions}
\label{sec:fermi_polaron}

\begin{figure}
\begin{center}
\includegraphics[width=0.8\columnwidth]{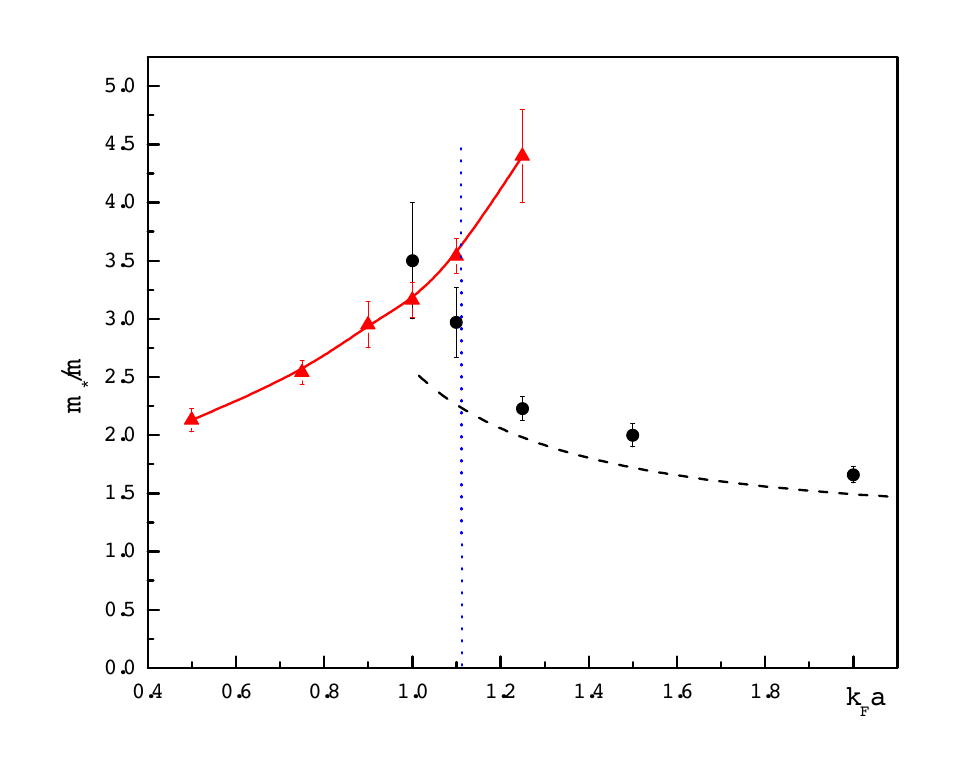}
\caption{Polaron (black circles) and molecule (red triangles) effective mass as function of $k_F a$. The vertical dotted line stands for $(k_F a)_c = 1.11$, where the polaron energy becomes lower than the molecular energy. The dashed line is the contribution from the first order diagram~\cite{Combescot07}. 
Reprinted figure with permission from Ref.~\cite{Prokofev08_fermipolaron}. Copyright (2008) by the American Physical Society.}
\label{fig:diagmc_fermipolaron}
\end{center}
\end{figure}

With 'Fermi polaron', a spin-down fermion resonantly interacting with a sea of non-interacting spin-up fermions is understood. It is an idealization of a Fermi mixture with strong imbalance. The Fermi polaron problem was crucial in understanding the difference between the Rice and MIT (and ENS) experiments for imbalanced Fermi gases. We refer to Ref.~\cite{Chevy2010} for a recent review on this interpretation, as well as for descriptions of the equation of state of imbalanced Fermi gases both on the BCS and the BEC side. We will restrict ourselves to the physics of the Fermi polaron problem at zero temperature.

For the Fermi polaron problem near unitarity, the nature of the interaction potential is irrelevant. Diagramatically, this implies that the sum  $\Gamma(\tau, p)$ of all ladder diagrams for the interaction potential has to be considered as a separate diagrammatic element, which also takes care of the UV divergences. A closed form for $\Gamma$ in universal form exists~\cite{Varenna},
\begin{eqnarray}
\Gamma^{-1}(\eta, {\bm p}) & = & \frac{m}{4\pi a} - \frac{m}{8 \pi} \sqrt{p^2 - 4 m \eta} - \bar{\Pi}(\eta, \bm{p}), \\
\bar{\Pi}(\eta, \bm{p}) & = & \int_{q \le k_F} \frac{d^3 \mathbf{q} }{ (2\pi)^3} \frac{1}{q^2/ (2m) + ({\mathbf{p}} -{ \mathbf{q}})^2/(2m) - \eta},
\end{eqnarray}
with $\eta = \omega + \epsilon_F + \mu + i0^{+}$. In Refs.~\cite{Prokofev07_fermipolaron, Prokofev08_fermipolaron} the T-matrix was calculated by applying bold diagrammatic Monte Carlo instead of converting the above formulae to the imaginary time domain.

Deep in the BEC limit (repulsive side) the impurity atom will form a polaron quasi-particle dressed by the majority fermions. Deep in the BCS limit (attractive) side, the impurity atom will form a bound state (a bosonic molecule) with a majority atom. Hence, one expects a transition between the polaronic and molecular regimes in the unitary regime.

On the BCS side, Chevy wrote down a variational Ansatz for the polaron wavefunction in the subspace of single particle-hole excitations created by the minority atom~\cite{Chevy2006},
\begin{equation}
\vert \psi \rangle = \left( \phi_0 d_0^{\dagger} + \sum_{\bm {k,q}} \phi_{\bm {k,q}} d^{\dagger}_{\bm{k} - \bm{q}} u_{\bm k}^{\dagger} u_{\bm{q}}  \right) \vert \rm{FS} \rangle,
\end{equation}
where the $\phi$ are variational parameters, the sum over $k$ ($q$) is restricted to be above(below) the Fermi surface (FS), $d$ annihilates the spin down atom and $u$ a spin up atom.
The quasiparticle dispersion can then be written  for small momenta ${\bm k}$ as
\begin{equation}
E_k^{\rm pol} = A E_F + \frac{k^2}{2m^*},
\label{eq:fermi_polaron_variational}
\end{equation} 
in analogy to the one for a free atom, but with renormalized (variational) parameters $A_{\rm var} \approx -0.6$ and $m^*_{\rm var} = 1.17m$ for $a \to \infty$~\cite{Chevy2006, Combescot07}. 

If a quasiparticle is well defined, the polaron energy and dispersion can be found in diagrammatic Monte Carlo simulations based on Eq.~\ref{eq:polaron_green}, while the molecular energy follows from a similar equation for the 2-particle propagator. In Refs.~\cite{Prokofev07_fermipolaron, Prokofev08_fermipolaron} the polaronic and molecular channel were  treated on equal footing. The diagrams consist of a backbone impurity propagator dressed by T-matrices. The T-matrices have  to be connected by majority fermion propagators in all possible ways, excluding (sub)diagrams that are already part of the T-matrix. A list of updates and the corresponding equations for detailed balance are described in detail in Ref.~\cite{Prokofev08_fermipolaron} and will not be repeated here. Expansion schemes based on bare and bold propagators were also discussed.

The transition between the polaron and molecule regime does not happen at unitarity, but at a slightly larger value $k_F a = 1.11(2)$~\cite{Prokofev07_fermipolaron, Prokofev08_fermipolaron} shown in  Fig.~\ref{fig:diagmc_fermipolaron}, where also the polaronic and molecular effective masses are shown. Diagrammatic Monte Carlo finds $A = -0.61(1)$ and $m^* = 1.20(1)m$ at unitarity. Variational fixed node Monte Carlo simulations find $A = -0.59(1)$ and $m^* = 1.09m$~\cite{Lobo06}. It is remarkable that the first order diagram on the polaronic side with selfconsistent polaron propagators~\cite{Chevy2006, Combescot07} (equivalent to the variational Ansatz) is surprisingly close to the full answer due to a remarkable cancellation of higher order diagrams. Note that the equation of state of a resonant Fermi gas on the BCS side is accurately described by a gas of polarons~\cite{Chevy2010}. On the other side of the polaron-to-molecule transition, the composed boson is interacting with the Fermi sea with a surprisingly accurate mean-field energy $g_{\rm ad} n_{\uparrow}$, where the atom-dimer coupling $g_{\rm ad}$ is related to the atom-dimer scattering length, $a_{\rm ad} = 1.18a$~\cite{STM}. Eq.~\ref{eq:fermi_polaron_variational} was extended to the molecular sector, where the dimer is dressed by single particle-hole pairs of the majority Fermi sea~\cite{Mora09, Punk09, Combescot09}. 

Experimentally, the parameter $A$ can be determined from rf-spectroscopy. This requires that the imbalance between majority and minority atoms is large enough so that no superfluid core is formed. The MIT experiments found values for $A$ in agreement with the Monte Carlo predictions~\cite{Schirotzek09}. The transition between the molecular and the polaronic transitions could not be determined because this transition is preempted by phase separation between the ideal Fermi gas and the polarized molecular superfluid. The effective mass can be determined from collective mode excitations. This was done in Ref.~\cite{Nascimbene09} where $m^* = 1.17(10)m$ was found, in close agreement with the Monte Carlo values.


\subsubsection{Hubbard model}

\begin{figure}
\begin{center}
\includegraphics[width=0.8\columnwidth]{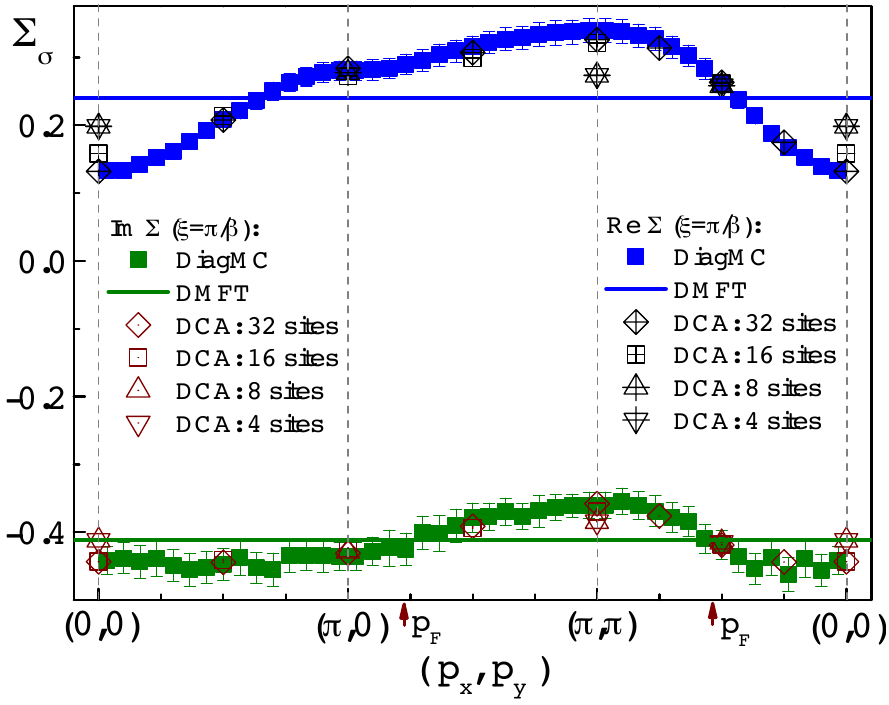}
\caption{Momentum dependence of the selfenergy at the lowest fequency for the 2D Hubbard model with $t=1, U=4,  \mu = 3.1,$ and $T=0.4$. Comparison is made with single-site DMFT and a cluster DMFT method, DCA. The mean-field contribution (the Hartree term $Un_{\sigma} = 2.3t$ ) was subtracted. Arrows indicate the position of the Fermi momentum $p_F$. The figure is taken from Ref.~\cite{Kozik2010}. }
\label{fig:diagmc_hubbard}
\end{center}
\end{figure}

For the Hubbard model, only a bare expansion for the selfenergy was explored thus far~\cite{Kozik2010}. This was done for $U/t  \le 4$ in the Fermi liquid regime, for temperatures down to $T/t = 1/40$. The momentum-dependent selfenergy is shown for the lowest frequency in Fig.~\ref{fig:diagmc_hubbard}, where comparison is made with the dynamical cluster approximation (DCA) method, see Sec.~\ref{sec:dmft}. A momentum cluster of size 32 is needed to obtain good agreement. Diagrams up to order 8 were generated, which is a limit set by the sign problem. Going to larger values of $U$ remains topic of investigation and will require a bold diagrammatic Monte Carlo approach, the use of a T-matrix (cf. Sec.~\ref{sec:fermi_polaron}), the combination with DMFT~\cite{Pollet_Anderson}, or a combination thereof.

\subsection{Continuous-time Auxiliary-Field methods}
\label{sec:auxiliary}

Auxiliary-field methods use a Hubbard-Stratonovich decomposition of the two-body propagator into a set of propagators of one-body potentials containing an auxiliary variable which should be integrated over. Discrete-time versions of this idea, such as the Blankenbecler-Sugar-Scalapino algorithm~\cite{Blankenbecler81} and the Hirsch-Fye algorithm~\cite{HirschFye86} algorithm have been developed long ago (cf.~\cite{Assaad2001}). 
The extrapotlation in the discrete time step is then mandatory, but can often be done reliably. Another advantage is the linear scaling of the discrete methods with $\beta$ (compared to $\beta^3$ for the continuous models). A fair comparison would also need to include the increase of the autocorrelation times, but no such systematic studies exist.
The first continuous-time method for fermionic lattice models was an auxiliary-field decomposition scheme presented by Rombouts {\it et al.}~\cite{Rombouts99} and applied to nuclear matter and small Hubbard lattices. It was reformulated 10 years later by Gull {\it et al.}~\cite{Gull08, Gull_review}, whose formulation we now follow.

A non-zero constant is added to the two-body part,
\begin{equation}
H_U = U \sum_i n_{\uparrow} n_{\downarrow} - K / \beta.
\end{equation}
By expanding the exponential in powers of $H_U$ and applying the auxiliary field decomposition~\cite{Rombouts99}
\begin{equation}
1 - \frac{\beta U}{K} \sum_i \left( n_{i\uparrow} n_{i\downarrow}  - \frac{n_{i\uparrow} + n_{i\downarrow}}{2} \right) = \frac{1}{2V} \sum_{i, s_i} e^{\gamma s_i ( n_{i\uparrow}  - n_{i \downarrow} )},
\end{equation}
where the last term on the left hand side denotes the usual shift in chemical potential, and with
\begin{equation}
\cosh(\gamma) = 1 + \frac{U \beta V}{2K},
\end{equation}
one arrives at a partition function written solely in one-body propagators 
\begin{eqnarray}
Z & = & \sum_{n=0}^{\infty} \prod_{j=1}^{n} \int_{\tau_{j-1}}^{\beta} d\tau_j \sum_{s_j = \pm 1} (\frac{K}{2 \beta V} )^n Z_n ( \{ s_j, \tau_j , x_j \} ), \nonumber \\
Z_n( \{ s_j, \tau_j, x_j \} )  & = &  {\rm Tr} \prod_{i=n}^{1}  \exp( - \Delta_i H_0) \exp(s_i \gamma (n_{x_j, \uparrow} - n_{x_j, \downarrow} ) ),
\end{eqnarray}
with $\Delta_i = \tau_{i+1} - \tau_i$ for $i<n$ and $\Delta_n = \tau_1 + \beta - \tau_n$. The time arguments are continuous variables and not regularly spaced on $[0 , \beta [$, unlike the discrete-time methods~\cite{Blankenbecler81, HirschFye86}.
Generalizing the derivation provided in Ref.~\cite{Georges1992} the weights are expressed as
\begin{eqnarray}
\frac{Z_n ( \{ s_j, \tau_j , x_j \} ) }{Z_0} = \prod_{\sigma} \det N_{\sigma}^{-1} ( \{ s_j, \tau_j , x_j \} ) \nonumber \\
 \det N_{\sigma}^{-1} ( \{ s_j, \tau_j , x_j \} ) \equiv e^{V_{\sigma}^{ \{ s_j \}} }  - G_{0 \sigma} ^{ \{ \tau_j, x_j \} } (  e^{V_{\sigma}^{ \{ s_j \}} }  - 1) \nonumber \\
e^{V_{\sigma}^{ \{ s_j \}} } \equiv {\rm diag} \left( \e^{\gamma (-1)^{\sigma} s_1} , \ldots , e^{\gamma (-1)^{\sigma} s-k}  \right), 
\end{eqnarray}
with the notations $(-1)^{\uparrow} = 1, (-1)^{\downarrow} = -1,$ and equal time evaluations are taken as $\tau = 0^{+}$. Rombouts used a fixed length representation, which leads to additional combinatorial factors in the weights.
The simplest possible set of updates that fulfill ergodicty are inserting and removing new auxiliary variables $\sigma$, which can be balanced against each other. If the new time and site are chosen arbitrarily, and if the auxiliary variable is also chosen uniformly among all variables for the reverse update, then the acceptance factor reads
\begin{equation}
R = \frac{K}{n+1} \frac{\det N_{\uparrow}(y) \det N _{\downarrow}(y) }{\det N_{\uparrow}(x) \det N _{\downarrow}(x)},
\end{equation}
where $x$ denotes the old configuration, $y$ the configuration after inserting the new auxiliary variable, and $n$ the expansion order in configuration $x$.
Efficient numerical schemes for such updates are discussed elsewhere~\cite{Rombouts99, Gull08, Gull_review}.

An application for this method is given in Sec.~\ref{sec:DCA_Hubbard} in combination with DMFT. We also draw attention to one other application, namely enhanced Pomeranchuk cooling schemes of a SU($2N$) ultra-cold fermionic in optical lattices at half filling~\cite{Cai2012} because of the large number of hyperfine-spin components.

\subsection{Dynamical Mean-Field Theory (DMFT)}
\label{sec:dmft}

Dynamical Mean-Field Theory (DMFT) provides an approximate solution to a many-body problem, unlike the other methods covered in this paper. There are many excellent introductory texts and reviews on DMFT available, which goes beyond the scope of this review~\cite{Georges1992, Maier2005, Kotliar2006, Gull_review}. DMFT  has however a close connection to diagrammatic methods, which we want to make clear. Like in any mean-field theory, it considers a single site coupled selfconsistently to the rest of the lattice,
but it additionally retains dynamical  information (retardation effects). In the diagrammatic language, DMFT sums up all skeleton diagrams contributing to the selfenergy built with purely local propagators.
Technically, this is done by mapping the many-body problem onto an impurity problem, for which efficient numerical procedures are known to solve it~\cite{Gull_review}.
There exist apart from approximate methods (such as the iterative perturbation theory (IPT) and the non-crossing approximation (NCA)), also controllable and/or exact methods such as exact diagonalization, the numerical renormalization group, and Monte Carlo solvers (such as the weak-coupling expansion method (see Sec.~\ref{sec:determinant}), the auxiliary field method (see Sec.~\ref{sec:auxiliary}), and a strong coupling expansion method~\cite{Werner2006}). They are all discussed in detail in Ref.~\cite{Gull_review}. Crucially, for a single site impurity problem, the sign problem is absent. 

\subsubsection{Formalism}

The basic set of equations, specified here for the Hubbard model, can be formulated as follows. Let's introduce the full local Green function obtained by integrating $G({\bf k},i \omega)$ over the Brillouin zone,
\begin{equation}
G_{\rm loc}(i \omega) = \int \frac{d {\bf k}}{ (2\pi)^d} G( {\bf k}, i \omega).
\label{eq:dmft_Gloc}
\end{equation}
We also introduce the functional integral representation of the partition function,
\begin{equation}
Z = \int \mathcal{D} \psi_{\sigma}^{\dagger} \mathcal{D}\psi_{\sigma} e^{-S^{\rm imp}},
\end{equation}
where the impurity action in imaginary time representation reads,
\begin{eqnarray}
S^{\rm imp} & = & \int_0^{\beta} d\tau \psi_{\sigma}^{\dagger} (\tau) (\partial_{\tau} - \mu ) \psi_{\sigma}(\tau) + U n_{\uparrow}(\tau) n_{\downarrow}(\tau)  \nonumber \\
{} & {} & + \int_0^{\beta} d\tau \int_0^{\beta} d\tau' \psi_{\sigma}^{\dagger}(\tau) \Delta(\tau - \tau') \psi_{\sigma}(\tau'),
\label{eq:dmft_impurity_action}
\end{eqnarray}
with $\Delta(\tau - \tau')$ an unknown 'hybridization' function. 
DMFT consists of solving the impurity problem $\Sigma^{\rm imp}[G_{\rm loc} (i \omega)]$ as a functional of the full local Green function such that the selfconsistency equation
\begin{equation}
G(\bm{k}, i\omega)^{-1} = G_0({\bm k}, i\omega)^{-1} - \Sigma^{\rm imp}(i\omega),
\label{eq:dmft_G}
\end{equation}
is fulfilled. Here, $G_0(\bm{k}, i \omega) = (i \omega - \mu + \epsilon_k )^{-1}$ is the non-interacting Green function of the many-body problem, with $\epsilon_k$ the dispersion.
There are many ways to solve a selfconsistency equation, but in practice an iteration scheme is used: From an initial guess for the impurity selfenergy, a first guess for the (full) Green function is obtained via Eq.~\ref{eq:dmft_G} and the local Green function $G_{\rm loc}$ is computed with Eq.~\ref{eq:dmft_Gloc}. 
The sum over the Brillouin zone in Eq.~\ref{eq:dmft_Gloc} is usually replaced by a one-dimensional integral over the density of states. The hybridization function is determined from the Dyson equation for the impurity problem, $\Delta(i \omega)^{-1} = G_{\rm loc}^{-1}(iw) + \Sigma^{\rm imp}(i \omega)$. 
The impurity problem for a given hybridization function $\Delta$ must then be solved, and a new $\Sigma_{\rm imp}$ is obtained, after which the scheme is repeated until convergence is reached. In this iteration process, the hybridization function (which is just an auxiliary function irrelevant for the underlying Hubbard model) is also determined. 


DMFT is a widely used method and still being developed  further. We mention two main directions: cluster dynamical mean field theory both for real-space clusters and momentum clusters (we refer to Ref.~\cite{Maier2005} for an extensive review) and the dual fermion approach, in which the irreducible vertex can be treated by using a dual set of variables~\cite{Rubtsov2008}. It has been used in electronic structure calculations in combination with density functional theory methods~\cite{Kotliar2006}. 
Close connections between cellular ({\it i.e.}, a real-space cluster) DMFT and cluster perturbation theory have also been revealed in the framework of self-energy functional methods ~\cite{Potthoff2003a, Potthoff2003b}.

 \subsubsection{Application: cluster DMFT for the 3D Hubbard model}
 \label{sec:DCA_Hubbard}
 
 \begin{figure}
\begin{center}
\includegraphics[width=0.8\columnwidth]{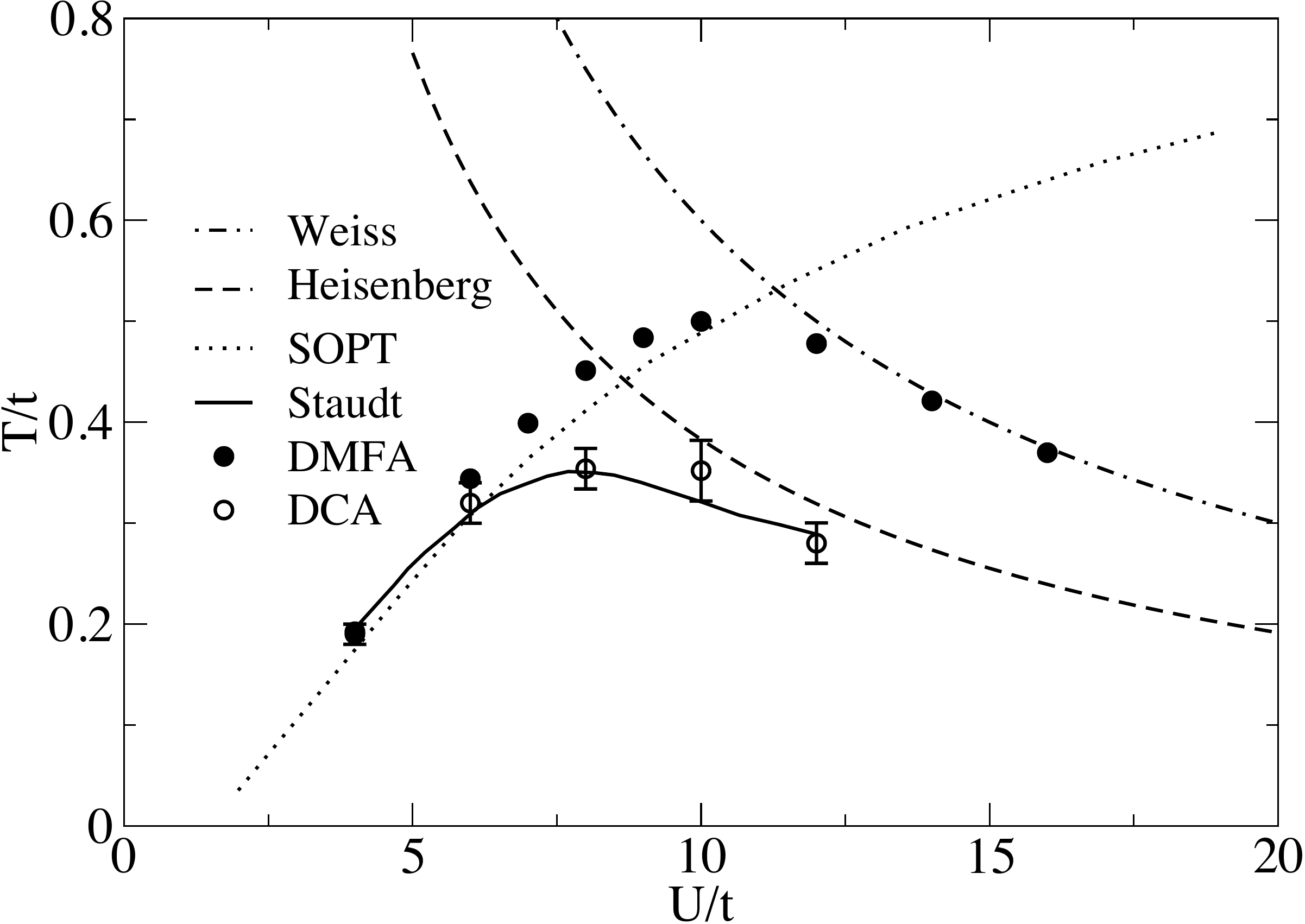}
\caption{Phase diagram of the 3D Hubbard model at half filling. Shown are the predictions form second order perturbation theory (SOPT), Weiss mean-field theory, the Heisenberg limit prediction,  dynamical mean-field theory (DMFA), dynamical cluster approximation (DCA)~\cite{Kent2005}, and the determinantal lattice quantum Monte Carlo simulations by Staudt {\it et al.}~\cite{Staudt2000}.    
Reprinted figure with permission from Ref.~\cite{Kent2005}. Copyright (2005) by the American Physical Society.}
\label{fig:hubbard_halffilling}
\end{center}
\end{figure}

\begin{figure}
\begin{center}
\includegraphics[width=0.8\columnwidth]{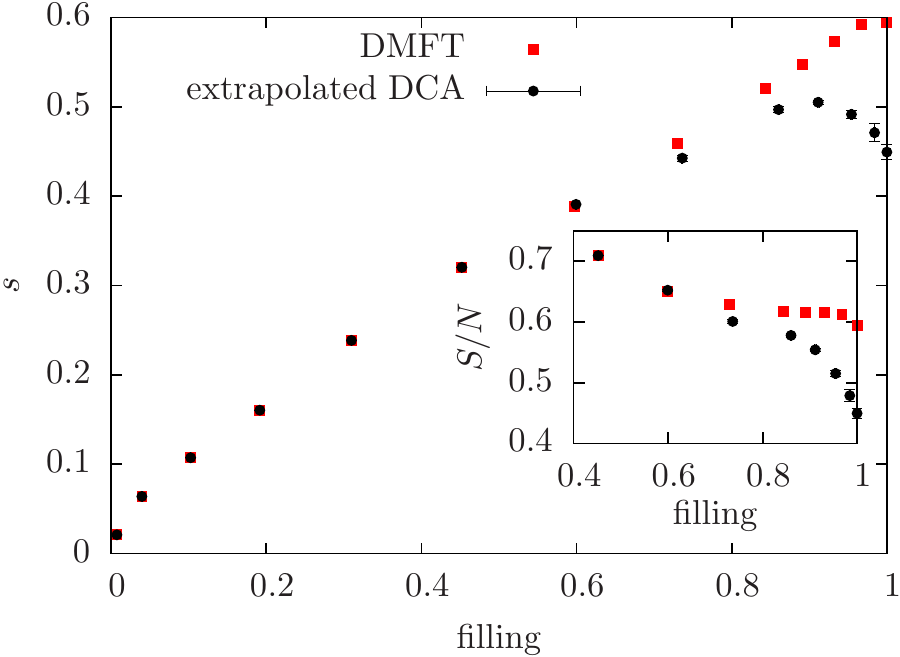}
\caption{Entropy per volume as a function of filling for a 3D Hubbard model at $T = 0.35t \approx T_N$ for $U/t = 8$. The inset shows the entropy per particle. Shown are the single site DMFT results and the DCA results extrapolated in the cluster size. The error bars are dominated by the extrapolation.  
Reprinted figure with permission from Ref.~\cite{Fuchs2011}. Copyright (2011) by the American Physical Society.}
\label{fig:dca_hubbard}
\end{center}
\end{figure}

Since the first experimental signatures of the Mott insulator in  the 3D Fermi Hubbard model have been observed~\cite{Joerdens2008, Schneider2008},  the questions of when and how to reach the antiferromagnetic transition became important, as well as how to observe such a phase~\cite{DeLeo2008, McKay2011}. The phase diagram at half filling, shown in Fig.~\ref{fig:hubbard_halffilling}, was known from the sign-free auxiliary field Monte Carlo calculation of Ref.~\cite{Staudt2000}, which was confirmed by a cluster DMFT (DCA) calculation by Kent {\it et al.}~\cite{Kent2005}. The latter showed that so-called periodic Betts clusters~\cite{Betts1997, Betts1999} of size 48 were as accurate as the simulations on  real-space clusters of size up to $L=10^3$~\cite{Staudt2000}, and are optimal for finite size scaling.

The lowest entropies per particle in present experiments are about $S/N \approx \log(2)$, but typically they are a bit higher, about $S/N \approx 1.2$~\cite{Joerdens2010}. At these temperatures, the local physics becomes a very good approximation: DMFT and high temperature series expansions~\cite{Oitmaa} give essentially identical answers~\cite{DeLeo2011, Fuchs2011, Scarola2009}.

The entropy per spin at the N{\'e}el temperature in the Heisenberg model (that is, the infinite $U$ case at half filling in the Hubbard model) is $S/N = 0.35$~\cite{Werner2005, Wessel2010}; about half of $\log(2)$. Can this value be substantially higher at lower values of $U$?

Fuchs {\it et al.} provided the full thermodynamics of the 3D Hubbard model for temperatures approaching the N{\'e}el temperature (without breaking the symmetry), for any filling, and interaction strengths up to $1.5$ times the bandwidth by using DCA with an auxiliary-field Monte Carlo impurity solver. From entropy curves such as the one shown in Fig.~\ref{fig:dca_hubbard} they could construct in the local density approximation the total entropy of a trapped system. They found that the maximal critical entropy per particle $S/N = 0.65(6)$ is found for $U/t = 8$, about 1.5 times as high as without a parabolic trapping potential, $S/N = 0.41(3)$ for $U/t=8$. So, just as in the case of 3D bosons, the liquid in the edges acts as a big entropy reservoir, primarily because of larger volume fractions in the edges compared to the middle of the trap. The value of the critical entropy was only weakly dependent on the value of $U/t$. We note that such entropies are nowadays easily reached in bosonic lattice systems. Fuchs {\it et al.} also found that the double occupancy changes little with temperature, while the nearest-neighbour spin-spin correlation functions show a stronger signal around the N{\'e}el temperature. All these findings were confirmed in Ref.~\cite{Paiva2011}, who employed a determinantal lattice Monte Carlo algorithm, and extrapolated in lattice size and Trotter time discretization step. They also went to lower temperatures inside the broken phase. In Ref.~\cite{Gorelik2010} a real-space extension of single site DMFT was used, focusing on the double occupancy, which for large values of $U/T$ shows a strong increase in the broken phase when temperature is lowered from $T_N$ to zero. In two dimensions, the temperature and entropy scales for observing antiferromagnetism have also been determined~\cite{Paiva2010}.

\subsubsection{Incorporating DMFT in diagrammatic Monte Carlo}

\begin{figure}
\begin{center}
\includegraphics[width=0.8\columnwidth]{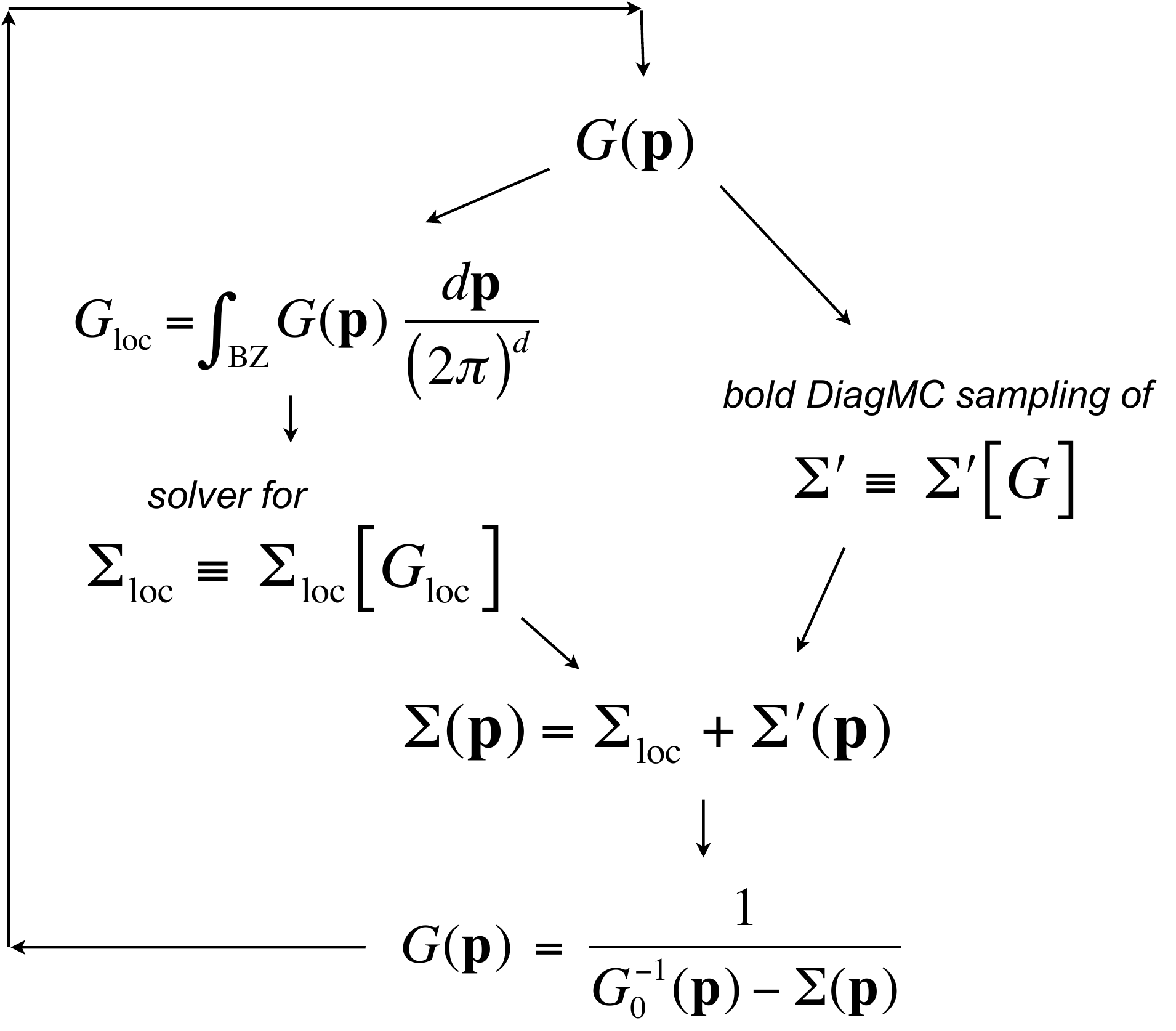}
\caption{Scheme illustrating how DMFT can be used to construct locally fully bold propagators as initial propagators for a general diagrammatic Monte Carlo simulation. The Matsubara frequency index has been suppressed to ease the notation. From a certain knowledge of the full Green function $G({\bm{p}})$ the local Green function is constructed (left). The impurity problem is then solved (it is assumed that this can be done in an efficient way), providing the selfconsistent solution for the local selfenergy. On the right, diagrammatic Monte Carlo samples all non-local diagrams contributing to the selfenergy. The local and non-local contributions to the selfenergy are then merged. The Dyson equation then gives us more information on the full Green function. This scheme is repeated until convergence is reached. 
Reprinted figure with permission from Ref.~\cite{Pollet_Anderson}. Copyright (2011) by the American Physical Society.}
\label{fig:anderson}
\end{center}
\end{figure}


By explaining DMFT in terms of Feynman diagrams, as done here and in Ref.~\cite{Kotliar2006}, it follows that it is possible to combine DMFT with bold diagrammatic Monte Carlo: Since DMFT sums up all local skeleton diagrams of the self energy, it can be used to construct initial 'bold' propagators for diagrammatic Monte Carlo, as is illustrated in Fig.~\ref{fig:anderson}. This was demonstrated in Ref.~\cite{Pollet_Anderson} for Anderson's model of localization~\cite{Anderson58}. It is a one-particle quantum  effect where destructive interference between all possible paths of the particle scattering off impurities can localize the particle.  Anderson localization depends strongly on the dimension. In one dimension the wave function is always localized. In three dimensions, there is a mobility edge separating extended states from localized ones.  Because of the single-particle Hamiltonian, Anderson localization is a problem that can be diagonalized and does not have an exponentially growing Hilbert space. In practice, the lattice sizes that can be fully diagonalized are small, and Anderson localization remains a tricky problem since it has no small parameter.

In Ref.~\cite{Pollet_Anderson} non-interacting fermions on a 3D lattice  with on every site a chemical potential distributed according to a (quenched) delta-correlated Gaussian distribution (which facilitates the diagrammatic technique) was studied. Spatial coordinates on a lattice and real time at zero temperature were used. The topology of the diagrams for the Green function built from bare propagators is then the same as in Fig.~\ref{fig:froehlich_diagram}.  The DMFT part reduces in this case to an algebraic equation and provides a solution very close to the correct answer: The magnitude of the contributions to the selfenergy that are not built from purely local propagators, is very small. Up to 50 expansion orders could be sampled. For strong disorder, expansion times up to $3-4$ times the hopping could be reached; for weak disorder expansion times up to $10$ times the hopping could be reached.
In order to find the mobility threshold, vertex corrections are needed. This has not been implemented yet for this model.
Another obvious extension of this idea would be to combine DMFT with diagrammatic Monte Carlo for the Hubbard model in the parameter regime of large values of $U$, so that the local physics can be summed up from the outset in the DMFT loop (also within the DMFT framework $U$ can be made non-local in imaginary time). Whether this approach would allow to provide controlled error bars for large values of $U$ has not been tried yet.

The cold gas community has also shown interest in Anderson localization in order to demonstrate localization with matter waves. The first experiments were done around 2007 with bosons in one dimension without lattice at low enough density such that interactions are negligible~\cite{Billy08}. The disorder is generated with optical speckles. The longitudinal trap is switched off, and the BEC starts expanding. Then the expansion rapidly stops, and the density in the wings is exponential, typical for Anderson localization~\cite{Billy08, SanchezPalencia07}. At the same time there were also experiments performed with one-dimensional quasi-periodic lattices (the Aubry-Andr{\'e} model), a system which features a crossover between extended and exponentially localized states~\cite{Roati08}. Localization was demonstrated by investigating transport properties, and spatial and momentum distributions. Anderson localization has also been studied with bosons in 3D~\cite{Jendrzejeweski11} and with fermionic, spin-polarized $^{40}$K atoms~\cite{Kondov11}. In the latter experiment, the cloud has a mobile component that expands ballistically, but more rapidly than a thermal gas. It also has a localized component that becomes fixed after a rapid initial expansion. A mobility edge was defined as the energy below which particles are localized. Although it increases with the disorder strength, it does not follow the self consistent Born prediction or predictions from weak scattering theory. We note that many aspects of this experiment remain unexplained, but in the absence of a small parameter, any theoretical description is difficult. It remains to be seen if diagrammatic Monte Carlo can provide more insight into this problem.

\subsubsection{Comments on real-time dynamics}

Non-equilibrium dynamics remains extremely difficult to describe accurately numerically. For one-dimensional systems, time-adaptive DMRG can be used~\cite{White2004, Daley2004}, but this fails in higher dimensions. We refer to Ref.~\cite{Eckstein2010} for a review of the non-equilibrium flow equation method and DMFT. It also contains a discussion of an interaction quench in the Hubbard model with DMFT using Monte Carlo methods. Instead of a rapid thermalization, an intermediate prethermalized state was found. Gull {\it et al.} have formulated a diagrammatic Monte Carlo method on the Keldysh contour for impurity models~\cite{Gull2011}. They pre-summed the class of non-crossing diagrams, sampled corrections to it, and could describe long-time and steady-state properties over a wide range of interaction strengths. 

\subsubsection{Bosonic DMFT}
\label{sec:bdmft}

\begin{figure}
\begin{center}
\includegraphics[width=0.8\columnwidth]{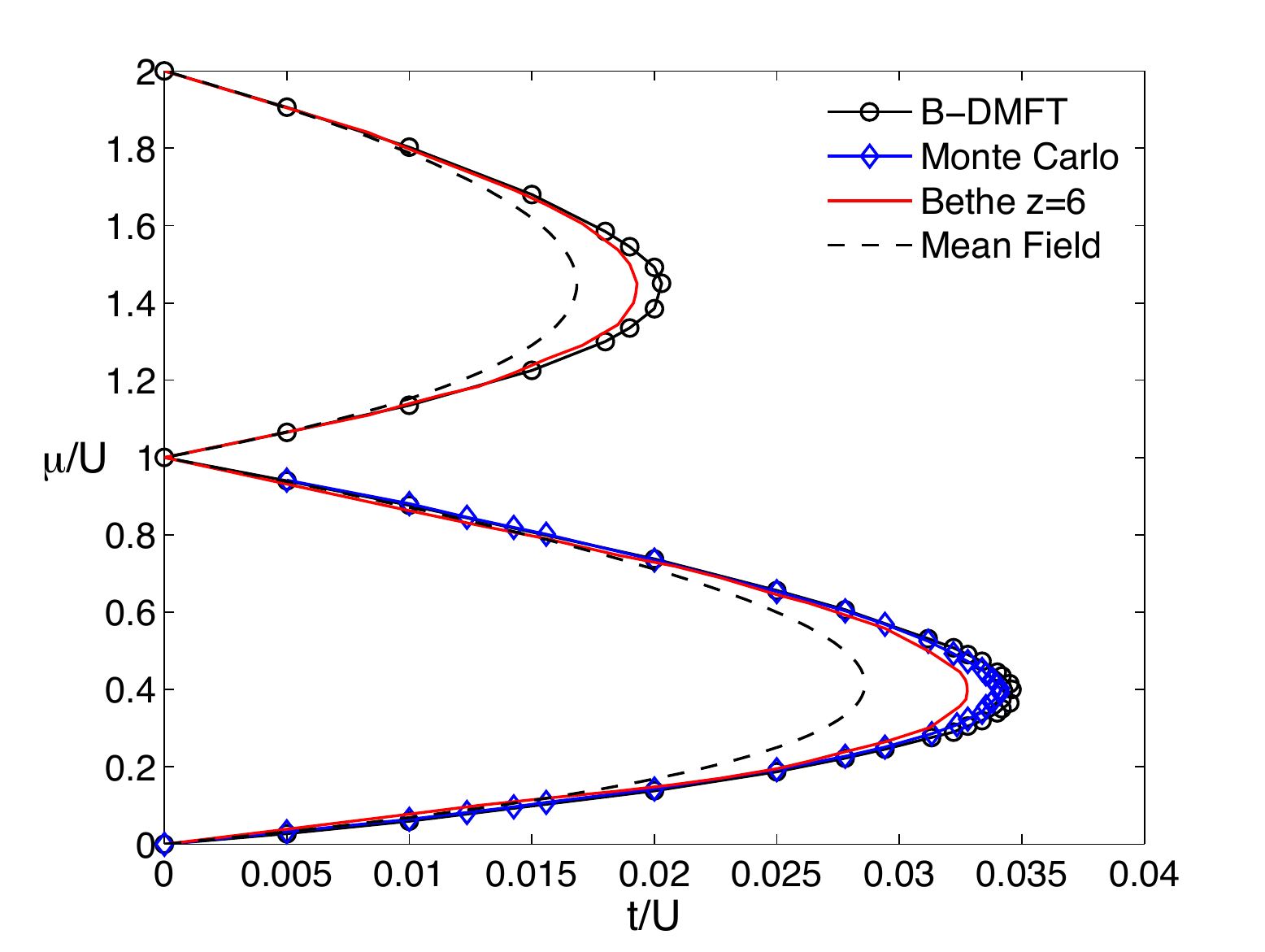}
\caption{Ground state phase diagram of the 3D Bose-Hubbard model. The ground state phase diagram in computed in the mean-field approximation ('Mean Field')~\cite{Fisher1989}, exactly with worm-type Monte Carlo path integral simulations ('Monte Carlo')~\cite{CapogrossoSansone07}, and using B-DMFT with the density of states of a cubic lattice ('B-DMFT') and on a Bethe lattice with coordination number 6 ('Bethe $z=6$'). 
Reprinted figure with permission from Ref.~\cite{Anders2010}. Copyright (2010) by the American Physical Society.}
\label{fig:bdmft}
\end{center}
\end{figure}

The bosonic dynamical mean-field theory (B-DMFT) can be developed along the same lines as in Sec.~\ref{sec:dmft}. The only point of attention is that bosons can condense, which requires a similar extension of DMFT as in Ref.~\cite{Chitra2001} (so-called EDMFT). As in any diagrammatic method, studying the condensed phase $\langle b \rangle  = \phi \neq 0$ requires that one allows to break the symmetry by writing a generalized action for the impurity problem. The action consists of all the local terms (chemical potential and potential energy terms), to which the mean-field contribution is added, $-\kappa \phi \int d \tau (b(\tau) + b^{\dagger}(\tau))$ for a real, homogeneous and time-independent condensate, where $\kappa = zt$ is the coordination number times the hopping amplitude. Recalling the theory of the weakly interacting Bose gas,  the introduction of normal and anomalous propagators is needed in the presence of a condensate. Switching to the Nambu-Gor'kov notation, $\bm{b^{\dagger}} = (  b^{\dagger}, b)$ and $\bm{\Phi} = (\phi, \phi)$, the hybridization part in the action can be written as,
\begin{equation}
S_{\rm hyb} = -\frac{1}{2} \int_0^{\beta} \int _0^{\beta} d\tau d\tau' (\bm{b^{\dagger}}(\tau) - \bm{\Phi})  \bm{\Delta}(\tau - \tau') (\bm{b}(\tau) - \bm{\Phi}),
\end{equation}
where $\Delta_{11}(\tau) = \Delta_{22}(-\tau)$ and $\Delta_{12}(\tau) = \Delta_{21}(\tau)$ are real functions describing the normal and anomalous hybridization functions.
The above terms linear in $\Phi$ can be combined with the mean-field decoupling changing $\kappa$ to $\kappa = zt - \Delta_{11}(i \omega = 0) - \Delta_{12}(i \omega_n = 0)$ and bringing the hybridization action into a form suitable for Monte Carlo simulations since it contains the full $\bm{b}$ (and not $\bm{b}$  - $\bm{\Phi}$). 
A Monte Carlo solver based on a strong coupling expansion was introduced in Ref.~\cite{Anders2010} for the impurity problem, which allows to compute the condensate, normal and anomalous Green function. The B-DMFT equations are closed by setting $\phi = \langle b \rangle$ calculated on the impurity site, extracting the connected Green functions (the ones for the depleted bosons) and selfenergies, from which a new hybridization matrix can be computed by the same selfconsistency relations as for fermions (a $2 \times 2$ matrix needs to be inverted because of the Nambu-Gor'kov formalism). The value of $\kappa$ also needs  to be updated in every iteration step. Because the expansion is done in the hybridization and condensate terms, a sign problem occurs in the condensed phase because of the opposite sign of anomalous and normal Green functions. Full details of the algorithm can be found in Ref.~\cite{Anders2011}.

The above formalism works for any dispersion and interaction, and is stable for any phase. The phase diagram of the 3D Bose-Hubbard model was calculated with a precision better than $2\%$, as is shown in Fig.~\ref{fig:bdmft}. It does not describe universal behaviour of the phase transition correctly (as expected), but  local physics are extremely well captured. Other results include the finding of non-universal critical exponents, the successful study of the weakly interacting Bose gas, the failure of the Hugenholtz-Pines relation, and the derivation of the  DMFT equations in three different ways, as well as the connection with cavity methods for Bethe structures (see Ref.~\cite{Zamponi} and the appendix in Ref.~\cite{Anders2011}). For completeness, we mention that other authors claimed 
 the correct B-DMFT formalism earlier~\cite{Byczuk2008, Hubener09_twocomponent,  Hu2009}, despite apparent differences in formalism with Refs.~\cite{Anders2010, Anders2011}, and failing to publish results in the broken phase for arbitrary system parameters~\cite{Byczuk2008}. The importance of the development of B-DMFT lies in possible extensions to real-time dynamics, Bose-Fermi mixtures, spinful bosonic systems with spin $F \ge 2$, and perhaps as a starting point for more general diagrammatic Monte Carlo simulations of bosonic systems. Just as in the fermionic case, there is a close connection with variational cluster approximations~\cite{Knap2011, Arrigoni2011}, which have been extended to non-equilibrium dynamics formulated on the Keldysh contour already~\cite{Knap2011b}. Two-component bosonic mixtures were also studied in the DMFT approximation (but with an exact diagonalization solver for the impurity problem) in Refs.~\cite{Li2011a, Li2011b} .


\section{Diffusion Monte Carlo}
\label{sec:diffusion}

In the last section of this review we switch to diffusion Monte Carlo. It is not a diagrammatic method, although it shares some similarities with path integral Monte Carlo: in diffusion Monte Carlo a number of walkers are propagated forward in imaginary time in order to project out the ground state. The method has been reviewed in detail in Refs.~\cite{Bajdic2009, Kolorenc2011} and in relation to cold atoms in Ref.~\cite{Giorgini2008}. We will therefore be rather schematic for the method, and focus on a single application, namely the controversial issue of Stoner ferromagnetism with cold fermionic atoms on the upper branch of the Feshbach resonance. For the BCS-BEC crossover which was studied much more intensely with this method, we refer to Ref.~\cite{Giorgini2008}. 

\subsection{Methods}

The position of the $N$ walkers is given by $\bm{R} = (\bm{r}_1, \ldots, \bm{r}_N)$, defined at every time $\tau = j \delta \tau, j = 1,2, \ldots$ where $\delta \tau$ is the time step. The ground state satisfies $\psi_0({\bm{r}}) = \lim_{j \to \infty} \langle \delta(\bm{r}_j - \bm{r}) \rangle$. The method is used almost exclusively in combination with importance sampling; one defines $f(\bm{R}, \tau) = \psi_T(\bm{R}) \psi(\bm{R}, \tau)$ as the product of the wave function $\psi$ with a time-independent trial wave function $\psi_T({\bm{R}})$. The trial wave function encodes physical knowledge we have about the system before the start of the simulation. It is not unique, but the closer it is to the true (unknown) ground state the more the simulation is enhanced. It should not be orthogonal to the ground state. The Schr{\"o}dinger equation in imaginary time for $f(\bm{R}, \tau)$ is
\begin{eqnarray}
- \frac{ \partial f( {\bm{R}}, \tau)}{\partial \tau} & = & - \frac{\hbar^2}{2m} \big( \nabla_{\bm{R}}^2 f(\bm{R}, \tau) - \nabla_{\bm{R}} [ \bm{F}(\bm{R}) f(\bm{R}, \tau)] \nonumber \\
{} & {} & + (E_L - E_{\rm ref}) f(\bm{R}, \tau) \big),
\end{eqnarray}
with $E_L(\bm{R}) = \psi_T(\bm{R})^{-1} H \psi_T({\bm{R}})$ the local energy, $E_{\rm ref}$ a reference energy introduced to stabilize the numerics, and $\bm{F}({\bm{R}}) = 2 \psi_T^{-1}({\bm{R}}) \nabla_{\bm{R}} \psi_T({\bm{R}})$ the quantum drift term. The energy can be calculated as 
\begin{equation}
E = \frac{\int d\bm{R} E_L(\bm{R}) f(\bm{R}, \tau \to \infty)}{\int d\bm{R} f( \bm{R}, \tau \to \infty)}.
\end{equation}
In every step, walkers propagate according to the drift term and a random diffusive term with variance $\hbar \delta \tau / m$. The potential energy is then evaluated, which modifies the weight of the walker. In order not to spend computer time on configurations with walkers that have an exponentially small weight, a killing and rebirth step of walkers is built in the algorithm in such a way that the average number of walkers remains constant~\cite{Bajdic2009, Kolorenc2011}. 
(note: the bias coming from the size of the population has never been systematically addressed~\cite{Cuervo2005}. There exist however diffusion Monte Carlo variants such as PIGS (path integral ground state) that do not suffer from a finite population bias, which are close in spirit to path integral Monte Carlo (at finite temperature) and which have  superior convergence properties than diffusion Monte Carlo~\cite{Sarsa2000, Cuervo2005, Rossi2009}.  Such a basic algorithm (with an appropriately chosen positive trial wave function) suffices for a bosonic system, and may be considered an alternative to path integral Monte Carlo. In our opinion path integral Monte Carlo is preferable since the bias coming from the trial wave function is not easy to filter out in practice. Another ground state method is reptation Monte Carlo which has been formulated both on the continuum~\cite{Baroni1999} and on the lattice~\cite{Carleo2010}. Methods such as PIGS and reptation Monte Carlo have not found widespread use in the field of cold gases however.
For fermions, the infamous sign problem occurs again~\cite{Troyer2005}. In such cases, the nodal surface is built into the trial wave function such that $f(\bm{R}, \tau) > 0$. Walkers should then (ideally) not cross the nodal surface, where the drift term is infinite and pushes the walkers away. Diffusion Monte Carlo is in such cases a variational method: if the exact knowledge of the nodal surface were known, the exact ground state energy can be found, while for any approximation of the nodal surface the obtained energy will be higher than the true ground state energy~\cite{Reynolds82}. There also exist methods with nodal relaxation.
It is crucial to obtain a good trial wave function and a good nodal surface. The most general trial wavefunction used in the studies of fermionic cold gases has the form $\psi_T({\bm{R}}) = \psi_J(\bm{R}) \psi_{\rm BCS}(\bm{r})$, with the Jastrow from $\psi_J(\bm{R}) = \prod_{i,j} f_{\sigma, \sigma'}( \vert \bm{r}_{i \sigma} - \bm{r}_{j \sigma'} \vert)$ describing the short-range correlations between particles of different spins at different positions, and the BCS part taken as a Slater determinant of orbitals $\phi(\bm{r}) = \alpha \sum_{k_{\alpha} < k_{\rm max}} e^{i \bm{k}_{\alpha}\cdot \bm{r}} + \phi_s(\bm{r})$, where $\alpha$ is a variational parameter (see Ref.~\cite{Giorgini2008, Astrakharchik05} and references therein). This can describe Fermi liquid regimes $(\phi_s = 0)$ and s-wave paired phases $(\phi_s \neq 0)$.  In practice, simulations are done for $N=14$ to $N=64$ particles and results are then extrapolated to the thermodynamic limit.



\subsection{Application: the Stoner model with atoms on the repulsive branch}

\begin{figure}
\begin{center}
\includegraphics[width=0.8\columnwidth]{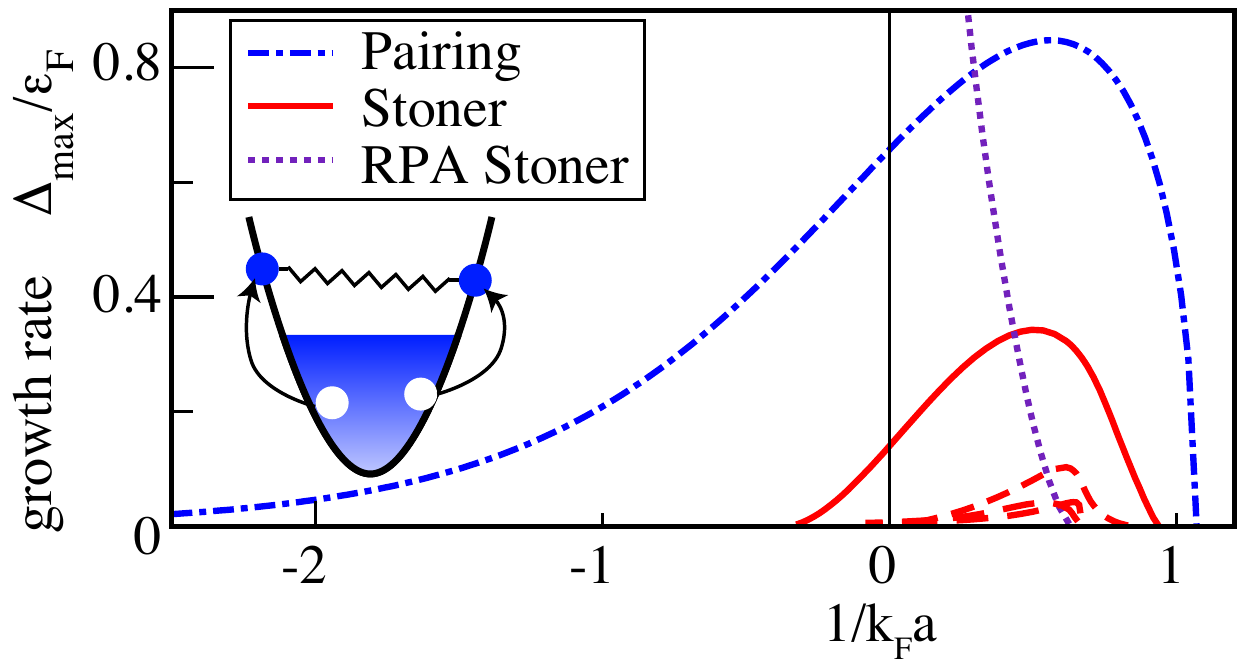}
\caption{Growth rate of the pairing and Stoner ferromagnetic instabilities after a quench as a function of the final interaction strength $1/k_F a$. Final interactions with negative (positive) values of $1/k_F a$ correspond to the BCS (BEC) side of the Feshbach resonance. The Stoner instability simultaneously occurs in multiple channels. The most unstable channel is indicated by the solid red line, the others by dashed red lines. The "RPA Stoner" instability corresponds to the RPA result~\cite{Babadi09} with bare as opposed to Cooperon-mediated interactions~\cite{Pekker2011}. Inset: Schematic diagram of the pair creation process showing the binding energy (spring) being absorbed by the Fermi sea (arrows). 
Reprinted figure with permission from Ref.~\cite{Pekker2011}. Copyright (2011) by the American Physical Society.}
\label{fig:Stoner}
\end{center}
\end{figure}

Itinerant ferromagnetism, known from the transition metals such as Co, Ni and Fe, is explained in textbooks in terms of the  Stoner mean-field criterion~\cite{Stoner}: when the density of states at the Fermi level times the coupling $U$ becomes unity then the RPA spin susceptibility has a pole signaling a transition to a ferromagnetic phase. However, the mean-field treatment breaks down for strong interactions and mean-field is thus applied beyond its range of validity. The Stoner picture is consequently not without criticism. Kanamori argued that screening should be taken into account, which may even prevent the transition~\cite{Kanamori}. The Stoner criterion also predicts a transition in one dimension where it violates the Lieb-Mattis theorem~\cite{LiebMattis}. Although ferromagnetism is known since ancient times the basic model remains partly unsolved. A quantum simulation of the Stoner model with cold gases would hence answer a fundamental question in condensed matter physics.

The equilibrium state of cold atoms swept across the Feshbach resonance to the repulsive side is however a gas of BEC molecules. The only possibility to observe the Stoner transition is to quench the atoms across the transition such that they remain on the upper branch of the Feshbach resonance. A ferromagnetic state is then created dynamically if the rate at which ferromagnetic correlations develop is sufficiently faster than the rate of molecule formation, {\it i.e.,} there might be a window where spin domains can be observed because of the two different time scales for the competing pairing and magnetic instabilities. 

There have been a number of theoretical studies of the repulsive two-component Fermi gas~\cite{Houbiers1997, ZhangDasSarma2005, Duine2005, Conduit2009a, Conduit2009b, Conduit2009c} assuming the (meta)stability of the repulsive gas. Mean field predicts a second order transition for $k_F a = \pi/2$ for a homogeneous system, with $k_F = (2 \pi^2 (n_{\uparrow} + n_{\downarrow}))^{1/3}$. Studies including next order corrections predict a smaller value of the critical density ($k_F a = 1.054$) and a discontinuous jump in the magnetization.

The MIT experiments on $^6$Li obtained indirect evidence of a Stoner transition~\cite{Jo2009}: a minimum in the kinetic energy, a maximum in the volume and a maximum of atom loss rate (around $k_F a \approx 2.2$, larger than the mean-field prediction). No spin domains, which would be direct evidence, were observed however.

Monte Carlo simulations can only be done in thermal equilibrium. In Ref.~\cite{Pilati2010} several interaction types are used: hard spheres, soft spheres and attractive square well potentials. The absence of the molecular bound states for the latter are implemented by choosing the Jastrow correlation term to be the scattering solution of the square well potential corresponding to positive energy. They calculated the equation of state for the unpolarized gas and found resuls independent of the interaction potential only for $k_F a < 0.4$. When the energy exceeds the energy of a phase separated gas, the gas is unstable to ferromagnetism. A partially ferromagnetic phase was also found in case of spin imbalance. The quantitative determination of the phase diagram is strongly model dependent.  In Ref.~\cite{Chang2010} model-specific backflow corrections, known to be important from electron gas and $^3$He studies, were added to the plane wave orbitals used to construct the  Slater determinants in the trial wave function. They substantially lowered the energy of a hard-sphere gas, but mattered less for the upper branch. We recall that for $^3$He sophisticated trial wave functions have been developed~\cite{Holzmann06}, otherwise wrong phases (overestimating polar fluids) and bad quantitative agreement with experiment is found. We also note that the variational and diffusion Monte Carlo simulations in 2D of Ref.~\cite{Drummond2011} found no ferromagnetic fluid between the paramagnetic fluid and a crystalline structure for hard-sphere interactions. Adding a $r^{-3}$ interaction did not significantly alter the phase diagram.

The competing magnetic and pairing instabilities were studied in a linear stability analysis of collective excitations~\cite{Pekker2011}. When the Cooperon is taken into account, it can be seen in Fig.~\ref{fig:Stoner} that the pairing instability is dominating over the ferromagnetic instability on both sides of the resonance. In such a study the minimum in kinetic energy is also found in the vicinity of the maximum of the pairing rate. It seriously questions the interpretation of the MIT experiments in terms of a Stoner transition. Other papers also question the validity of the MIT interpretation~\cite{Zhai2009, ZhangHo2011, Barth2011}, the latter claiming that ferromagnetism on the upper branch with zero range interactions violates the Tan relations in combination with a variational argument for a gas in equilibrium.

The MIT experiments were repeated and improved with a faster change of the scattering length and explicitly measuring spin fluctuations~\cite{Sanner2011}. No domains were visible however, even domains as small as consisting of 5 spins were absent. The molecule formation occurs very fast on a scale of $10 /\epsilon_F$ which is accompanied by strong local heating. The new MIT experiments rule out the existence of a ferromagnetic metastable phase  in agreement with ~\cite{Pekker2011, ZhangHo2011}.  Hence, a ferromagnetic phase will have to be to specially prepared or sought in other systems, for instance with a narrow resonance, or with different dispersions or for  mass or spin imbalance~\cite{Conduit2010a, Conduit2010b, Keyserlingk2011}.

On the lattice, ferromagnetism is well established. Adding a single hole to the Hubbard model at half filling for $U = \infty$ on a bipartite lattice leads to Nagaoka-ferromagnetism~\cite{Nagaoka}. The ferromagnetic phase extends to finite doping, but has a very low critical temperature ( of the order of a percent of the Fermi energy at most) and is very  sensitive to the dispersion according to a DMFT study~\cite{Park07}.  Similar conclusions are found in diffusion Monte Carlo (see Ref.~\cite{Carleo2011} for the latest study and references therein for older work).

As a final application of diffusion Monte Carlo we mention that is a tool of preferene to compute exchange functionals used in density functional theory. For cold gases, this was done and combined with the Kohn-Sham equations in Ref.~\cite{Tama_DFT}, where the issue of ferromagnetism in a weak optically lattice was also studied.

%


\section{Conclusion}\label{sec:conclusion}

We have given an overview of the interplay between quantum simulation in the traditional sense by performing simulations on classical computers of quantum problems, and quantum simulation in the atomic-physics-quantum-optics meaning where an experiment simulates a prototypical model of condensed matter physics which is intractable numerically. We have provided a roadmap showing how different expansions lie at the heart of different types of algorithms, and provided references to the literature for a detailed description of each algorithm. We have seen how large-scale path-integral Monte Carlo simulations of bosonic systems have culminated in excellent agreement between theory and experiment for up to a million particles at experimentally relevant temperatures. For systems with long-range interactions and particles with an internal spin degree of freedom, questions remain however.
For fermionic systems, no method with a positive-definite expansion exists. One resorts then to approximations such as density mean-field theory (DMFT), or tries to sample all possible Feynman diagrams (diagrammatic Monte Carlo) and hopes for fast convergence, possibly after analytical manipulations such as series resummations. We have seen examples of both (large scale DCA simulations in order to provide benchmarking for  the 3D Hubbard model and the Fermi-polaron problem, to name just a few), and we showed that DMFT methods form  a subclass of diagrammatic Monte Carlo methods from the diagrammatic point of view. DMFT, thanks to its widespread use and technical advantages, can hence be used as a promising starting point for diagrammatic Monte Carlo, which seems an interesting avenue for future research. We looked at diffusion Monte Carlo in relation to the controversial issue of ferromagnetism for atoms on the upper branch of a Feshbach resonance.\\

{\it Acknowledgements} I wish to thank all my teachers, coworkers and colleagues who selflessly explained me the physics of cold gases and Monte Carlo simulations over the past decade. In particular, I wish to mention Peter Anders, Waseem Bakr, George Batrouni, Gianni Blatter,  Immanuel Bloch (and coworkers), Massimo Boninsegni, Hanspeter B{\"u}chler, Evgeni Burovski, Barbara Capogrosso-Sansone, Philippe Corboz, Andrew Daley, Eugene Demler, Tilman Esslinger (and coworkers), Simon F{\"o}lling, Sebastian Fuchs, Fabrice Gerbier, Thierry Giamarchi, Markus Greiner (and coworkers), Daniel Greif, Emanuel Gull, Fabian Hassler, Sebastian Huber, Gregor Jotzu, Wolfang Ketterle, Corinna Kollath, Evgeny Kozik, Stefan Kuhr, Anatoly Kuklov, Ping Nang Ma, David Pekker, Trey Porto, Sebastiano Pilati, Guido Pupillo, Nikolay Prokof'ev, Stefan Rombouts, Anders Sandvik, Vito Scarola, Radjeep Sensarma, Manfred Sigrist, Ulrich Schneider, Ulrich Schollw{\"o}ck, Boris Svistunov, Leticia Tarruell, Stefan Trotzky, Matthias Troyer, Felix Werner, Philipp Werner,  Wilhelm Zwerger, and Martin Zwierlein (and coworkers). This work was partly supported by a grant from the Army Research Office with funding from the DARPA OLE program. \\


\end{document}